\documentclass[twocolumn]{aastex62}

\usepackage[utf8]{inputenc}
\usepackage{tablefootnote}
\usepackage{natbib}
\usepackage{amsmath}
\usepackage{threeparttable}
\usepackage{xspace}
\usepackage{graphicx}
\usepackage{longtable}
\usepackage{multirow}

\newcommand{\rearth}{$R_\oplus$\xspace}
\newcommand{\Teff}{$T_\mathrm{eff}$\xspace}

\newcommand{\vsini}{$v$sin$i$\xspace}

\graphicspath{{./}{figures/}}

\accepted{\aj}

\shorttitle{Three Terrestrial Planets Transiting L 98-59}
\shortauthors{Kostov et al.}

\begin{document}

\title{The L 98-59 System: Three Transiting, Terrestrial-Sized Planets Orbiting a Nearby M-dwarf}



\author[0000-0001-9786-1031]{Veselin B. Kostov}
\affiliation{NASA Goddard Space Flight Center, Greenbelt, MD 20771, USA}
\affiliation{SETI Institute, 189 Bernardo Ave, Suite 200, Mountain View, CA 94043, USA}
\author[0000-0001-5347-7062]{Joshua E. Schlieder}
\affiliation{NASA Goddard Space Flight Center, Greenbelt, MD 20771, USA}

\author[0000-0001-7139-2724]{Thomas Barclay}
\affiliation{NASA Goddard Space Flight Center, Greenbelt, MD 20771, USA}
\affiliation{University of Maryland, Baltimore County, 1000 Hilltop Cir, Baltimore, MD 21250, USA}

\author[0000-0003-1309-2904]{Elisa V. Quintana}
\affiliation{NASA Goddard Space Flight Center, Greenbelt, MD 20771, USA}
\author[0000-0001-8020-7121]{Knicole D. Col\'on}
\affiliation{NASA Goddard Space Flight Center, Greenbelt, MD 20771, USA}
\author[0000-0002-2072-6541]{Jonathan Brande}
\affiliation{NASA Goddard Space Flight Center, Greenbelt, MD 20771, USA}
\affiliation{University of Maryland, College Park, MD 20742, USA}
\affiliation{Sellers Exoplanet Environments Collaboration}
\author[0000-0001-6588-9574]{Karen A.\ Collins}
\affiliation{Harvard-Smithsonian Center for Astrophysics, 60 Garden St, Cambridge, MA, 02138, USA}
\author[0000-0002-9464-8101]{Adina~D.~Feinstein}
\affiliation{Department of Astronomy and Astrophysics, University of
Chicago, 5640 S. Ellis Ave, Chicago, IL 60637, USA}
\author{Samuel Hadden}
\affiliation{Harvard-Smithsonian Center for Astrophysics, 60 Garden St, Cambridge, MA, 02138, USA}
\author{Stephen R. Kane}
\affiliation{Department of Earth and Planetary Sciences, University of California, Riverside, CA 92521, USA}
% ...

\begin{abstract}

We report the Transiting Exoplanet Survey Satellite (TESS) discovery of three terrestrial-sized planets transiting L 98-59 (TOI-175, TIC 307210830) -- a bright M dwarf at a distance of 10.6 pc. Using the Gaia-measured distance and broad-band photometry we find that the host star is an M3 dwarf. Combined with the TESS transits from three sectors, the corresponding stellar parameters yield planet radii ranging from 0.8${R_\oplus}$ to 1.6${R_\oplus}$. All three planets have short orbital periods, ranging from 2.25 to 7.45 days with the outer pair just wide of a 2:1 period resonance. Diagnostic tests produced by the TESS Data Validation Report and the vetting package {\textsf {DAVE}} rule out common false positive sources. These analyses, along with dedicated follow-up and the multiplicity of the system, lend confidence that the observed signals are caused by planets transiting L 98-59 and are not associated with other sources in the field. The L 98-59 system is interesting for a number of reasons: the host star is bright ($V$ = 11.7 mag, $K$ = 7.1 mag) and the planets are prime targets for further follow-up observations including precision radial-velocity mass measurements and future transit spectroscopy with the James Webb Space Telescope; the near resonant configuration makes the system a laboratory to study planetary system dynamical evolution; and three planets of relatively similar size in the same system present an opportunity to study terrestrial planets where other variables (age, metallicity, etc.) can be held constant. L 98-59 will be observed in 4 more TESS sectors, which will provide a wealth of information on the three currently known planets and have the potential to reveal additional planets in the system. 

\end{abstract}


\section{Introduction} \label{sec:intro}
The Transiting Exoplanet Survey Satellite \citep[TESS, ][]{Ricker2015}, a near all-sky transit survey that began science operations July 2018, is expected to find thousands of planets. This includes hundreds of small planets with radii ${\rm R<4\,R_\oplus}$, around nearby, bright stars \citep{Barclay2018, Huang2018a}. During the 2-year primary mission, TESS will monitor more than 200,000 pre-selected stars at 2-min cadence and will observe additional targets spread over most of the sky ($\approx$85\%) in 30-min cadence Full-Frame-Image (FFI) mode \citep{Ricker2015}. The spacecraft carries four identical wide-field cameras that combine to produce a nearly continuous $24^{\circ}\times96^{\circ}$ field-of-view (FOV). TESS uses this large FOV to observe thirteen partially overlapping sectors per ecliptic hemisphere, per year and started its survey in the southern ecliptic hemisphere. The spacecraft observes each sector for two consecutive orbits that cover an average time baseline of 27.4 days\footnote{The orbital period of TESS is not constant due to 3-body gravitational interactions between TESS, the Earth, and the Moon. This leads to slightly different baselines in each sector.}. The increasing overlap of sectors toward the ecliptic poles provides Continuous Viewing Zones (CVZs) surrounding the poles where targets receive $\approx$350 days of coverage. The long observing duration of the TESS CVZs will enable the detection of smaller and longer period planets. It will also overlap with the CVZs of the James Webb Space Telescope (JWST), providing key targets for detailed characterization. In about a hundred days of observations, TESS has already identified more than a hundred planet candidates, provided key observations to confirm several new planets, and provided new data on known transiting systems \citep{Huang2018,Gandolfi2018,Vanderspek2018,Wang2018,Nielsen2018,Shporer2018,Dragomir2019,Quinn2019,Rodriguez2019}.

Planets discovered around bright, nearby stars provide ideal targets for mass measurements via Doppler spectroscopy, emission and transmission spectroscopy for atmospheric characterization, and for precise stellar characterization. Multi-planet systems provide an additional layer of information on planet formation and evolution, orbital dynamics, planetary architectures \citep[e.g.][]{Lissauer2011,Fabrycky2014}, and in some cases mass measurements via transit-timing variations \citep[e.g. ][]{Hadden2016,Hadden2018}. While NASA's Kepler and K2 missions successfully discovered thousands of planets around stars in the Kepler field and in the vicinity of the ecliptic plane \citep[e.g.][]{Rowe2014,Morton2016,Livingston2018}, TESS will perform a nearly all-sky survey focused on stars in the solar neighborhood and find the touchstone planets that will be prime targets for observations with the Hubble Space Telescope (HST), JWST, and future ground-based observatories \citep{Louie2018,Kempton2018}.  

Here we report the TESS discovery of three small planets transiting the bright (K = 7.1 mag), nearby (10.6 pc) M3-dwarf L 98-59. This paper is organized as follows. In Section \ref{sec:TESS} we describe the TESS observations and data analysis, as well as our ground-based follow-up efforts. In Section \ref{sec:sys_param} we discuss the properties of the system, and draw our conclusions in Section \ref{sec:end}.

\section{Observations and Data Analysis}
\label{sec:TESS}

\subsection{TESS Observations and Stellar Parameters}

\begin{figure}
    \centering
    \plotone{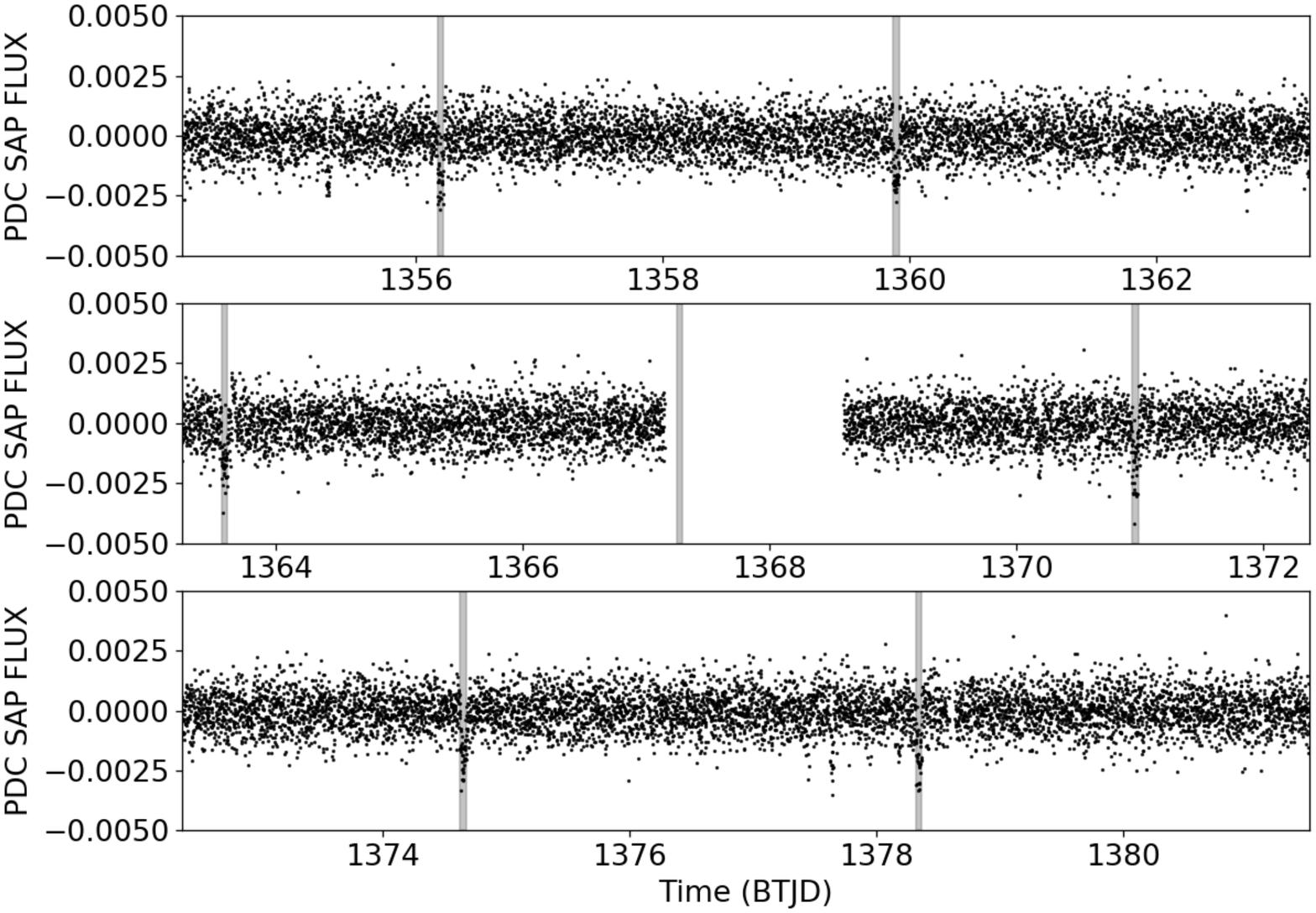}
    \plotone{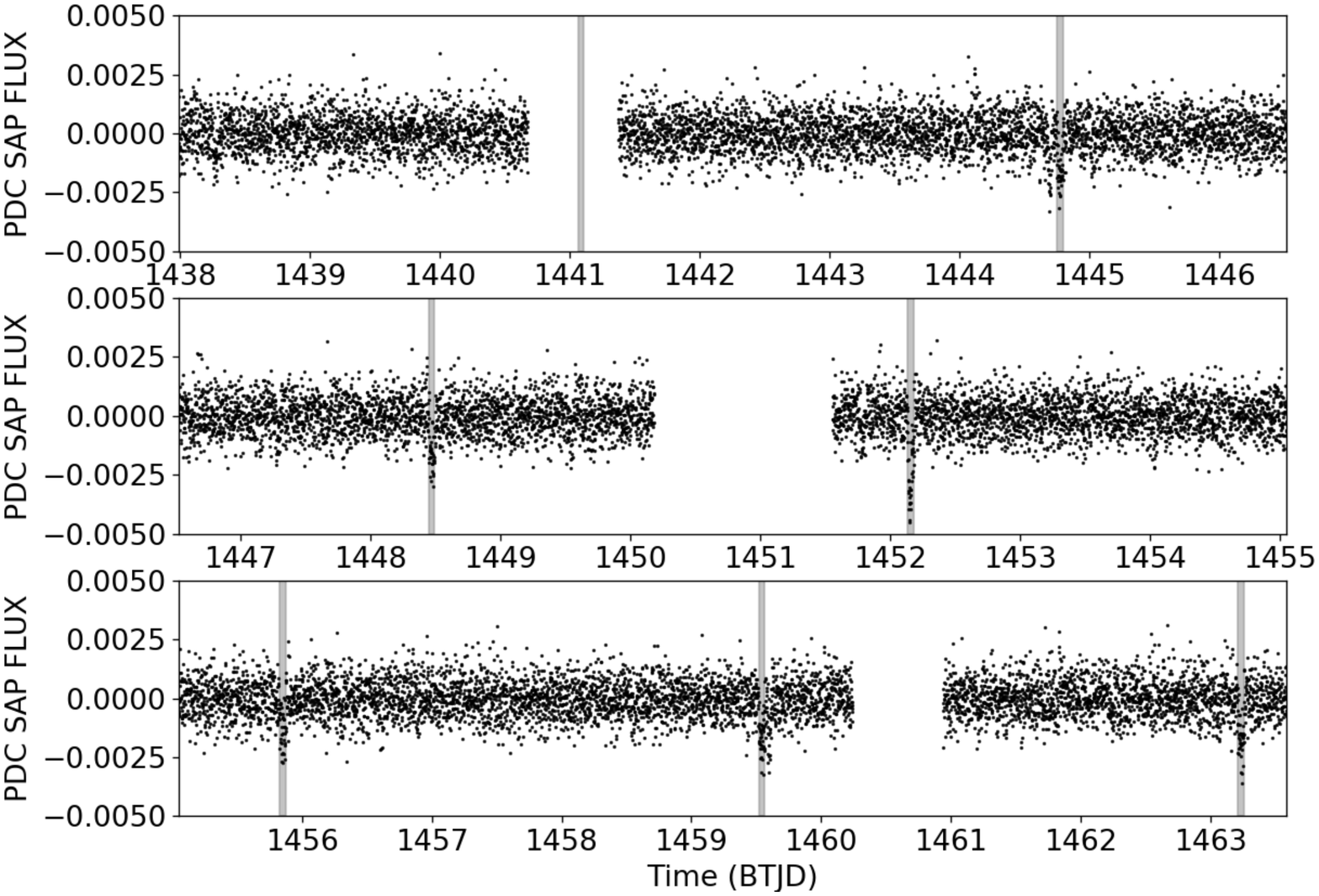}
    \plotone{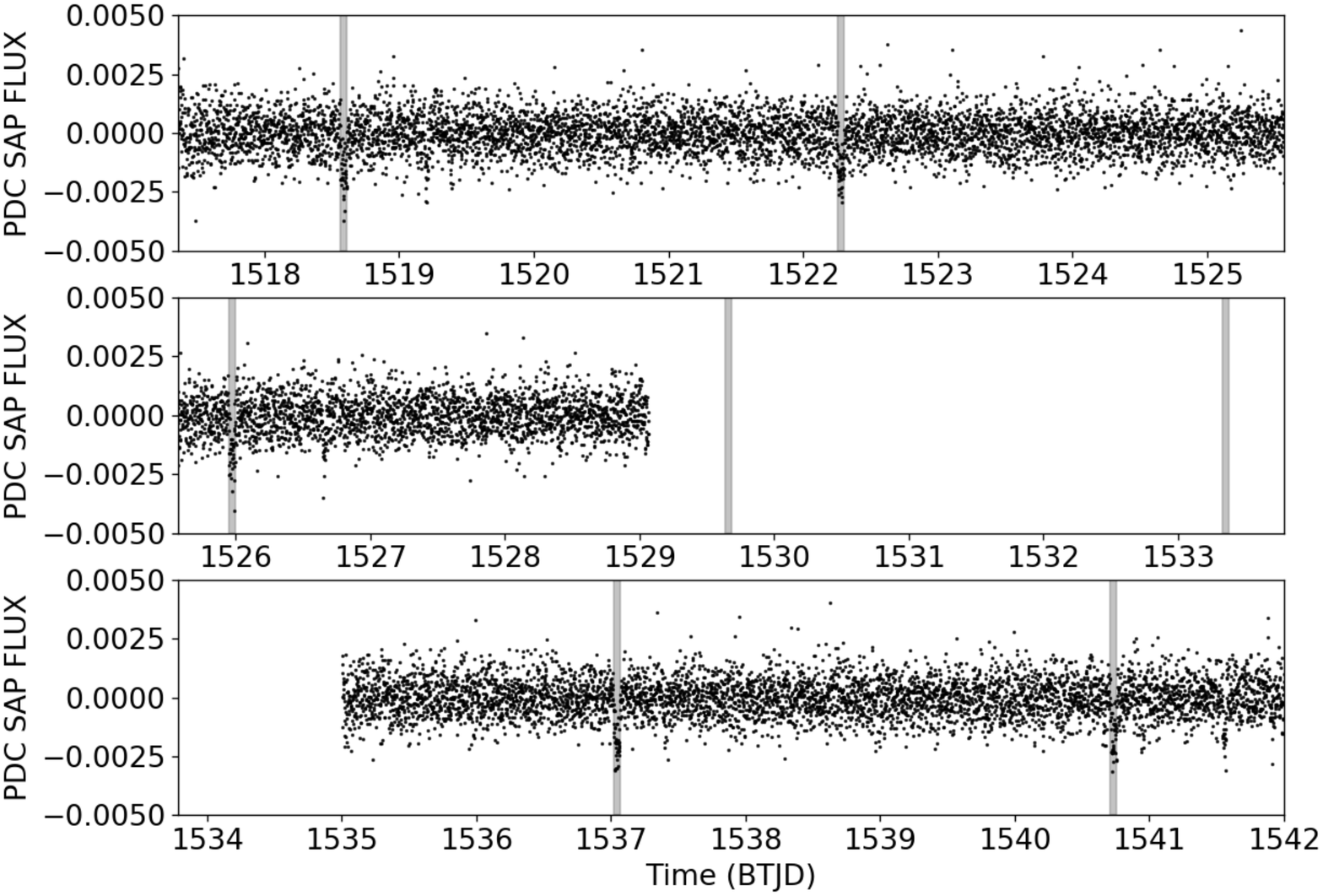}
    \caption{TESS PDCSAPFLUX lightcurves for L 98-59 as a function of time. Upper three panels represent sector 2 data, middle three panels sector 5, and lower panels---sector 8. For context, the transits of planet candidate L 98-59 c are indicated with vertical grey lines. As L 98-59 c and L 98-59 d are just wide of a 2:1 period resonance, their transits can sometimes occur close to each other (e.g. around days 1445 and 1459.5), and can even create a syzygy---like near day 1452.2.}
    \label{fig:TESS_data}
\end{figure}

TESS observed L 98-59 (TIC 307210830, TOI-175; RA = 08:18:07.62, Dec = -68:18:46.80 (J2000)) in Sectors 2, 5, and 8 with Camera 4. The target was added to the TESS Candidate Target List---a list of targets prioritized for short-cadence observations \citep{Stassun_CTL_2018}---as part of the specially curated Cool Dwarf list \citep{Muirhead2018}. The TESS data were processed with the Science Processing Operations Center Pipeline \citep[SPOC;][]{Jenkins2016} and with the MIT Quick Look Pipeline. The three candidates identified by the SPOC pipeline passed a series of data validation tests \citep[]{Twicken2018, Li2019} summarized below, and were made publicly available on the MIT TESS Data Alerts website\footnote{\url{https://tess.mit.edu/alerts/}} and the Mikulski Archive for Space Telescopes (MAST) TESS alerts page\footnote{\url{https://archive.stsci.edu/prepds/tess-data-alerts/}} as TOI-175.01, -175.02 and -175.03. These candidates had periods P = 3.690613 days, 7.451113 days, and 2.253014 days and transit epochs (BTJD) = 1356.203764, 1355.2864 and 1354.906208, respectively\footnote{BTJD = BJD - 2457000}, and are referred to in the rest of the manuscript as L 98-59 c, L 98-59 d, and L 98-59 b respectively. The SPOC simple aperture photometery (SAP) and pre-search data conditioned (PDCSAP) lightcurves \citep[]{Smith2012, Stumpe2014} of L 98-59 are shown in Figure \ref{fig:TESS_data}. 

We use the methods appropriate for M dwarfs previously used by \citep{Berta2015,Dittmann(2017a),Ment(2018a)} to determine the stellar parameters of the host star, and adopt these parameters throughout our analysis. We estimate the mass of the star using the mass-luminosity relation in the $K-$band from \citet{Benedict(2016)} to be 0.313$\pm$0.014 M$_{\odot}$. We then use single star mass-radius relations \citep{Boyajian(2012)} to find a stellar radius of 0.312$\pm$0.014 R$_{\odot}$. We calculate the bolometric correction in $K$ from \citet[erratum]{Mann2015} to be 2.7$\pm$0.036 mag, resulting in a bolometric luminosity for L~98-59 of 0.011$\pm$0.0004 L$_{\odot}$. We calculate the correction in $V$ from \citet{Pecaut(2013)} to be -2.0$\pm$0.03 mag\footnote{We assume the uncertainty on the bolometric correction in $V$ is that of the ($V-K$) color.}, resulting in a bolometric luminosity of 0.0115$\pm$0.0005 L$_{\odot}$. We adopt the mean of the two bolometric luminosities from which we calculate the luminosity of the host star to be 0.0113$\pm$0.0006 L$_{\odot}$ (i.e.~4.31e24 W). From the Stephan-Boltzmann Law, we find an effective temperature ${\rm T_{eff} = 3367\pm150 K}$. As a comparison, we also used the relations in \citet{Mann2015} to determine an effective temperature of 3419$\pm$77 K for L~98-59, in agreement with the $T_{eff}$ derived from the Stefan-Boltzmann Law.

\begin{figure}[htb]
    \centering
    \includegraphics[width=1\linewidth]{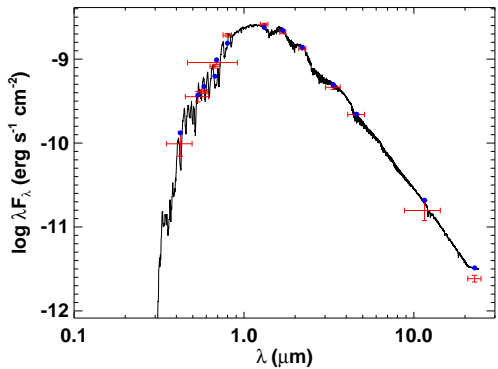}
    \caption{Spectral energy distribution (SED) fit, yielding $T_{\rm eff}$ and [Fe/H]. With the {\it Gaia\/} DR2 parallax, the $F_{\rm bol}$ from integrating SED then gives the stellar radius. Solid curve is the best fitting NextGen atmosphere, red symbols are the observed broadband fluxes, and blue points are the integrated model fluxes.}
    \label{fig:sed}
\end{figure}

In addition, following the procedures described in \citet{StassunTorres:2016} and \citet{Stassun:2017} to fit a NextGen stellar atmosphere model \citep{Hauschildt1999} to broadband photometry data from {\it Tycho-2}, \citet{Winters2015}, {\it Gaia}, 2MASS, and WISE, we performed a full fit of the stellar spectral energy distribution (SED) to estimate the stellar $T_{\rm eff}$ and [Fe/H] which, together with the {\it Gaia\/} DR2 parallax, provides an estimate of the stellar radius. The free parameters of the fit were $T_{\rm eff}$ and stellar metallicity [Fe/H], 
and we set the extinction $A_V \equiv 0$
due to the very close distance of the system. 
The resulting best fit is shown in Figure~\ref{fig:sed}, with a reduced $\chi^2$ of 3.8 for 8 degrees of freedom. The best fit parameters are $T_{\rm eff} = 3350 \pm 100$~K and [Fe/H] $= -0.5 \pm 0.5$. Integrating the SED gives the bolometric flux at Earth as $F_{\rm bol} = 2.99 \pm 0.18 \times 10^{-9}$~erg~s$^{-1}$~cm$^{-2}$. Finally, adopting the {\it Gaia\/} DR2 parallax and the correction of 80~$\mu$as from \citet{StassunTorres:2018}, we calculate a stellar radius of $0.305 \pm 0.018$~R$_\odot$ using the Stefan-Boltzmann law. These are consistent with the adopted parameters discussed above.

We also estimated a prior on the stellar density ($\rho_*$) by estimating the stellar mass. Here we used the empirical relations of \citet{Mann2015}, which provides $M_{star} \approx 0.32 M_\odot$ from the absolute $K_S$ magnitude ($M_{K_S}$) determined from the observed 2MASS $K_S$ magnitude and the {\it Gaia\/} DR2 parallax \citep[corrected for the offset from][]{StassunTorres:2018}. The quoted uncertainty in the Mann et al. empirical relation is $\sim$3\%; here we conservatively adopt an uncertainty of 10\%. Together with the radius determined above from the SED and parallax, we obtain $\rho_* = 15.9 \pm 3.3$~g~cm$^{-3}$ (this value is used as a prior in the transit model). We also determined the stellar temperature and radius using empirical relations calibrated using low-mass stars with interferometrically measured radii and precise distances \citep[see][and references therein]{feinstein2019}. These alternative stellar parameter estimates were consistent with those determined from the empirical M-dwarf relations and from the SED fitting. An additional set of stellar parameters for L 98-59 were previously derived from a medium-resolution optical spectrum in the CONCH-SHELL survey \citep{gaidos2014}. This work also provides parameters consistent with our estimates. We compile the stellar parameters used in subsequent analyses and other identifying information for L 98-59 in Table~\ref{tab:stellarparameters}. While it is difficult to pin down the ages of old M dwarfs due to their long main-sequence lifetime, the lack of a rapid rotation signal in the TESS SAP light curve and the low activity of L 98-59 (see \S~2.4) indicate that it is likely an old M dwarf with an age $>$1 Gyr. The stellar parameters of L 98-59 are consistent with a spectral type of M3~$\pm$~1.

\begin{deluxetable}{l r r }[!ht]
\hspace{-1in}\tabletypesize{\scriptsize}
\tablecaption{  Stellar Parameters \label{tab:stellarparameters}}
\tablewidth{0pt}
\tablehead{
\colhead{Parameter} & \colhead{Value} & \colhead{Notes}
}
\startdata
\multicolumn{3}{c}{\em Identifying Information} \\
Name & L 98-59 & \\
TIC ID & 307210830 & \\
TOI ID & 175 & \\  \\
$\alpha$ R.A. (hh:mm:ss) & 08:18:07.62  & Gaia DR2 \\
$\delta$ Dec. (dd:mm:ss) & -68:18:46.80  & Gaia DR2 \\
$\mu_{\alpha}$ (mas~yr$^{-1}$) & $94.767 \pm 0.054$ & Gaia DR2 \\
$\mu_{\delta}$ (mas~yr$^{-1}$) & $-340.470 \pm 0.052$  & Gaia DR2 \\
Distance (pc) & $10.623 \pm 0.003$  & Gaia DR2\\
\multicolumn{3}{c}{\em Photometric Properties} \\
$B$ (mag) ..........  & $13.289 \pm 0.027$ & APASS DR9 \\
$V$ (mag) ..........  &  $11.685 \pm 0.017$ & APASS DR9 \\
$G$ (mag) ..........  & $10.598 \pm	0.001$  & Gaia DR2 \\
$g^{\prime}$ (mag) ..........  & $12.453 \pm 0.019$  & APASS DR9 \\
$r^{\prime}$ (mag) ..........  & $11.065 \pm 0.044$  & APASS DR9 \\
$T$ (mag) ..........        &  9.393  & TIC \\
$J$ (mag) ..........  & $7.933 \pm	0.027$		  & 2MASS \\
$H$ (mag) ..........  & $7.359 \pm	0.049$  & 2MASS \\
$K_s$ (mag) .........   &  $7.101 \pm 0.018$ & 2MASS \\
$W1$ (mag) .........   & $6.935 \pm	0.062$	& ALLWISE \\
$W2$ (mag) .........   & $6.767 \pm	0.021$  & ALLWISE \\
$W3$ (mag) .........   & $6.703 \pm	0.016$ & ALLWISE \\
$W4$ (mag) .........   & $6.578 \pm	0.047$ & ALLWISE \\
\multicolumn{3}{c}{\em Stellar Properties} \\
Spectral Type .......... & M3V $\pm$ 1 & This Work \\
\Teff\ (K) .......... & $3367 \pm 150$  & This Work\\
$\lbrack $Fe/H$ \rbrack$ .......... & $-0.5 \pm 0.5$ & This Work \\
$M_{star}$ ($M_\odot$) .......... & $0.313 \pm 0.014$  & This Work\\
$R_{star}$ ($R_\odot$) .......... & $0.312 \pm 0.014$ & This Work\\
$L_{star} (L_\odot)$ .......... & $0.0113 \pm 0.0006$ & This Work\\
\enddata
\tablenotetext{}{Gaia DR2 - \citep[]{gaia2018}, UCAC5 - \citep[]{Zacharias2017}, APASS DR9 - \citep[]{Henden2016}, 2MASS - \citep[]{Skrutskie2006}, ALLWISE - \citep[]{Cutri2013}}
\end{deluxetable}

\subsection{Lightcurve Analyses}
\begin{figure*}
    \centering
    \includegraphics[width=0.99\linewidth]{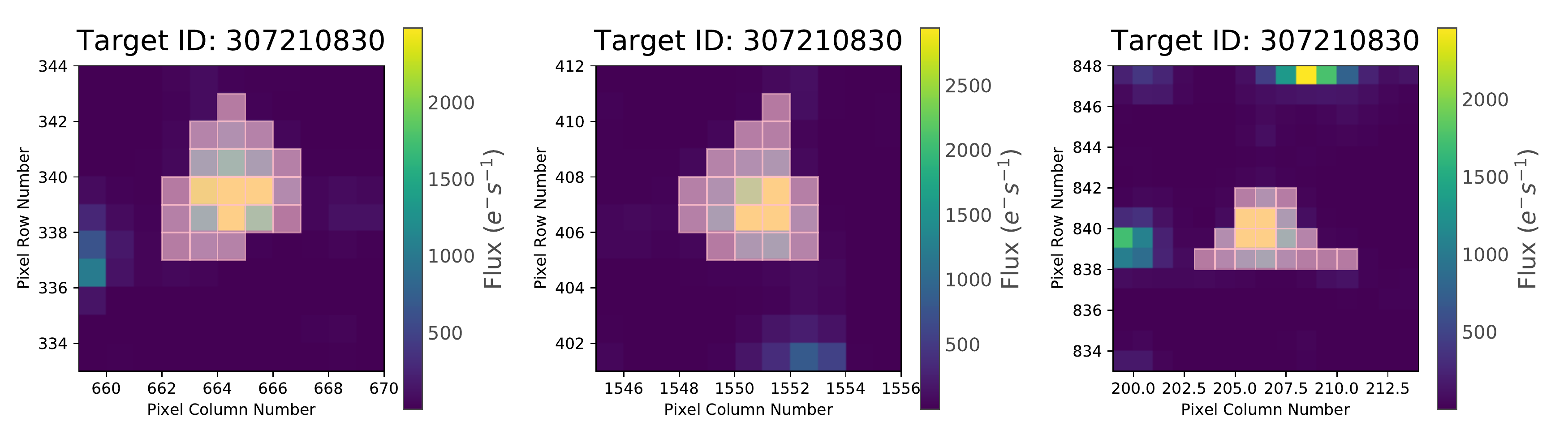}
    \caption{The pixel mask (pink squares) we used to create a lightcurve of L 98-59. The mask was created using the threshold method in the lightkurve extract aperture photometry tool. The three images show are from TESS Sectors 2, 5 and 8.}
    \label{fig:pixmask}
\end{figure*}

We opted to create our own apertures from Sectors 2, 5 and 8 target pixel file data to analyze the photometric time-series from TESS, instead of using the pipeline apertures used to first identify the transiting planet candidates. Our primary motivation for performing our own photometry is that we can avoid any attenuation to the transit signals by explicitly masking them during the systematic correction step. We first used the \textsf{lightkurve} package \citep{lightkurve}\footnote{\url{https://github.com/KeplerGO/lightkurve}} to extract lightcurves from each of the three sectors using the threshold method, which selects pixels that (a) are a fixed number of standard deviations above the background, and (b) create a contiguous region with the central pixel in the mask. We used a threshold value of 3$\sigma$; the corresponding mask shape is shown in Figure~\ref{fig:pixmask}. Thus produced, the resulting lightcurve still contains low-level instrumental systematic signals. To identify and subtract instrumental signals, we used a second order pixel-level de-correlation (PLD), which is a technique based on Spitzer and K2 analysis methods \citep{Deming2015,Luger2016}. During the PLD step, we masked out transits to avoid attenuating the signals. Finally, we normalized the lightcurve by dividing by the median and subtracting one to center the flux about zero. We did the PLD detrending separately for each sector.

We used the \textsf{exoplanet} toolkit for probabilistic modeling of the exoplanet transits \citep{exoplanet:exoplanet}. The model we built consisted of four elements: three planet transit components with Keplerian orbits and limb-darkened transits, and a Gaussian Process (GP) component that models residual stellar variability. The planet models were computed with \textsf{exoplanet} using \textsf{STARRY} \citep{exoplanet:luger18}, while the GP was computed using \textsf{celerite} \citep{exoplanet:foremanmackey17,exoplanet:foremanmackey18}. The GP component is described as a stochastically-driven, damped harmonic oscillator with parameters of $\log(S_0)$ and $\log(\omega_0)$, where the power spectrum of the GP is
\begin{equation}
    S(\omega) = \sqrt\frac{2}{\pi} \frac{S_0 \, \omega_{0}^{4}}{(\omega^{2} - \omega_{0}^{2})^2 + \omega^{2} \, \omega_{0}^{2} / Q^2},
\end{equation}
and a white noise term, with a model parameter of the log variance. We fixed Q to $1/\sqrt{2}$ and put wide Gaussian priors on $\log(S_0)$ and $\log(\omega_0)$ with means of the log variance, and one log of one tenth of a cycle, respectively, and a standard deviation on the priors of 10. For each Sector we used separate GP parameters.
This form of GP was chosen because of the flexible nature and smoothly varies (it is once mean square differentiable) enables us to use it to model a wide range of low frequency astrophysical and instrumental signals. The white noise term carried the same prior as $\log(S_0)$. Each sector of data had a separate parameter for the mean flux level.

The planet model was parameterized in terms of consistent limb darkening, log stellar density, and stellar radius for the three planets. Each individual planet was parameterized in terms of log orbital period, time of first transit, log planet-to-star radius ratio, impact parameter, orbital eccentricity and periastron angle at time of transit. 
The stellar radius had a Gaussian prior with mean 0.312 and 0.014 standard deviation, with solar units, and is additionally required to be positive. The log mean stellar density, in cgs units, had a Gaussian prior with a mean of $\log{15}$ and standard deviation of 0.2~dex (as per Section 2.1). The limb darkening followed the \citet{exoplanet:kipping13} parameterization.

The log orbital periods, time of first transits, and log planet-to-star radius ratio of the three planets had Gaussian priors with means at the values found in the TESS alert data, and standard deviations of 0.1, 0.1, and 1, respectively. The impact parameter had a uniform prior between zero and one plus the planet-to-star radius ratio. Eccentricity had a beta prior with $\alpha=0.867$ and $\beta=3.03$ \citep[as suggested by ][]{Kipping2013}, and was bounded between zero and one. The periastron angle at transit was sampled in vector space to avoid the sampler seeing a discontinuity at values of $\pi$.

We sampled the posterior distribution of the model parameters using the No U-turn Sampler \citep[NUTS, ][]{NUTS} which is a form of Hamiltonian Monte Carlo, as implemented in \textsf{PyMC3} \citep{exoplanet:pymc3}. We ran 4 simultaneous chains, with 5000 tuning steps, and 3000 draws in the final sample. The effective number of independent samples of every parameter was above 1000, and most parameters were above 5000. The Gelman–Rubin diagnostic statistic was within 0.005 of 1.000 for each parameter in the model. The impact parameter for the outer planet is relatively high, which caused this parameter along with the orbital eccentricity to be most time consuming to sample independently. 

\begin{figure*}
    \centering
    \includegraphics[width=.99\textwidth]{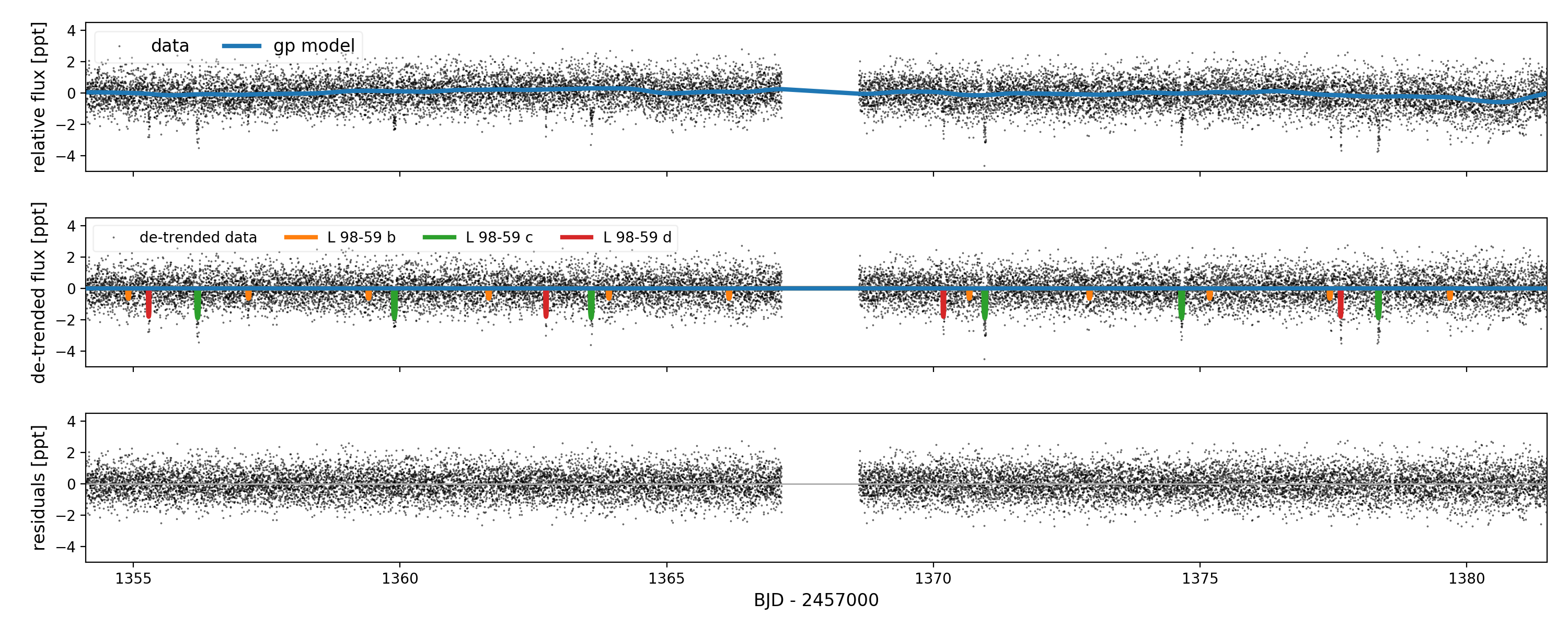}\\
    \includegraphics[width=.99\textwidth]{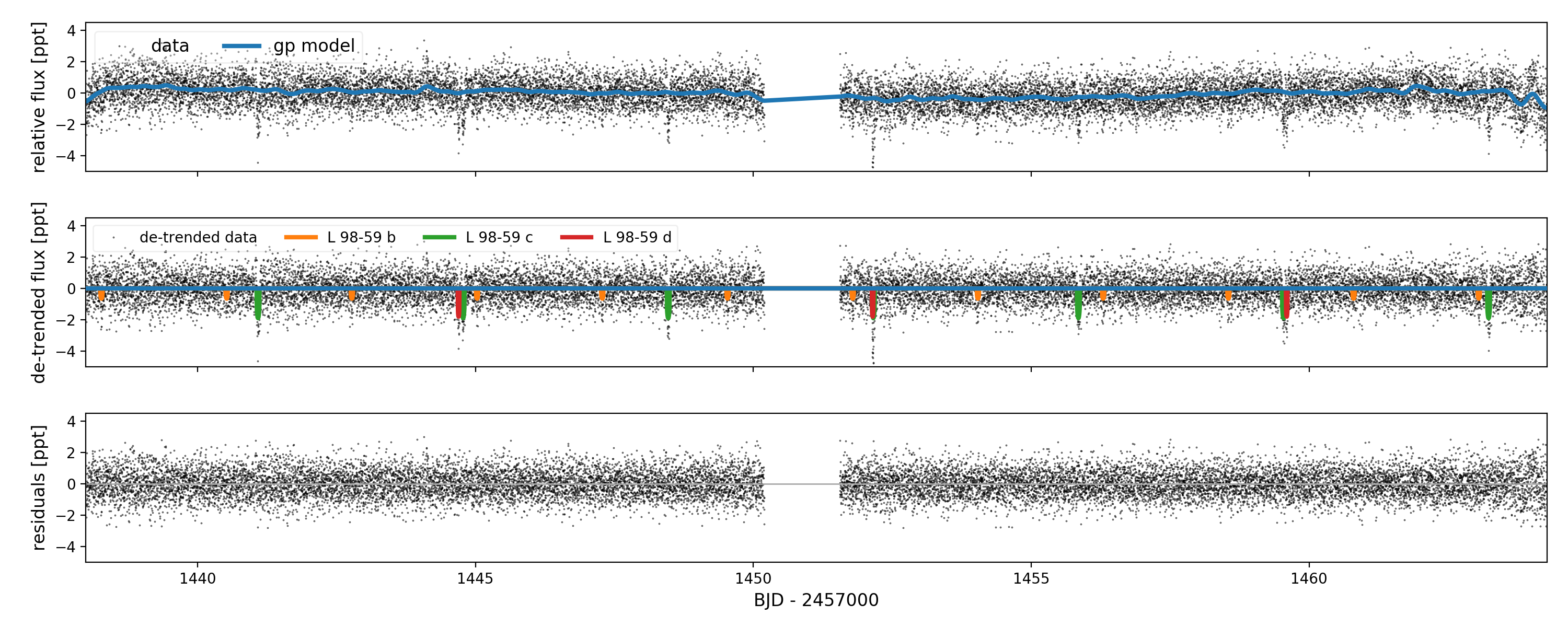}\\
    \includegraphics[width=.99\textwidth]{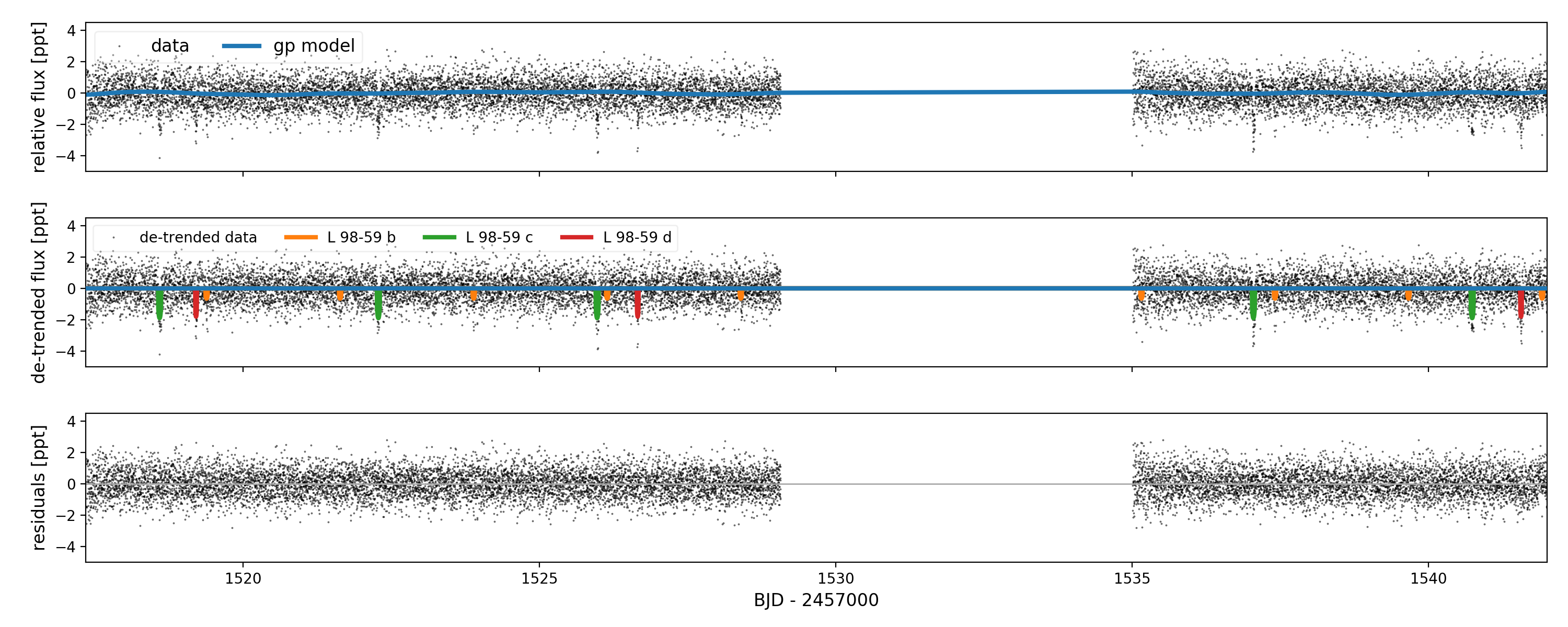}
    \caption{TESS data of L 98-59 from Sectors 2, 5 and 8. The top panels shows the data after it has been extracted from the TESS target pixel file, and detrended using the PLD algorithm. The green line shows the best fitting GP mean model. In the central panels, we show the data with a GP mean model subtracted (this subtraction is only performed for display purposes in this figure). The best fitting models for the three planets are also shown. The lower panels has the GP and planet models removed.}
    \label{fig:allfit}
\end{figure*}

\begin{figure}
    \centering
    \includegraphics[width=0.5\textwidth]{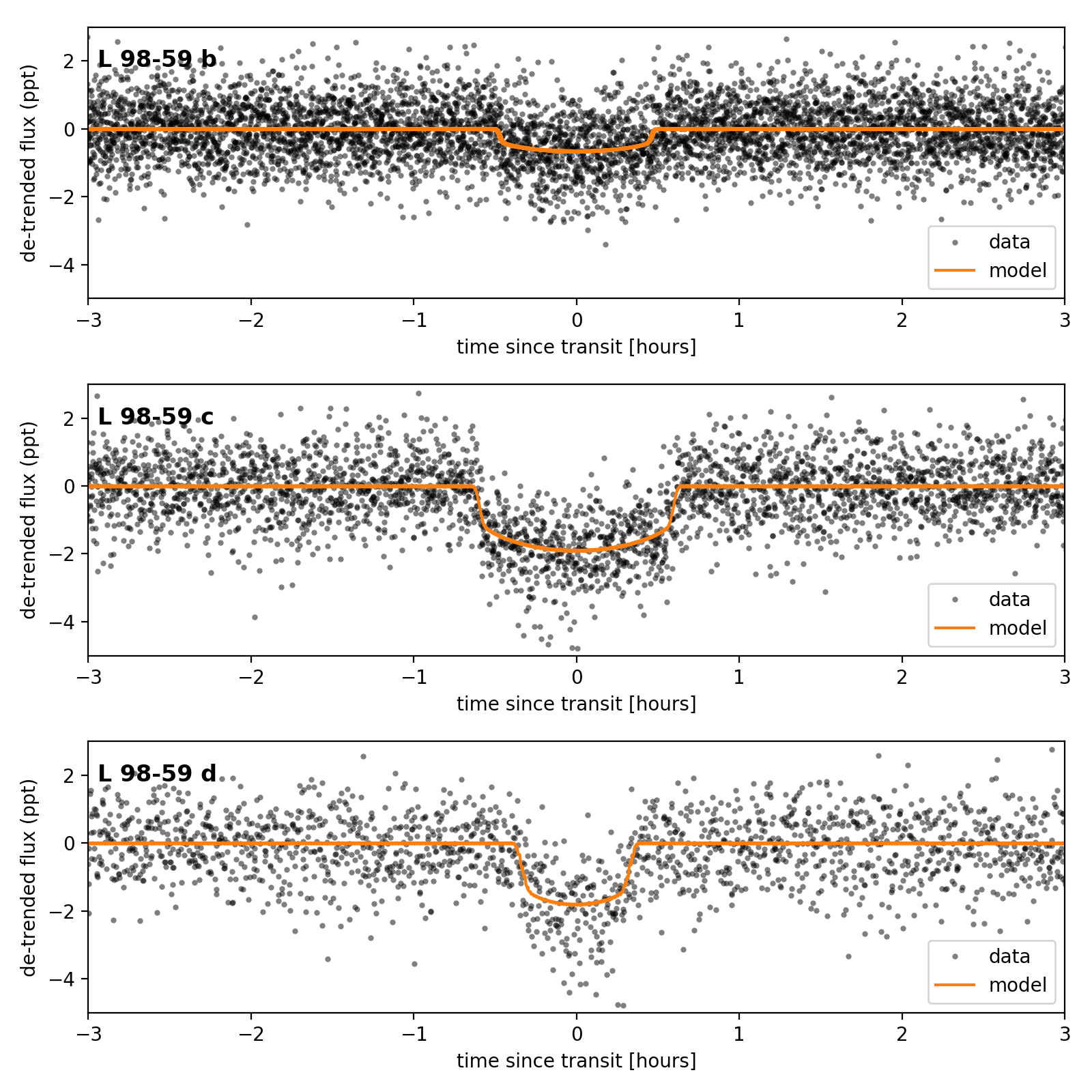}
    \caption{Phase-folded, three sector lightcurves for planet  L 98-59 b (upper panel), L 98-59 c (middle panel), and L 98-59 d (lower panel), along with the respective best-fitting models (orange). The corresponding transit parameters are listed in Table \ref{tab:TransitParameters}.}
    \label{fig:folded_transits}
\end{figure}

Figure~\ref{fig:allfit} shows the GP model of the low-level variability in the upper panels, and the best fitting transit model in the central panels. The phase-folded transits of the three candidates, along best-fitting transit models, are shown in Figure \ref{fig:folded_transits}, and the model parameters are provided in Table \ref{tab:TransitParameters}. The transit modeling reveals that the candidate planets have small radii ranging from 0.8 to 1.6 \rearth. A chain of small terrestrial-sized planets is common among M-dwarfs \citep{Muirhead2015}, and L 98-59 is reminiscent of other systems such as the TRAPPIST-1, Kepler-186 and Kepler-296 \citep{Gillon2017,Barclay2015,Quintana2014}. The stellar density obtained from the transit model is fully consistent with that determined from the stellar parameters in Section 2.1 (${\rm 15.8^{+2.6}_{-2.7}~g~cm^{-3}}$ for the former vs ${\rm 15.9\pm3.3~g~cm^{-3}}$ for the latter.

\begin{deluxetable}{l c c c}[htb]
\tablecaption{Planet Parameters \label{tab:TransitParameters}}
\tablewidth{0pt}
\tablehead{
\colhead{Parameter} & \colhead{-1$\sigma$} & \colhead{Median}  & \colhead{+1$\sigma$}
}
\startdata
\multicolumn{4}{c}{\em Model Parameters} \\
Star & & & \\
$\ln{\rho}$ [g~cm$^{-3}$] & 2.57 & 2.76 & 2.91 \\
Limb darkening $q_1$ & 0.41 & 0.65 & 0.83 \\
Limb darkening $q_2$ & -0.34 & -0.11 & 0.28 \\
{\bf L 98-59 b} & & & \\
${T_0}$ (BJD - 2457000)& 1366.1694 & 1366.1701& 1366.1707 \\
$\ln{Period}$ [days] & 0.812318 & 0.812326 & 0.812334 \\
Impact parameter & 0.13 &	0.36 &	0.55 \\
$\ln{R_p/R_*}$ & -3.79& 	-3.75& 	-3.72 \\
eccentricity & 0.03 & 0.10 & 0.27 \\
$\omega$ [radians] & -2.2& 	0.3& 	2.4 \\
 & & & \\
{\bf L 98-59 c} & & & \\
${T_0}$ (BJD - 2457000)& 1367.2752& 1367.2755& 1367.2759\\
$\ln{Period}$ [days] & 1.305791 &	1.305795 &	1.305798 \\
Impact parameter & 0.09 &	0.29 &	0.49 \\
$\ln{R_p/R_*}$ & -3.25& 	-3.23& 	-3.20 \\
eccentricity & 0.02 & 0.09 & 0.25 \\
$\omega$ [radians] & -2.5&	-0.4& 	2.2 \\
 & & & \\
{\bf L 98-59 d} & & & \\
${T_0}$ (BJD - 2457000)& 1362.7367 & 1362.7375 & 1362.7382\\
$\ln{Period}$ [days] & 2.008323 &	2.008329 &	2.008334 \\
Impact parameter & 0.75 & 0.89 & 0.93 \\
$\ln{R_p/R_*}$ & -3.16&	-3.07& 	-3.01 \\
eccentricity & 0.04 & 0.20 & 0.52 \\
$\omega$ [radians] & -1.9& 	0.7& 	2.3 \\
\hline
\multicolumn{4}{c}{\em Derived Parameters} \\
{\bf L 98-59 b} & & & \\
Period [days] & 2.25312 &	2.25314 &	2.25316\\
${R_p/R_*}$ & 0.0226 &	0.0234 &	0.0243\\
Radius ${[R_\oplus]}$ & 0.75& 	0.80& 	0.85\\
Insolation & 19.5 & 23.9& 29.2\\
${a/R_*}$ & 15.2& 16.2& 17.0\\
$a$~[AU] & 0.0216& 0.0233& 0.0250\\
Inclination (deg) & 88.0& 88.7& 89.5\\
Duration (hours) & 0.89& 1.02& 1.19\\
 & & & \\
{\bf L 98-59 c} & & & \\
Period [days] & 3.690607 &	3.690621 &	3.690634\\
${R_p/R_*}$ & 0.0388& 	0.0396& 	0.0407\\
Radius ${[R_\oplus]}$ & 1.28& 	1.35& 	1.43\\
Insolation & 10.1 & 12.4 & 15.2\\
${a/R_*}$ & 21.1& 22.5&, 23.6\\
$a$~[AU] & 0.0300& 0.0324& 0.0347\\
Inclination (deg) & 88.8& 89.3& 89.7\\
Duration (hours) & 1.07& 1.24& 1.36\\
 & & & \\
{\bf L 98-59 d} & & & \\
Period [days] & 7.45081 &	7.45086 &	7.45090\\
${R_p/R_*}$ & 0.0426& 	0.0462& 	0.0492\\
Radius ${[R_\oplus]}$ & 1.43& 	1.57& 	1.71\\
Insolation & 3.96& 4.85& 5.93\\
${a/R_*}$ & 36.2 & 37.4 & 38.5\\
$a$~[AU] & 0.048& 0.052& 0.056\\
Inclination (deg) & 88.0& 88.5& 88.7\\
Duration (hours) & 0.74& 0.91& 1.68\\
\enddata
\end{deluxetable}

We repeated this analysis using the systematics-corrected light curves from the TESS pipeline \citep[PDCSAP][]{Jenkins2016,Stumpe2014} rather than using the respective target pixel files. We found consistent results, aside from different GP parameters owing to the different systematics corrections applied. The transits depths were lower in the PDCSAP data at the ${\rm <1\sigma}$ level, which we attribute to masking out transits in the systematics-correction technique we applied. We did not include any flux contamination from nearby stars in our models because there are no bright nearby stars to contaminate our pixel mask -- the TIC estimates that the contamination fraction for L 98-59 is 0.002. Even if the TIC contamination is dramatically underestimated, it is highly likely that the stellar radius uncertainty will be the dominant term in the planet radius uncertainty, therefore we feel comfortable neglecting it.

There are significant impact parameter differences between the inner two planets and the outer planet. The outer planet transits close to the limb of the star, although it is not grazing. This manifests in the light curve as a shorter transit duration for the outer planet than the two inner planets and indicates a modest mutual inclination of at least one degree between the inner two and outer planets. 
All three planets are subject to significantly more flux than Earth receives from the Sun, and are therefore unlikely to be good astrobiology targets. However, with an insolation flux of $4.5\pm0.8$, the outer planet is a candidate Venus-zone planet \citep{Kane2014}. 

\subsection{Potential false positive scenarios}

The TESS Data Validation Report performs a series of tests designed to rule out various false positive scenarios. The results from these tests are as follows: \\
i) L 98-59 c and L 98-59 d pass the difference image centroiding test, which employs PSF-based centroiding on the difference images (expected to be more precise and accurate than a brightness-weighted moment on the difference images). While L 98-59 b does not quite pass the difference image centroiding test, its transits are much shallower compared to the other two candidates. Thus it is likely that the centroiding errors are underestimated to some degree due to the variable pointing performance at timescales less than the 2 minutes observation cadence. We expect the analysis of this candidate to improve with new data; \\
ii) All three candidates pass the odd-even difference tests;
iii) Secondary eclipses are ruled out at the 3.6-, 2.6- and 1.8-${\sigma}$ levels;\\
iv) A bootstrap analysis of the out-of-transit data is used to quantify the probability of false alarms due to stellar variability and residual instrumental systematics. In the case of L 98-59, the light curve is well behaved and the analysis excludes the possibility of a false alarm at the 2.45E-25, 6.6E-62, and 2.2E-25 levels (as extrapolations of the upper tail of the bootstrap distribution to the observed maximum Multiple Event Statistics (MES) that triggered the detections of these candidates in the pipeline \citep{Jenkins2017});\\
v) All three candidates pass a ghost diagnostic test, designed to flag instances of scattered light, other instrumental artifacts or background eclipsing binaries. 

For completeness, we also applied the vetting pipeline \textsf{DAVE} \citep{Kostov2019} to the TESS lightcurve of L 98-59. Briefly, \textsf{DAVE} evaluates whether detected transit-like events produced by the candidate are real or false positives by analyzing the data for (a) odd-even differences between consecutive transits; (b) secondary eclipses; (c) stellar variability mimicking a transit; and (d) photocenter shifts during transit; 

\begin{figure*}
\centering
    \includegraphics[width=0.66\textwidth]{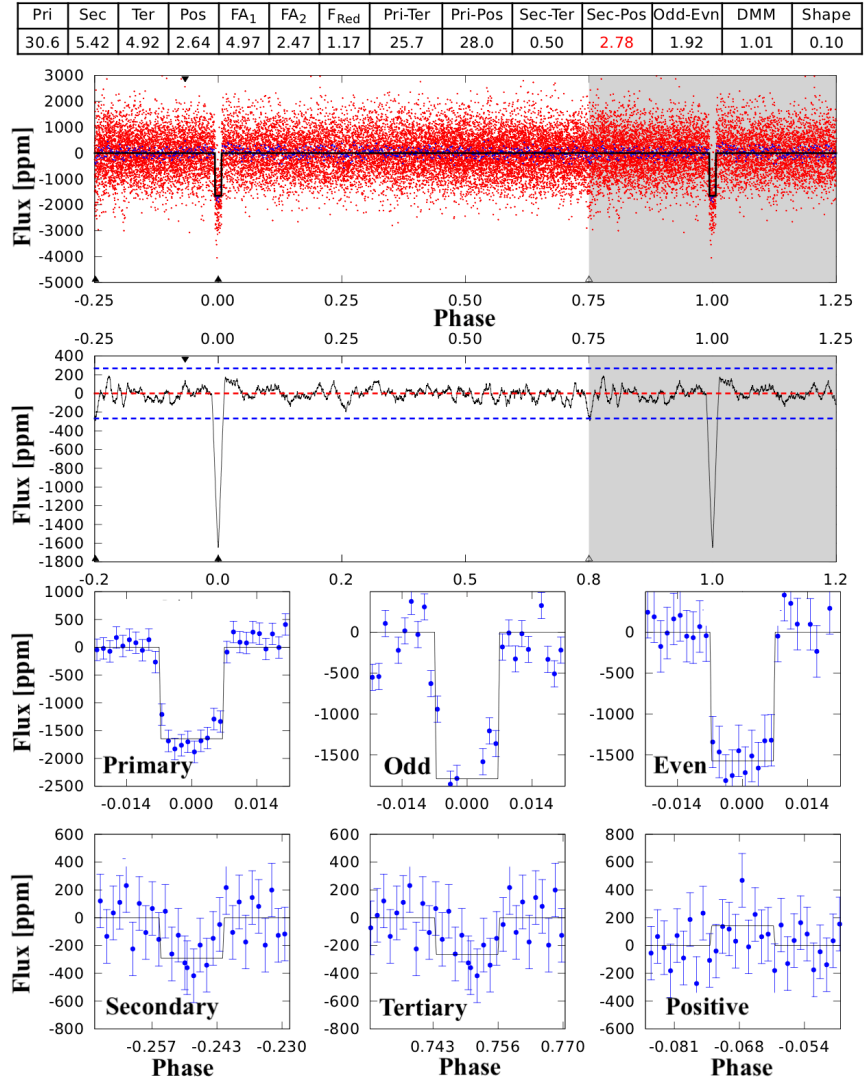}
    \caption{DAVE Modelshift analysis of L 98-59 c. The upper two rows represent the phase-folded lightcurve with the best-fit transit model (first row), and the phase-folded light convolved with the best-fit transit model (second row). The six panels in the lower two rows show all transits (label ``Primary''), all odd transits (``Odd''), all even transits (``Even''), most significant secondary  (``Secondary''), tertiary (``Tertiary''), and (``Positive'') features in the lightcurve. The table above the figure lists the significance of each feature (see text for details). There are no significant odd-even differences,  secondary eclipses or photocenter shifts indicating that the transit events are consistent with genuine planet candidates. }
    \label{fig:DAVE_c01}
\end{figure*}

\begin{figure*}
\centering
    \includegraphics[width=0.66\textwidth]{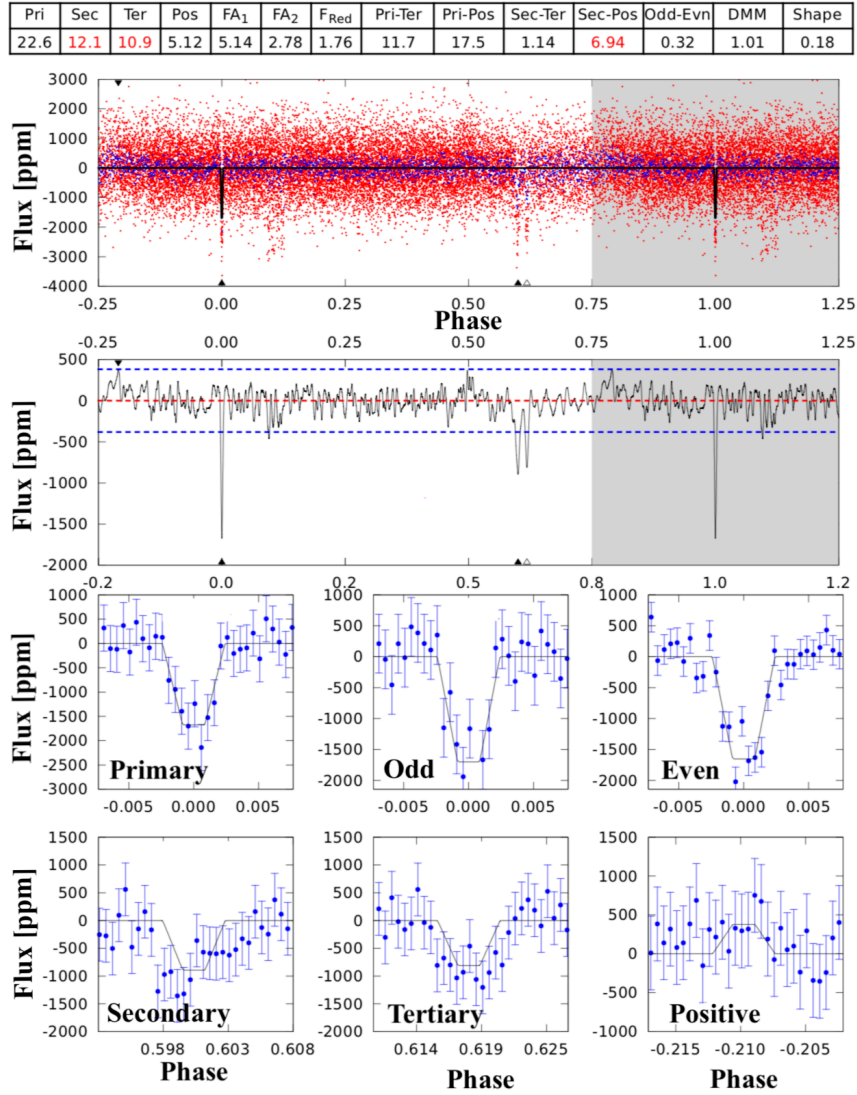}
    \caption{Same as Figure \ref{fig:DAVE_c01} but for L 98-59 d. The significant secondary and tertiary features detected are transits of 175-01.} 
    \label{fig:DAVE_c02}
\end{figure*}

\begin{figure*}
\centering
    \includegraphics[width=0.66\textwidth]{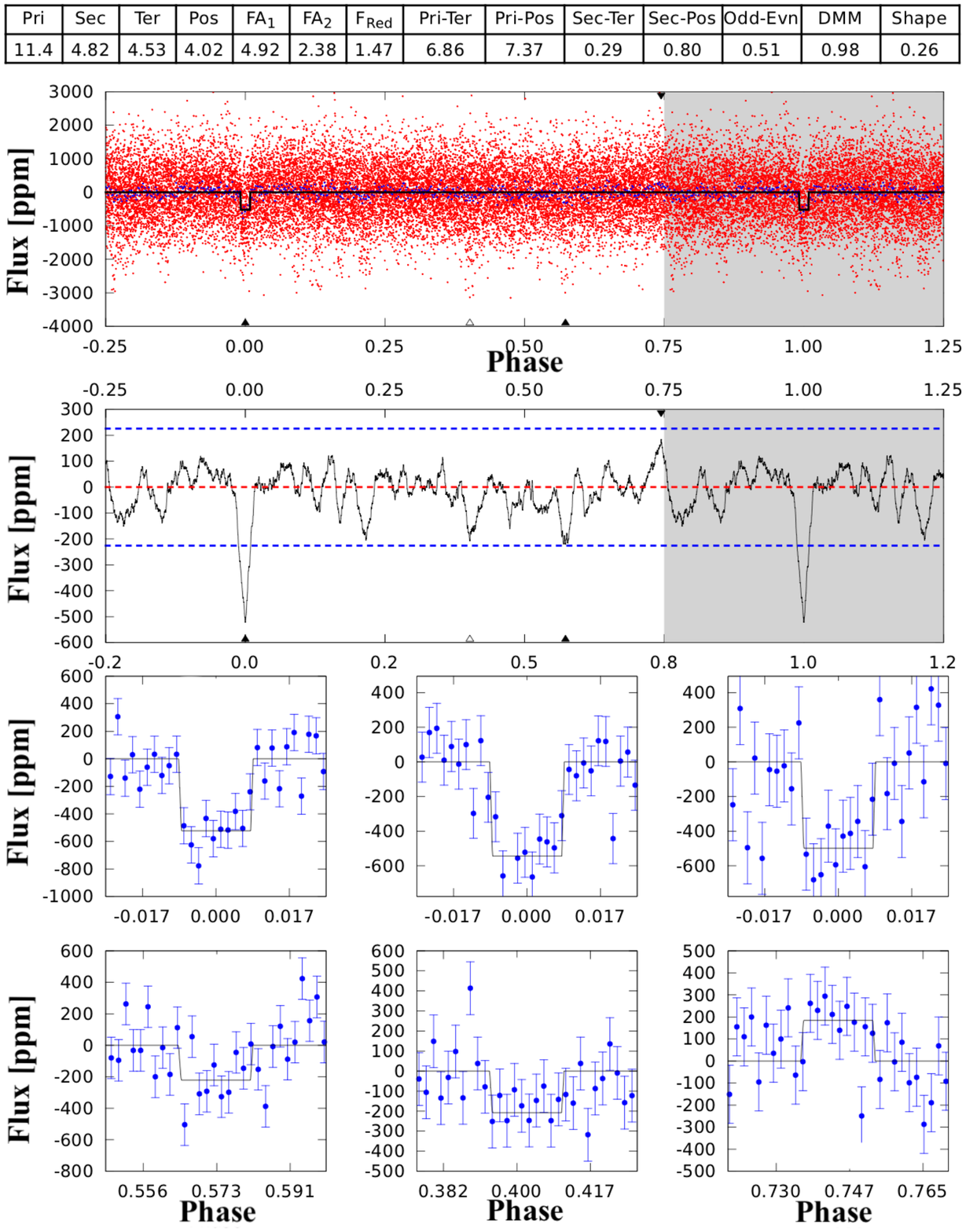}
    \caption{Same as Figure \ref{fig:DAVE_c01} but for L 98-59 b.} 
    \label{fig:DAVE_c03}
\end{figure*}

To perform the (a), (b) and (c) analysis we used the Modelshift module of \textsf{DAVE}---an automated package designed to emphasize features in the lightcurve that resemble the shape, depth and duration of the planetary transit but located at different orbital phase. To identify secondary eclipses and odd-even transit differences, or flares and heartbeat stars \citep[see e.g.]{Welsh2011}, Modelshift first convolves the lightcurve with the transit model of the planet candidate. The module then computes the significance of the primary transits, odd-even differences, secondary, tertiary and positive features assuming white noise in the lightcurve, and compares the ratio between each of these and the systematic red noise ${\rm F_{red}}$ to the false alarm thresholds ${\rm FA_1 = \sqrt{2}erfcinv(T_{dur}/(P\times N))}$ (assuming 20,000 objects evaluated), and ${\rm FA_2 = \sqrt{2}erfcinv (T_{dur}/P)}$ (for two events), where ${\rm T_{dur}, P,  N}$ are the duration, period and number of events (see Coughlin et al. 2014 for details). For example, a secondary feature is considered significant if ${\rm Sec /F_{red} > FA_1}$. The Modelshift results are shown in Figures \ref{fig:DAVE_c01}, \ref{fig:DAVE_c02}, and \ref{fig:DAVE_c03} where the panels show the phase-folded lightcurve (first row), the phase-folded lightcurve convolved with the best-fit transit model (second row), as well as the the best-fit to all primary transits, all odd and all even transits, the most prominent secondary, tertiary and positive features in the lightcurve (lower two rows). The tables above the figures list the individual features evaluated by the module: the significance of the primary (``Pri''), secondary (``Sec''), tertiary (``Ter'') and positive (``Pos'') events assuming white noise, along with their corresponding differences (``Pri-Ter'', ``Pri-Pos'', ``Sec-Ter'', ``Sec-Pos''), the significance of the odd-even metric  (``Odd-Evn''), the ratio of the individual depths' median and mean values (``DMM''), the shape metric (``Shape''), the False Alarm thresholds (``${\rm FA_1}$'', ``${\rm FA_2}$''), and the ratio of the red noise to the white noise in the phased light curve at the transit timescale (``Fred'').  Our analysis shows that there are no secondary eclipses or odd-even differences for any of the L 98-59 planet candidates. We note that the significant secondary and tertiary eclipses identified by \textsf{DAVE} for L 98-59 d (Figure \ref{fig:DAVE_c02}) are due to the transits of L 98-59 c and thus not a source of concern.

\begin{figure}
    \includegraphics[width=0.4\textwidth]{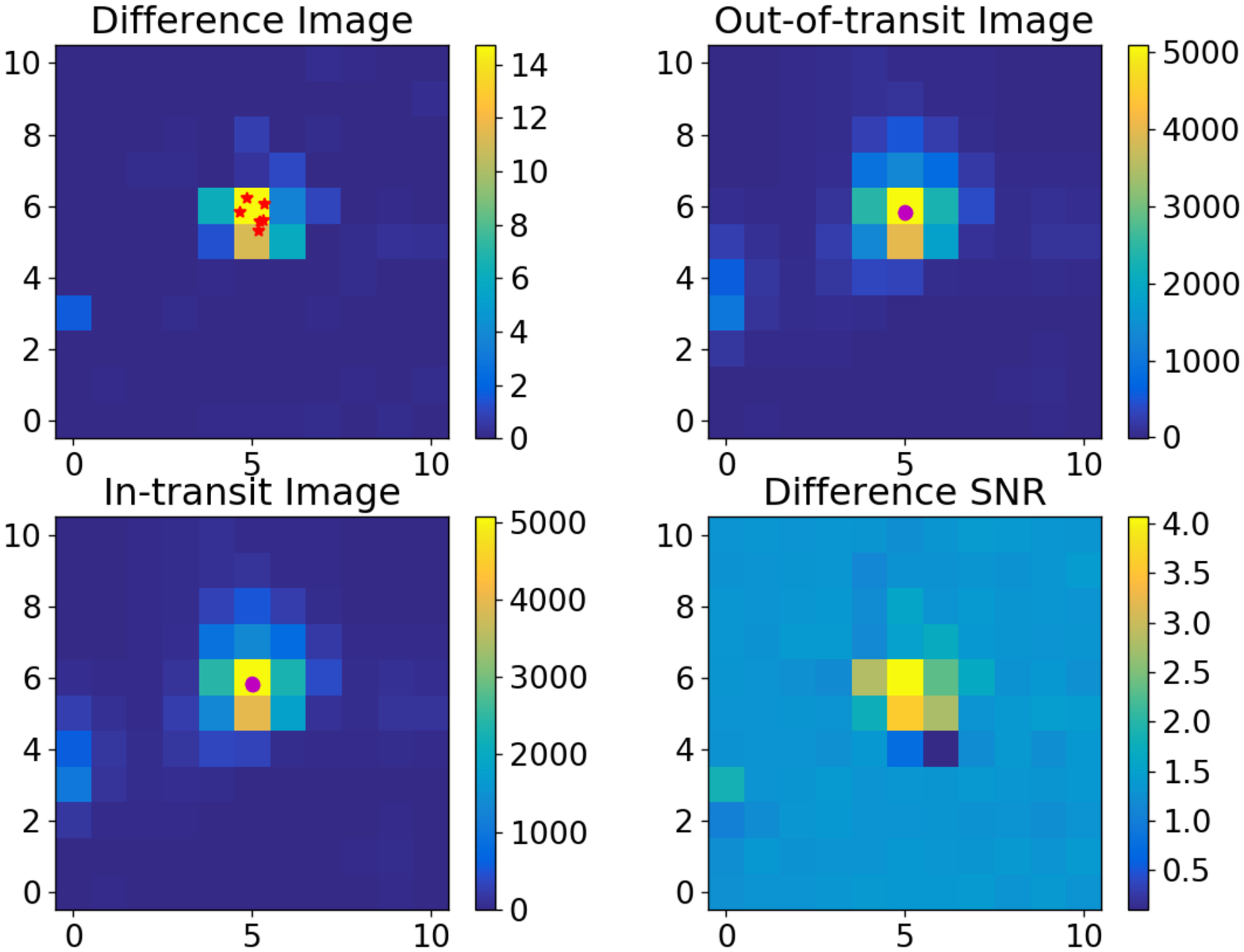}
    \includegraphics[width=0.4\textwidth]{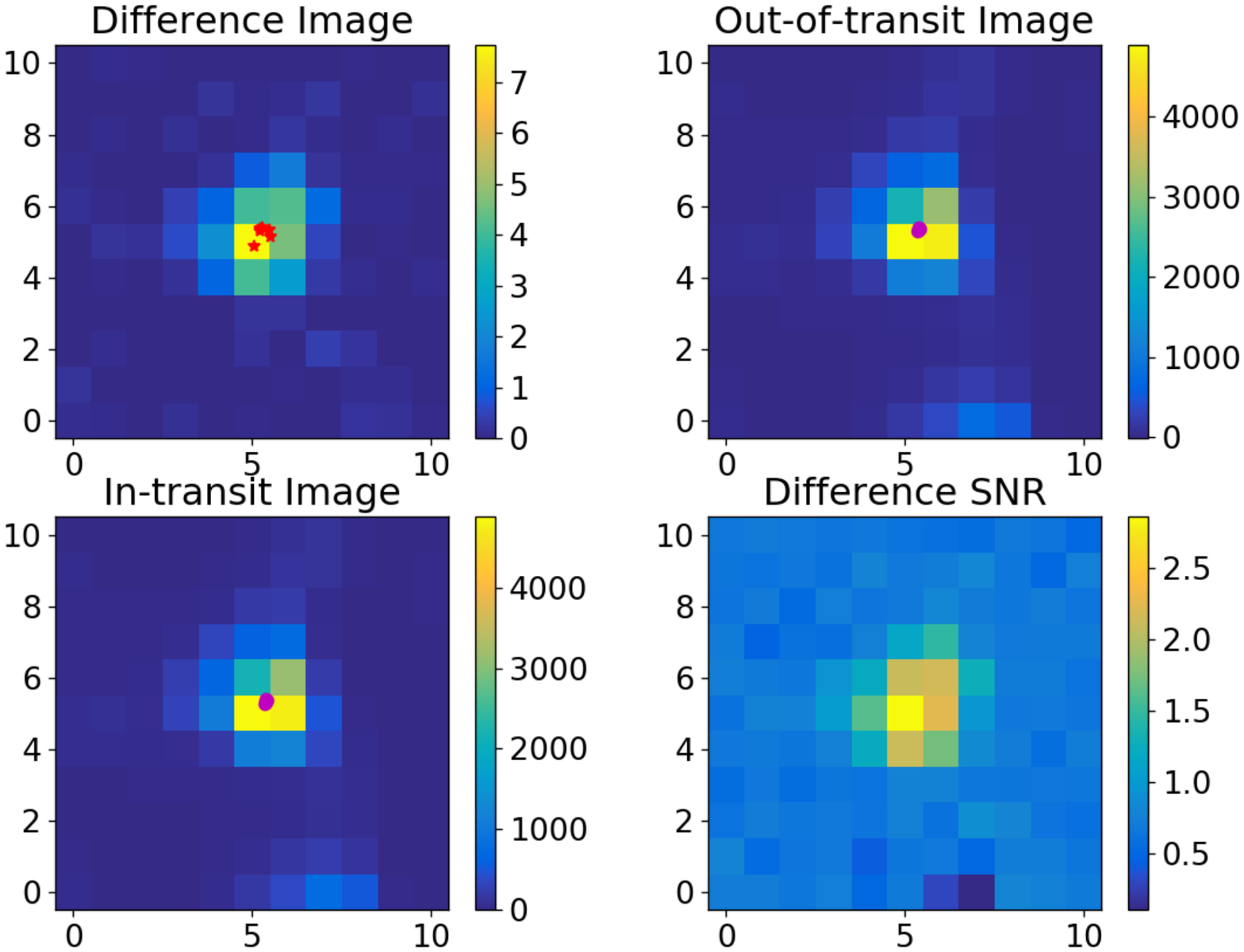}
    \includegraphics[width=0.4\textwidth]{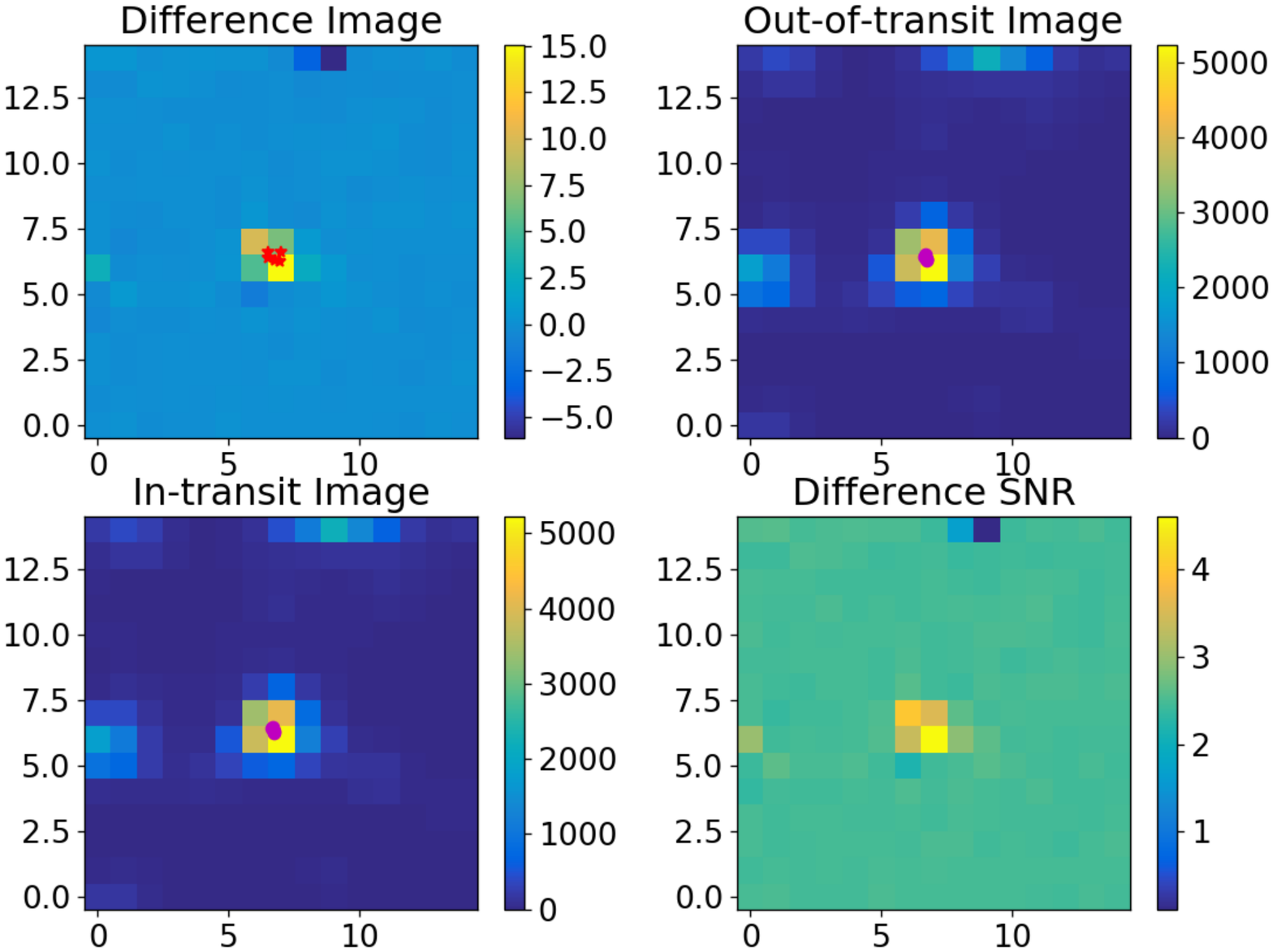}
    \caption{DAVE centroid analysis of L 98-59 c for Sector 2 (four upper left panels), Sector 5 (upper right panels), and Sector 8 (lower panels). The four panels shown are in the same format as in the Data Validation Report, i.e. mean difference image (upper left), mean out-of-transit image (upper right), mean in-transit image (lower left) and SNR of the mean difference image (lower right). The red circles and cyan stars represent the measured individual photocenter for each transit. We measure no significant photocenter shift between the difference and out-of-transit images, consistent with the transit signals originating from the target itself.}
    \label{fig:DAVE_centroid_01}
\end{figure}

\begin{figure}
    \includegraphics[width=0.4\textwidth]{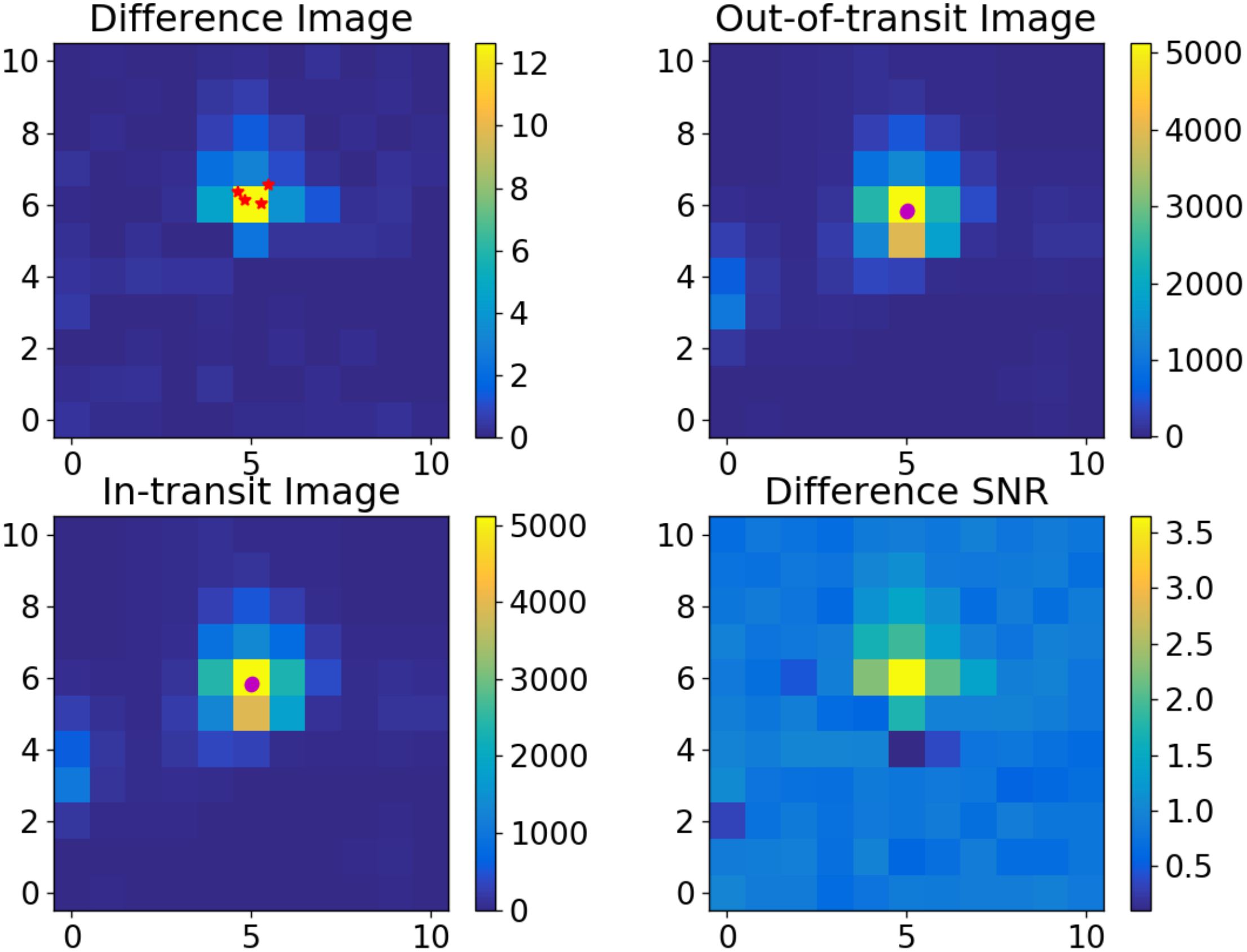}
    \includegraphics[width=0.4\textwidth]{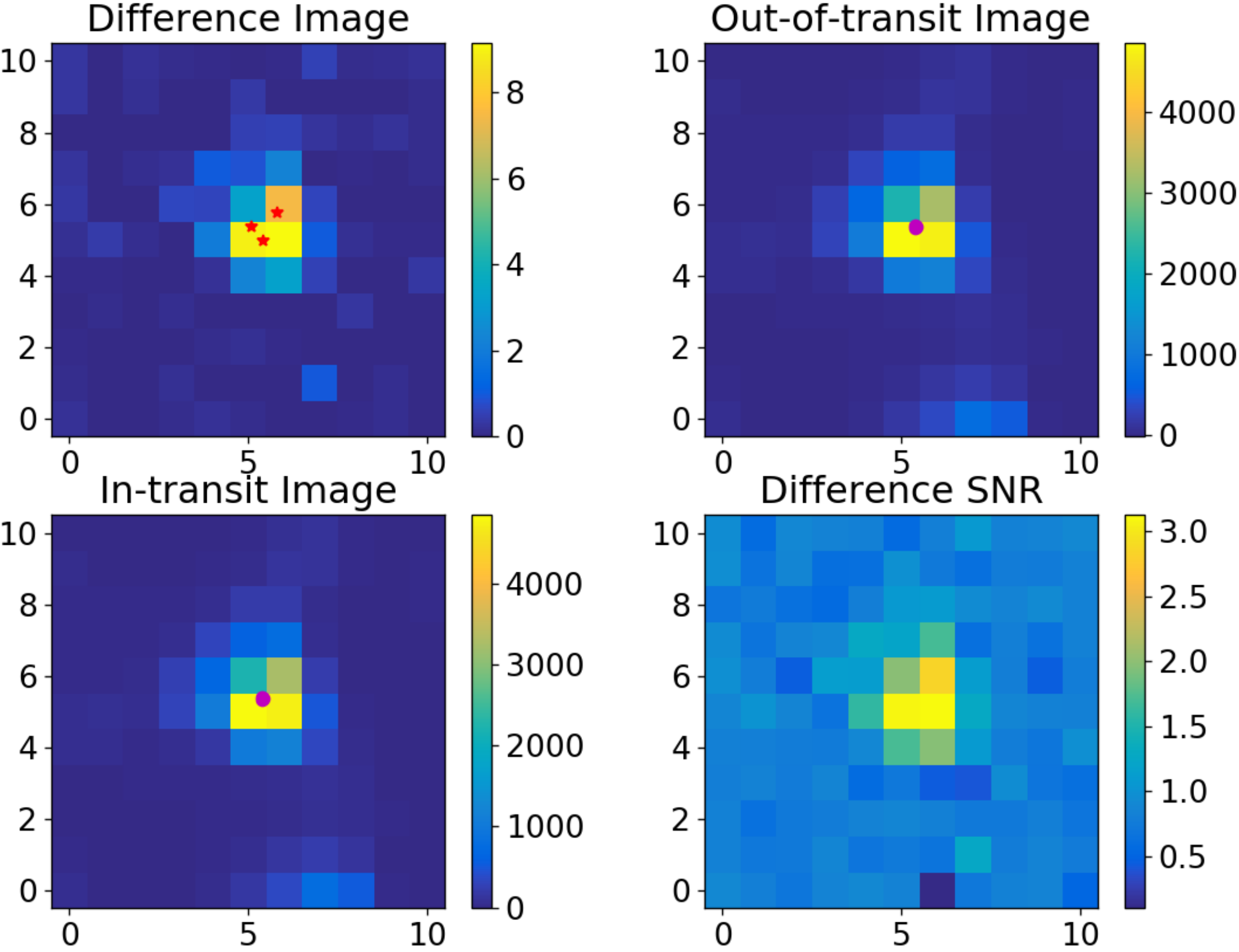}
    \includegraphics[width=0.4\textwidth]{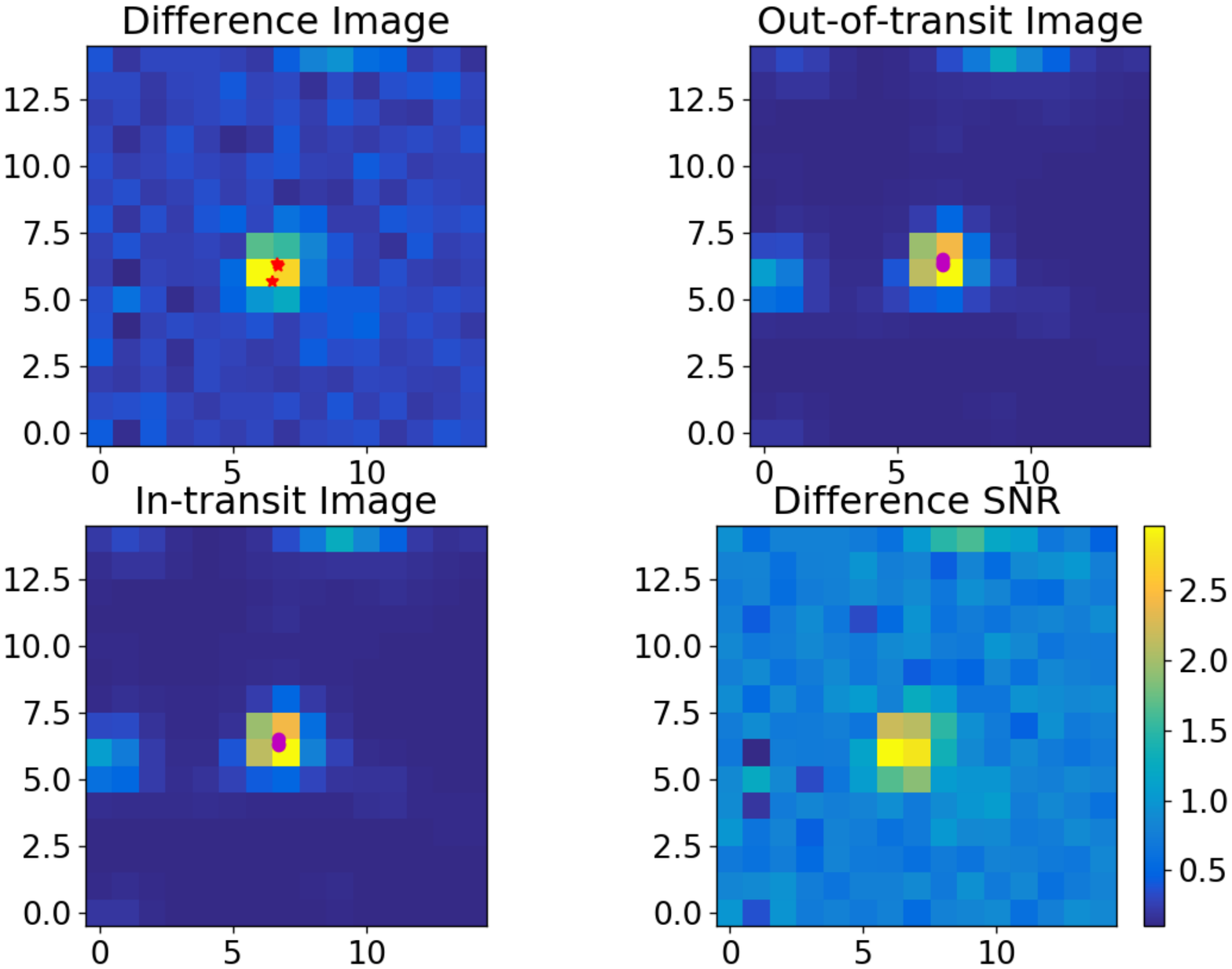}
    \caption{Same as Figure \ref{fig:DAVE_centroid_01} but for L 98-59 d.} 
    \label{fig:DAVE_centroid_02}
\end{figure}

\begin{figure}
\includegraphics[width=0.4\textwidth]{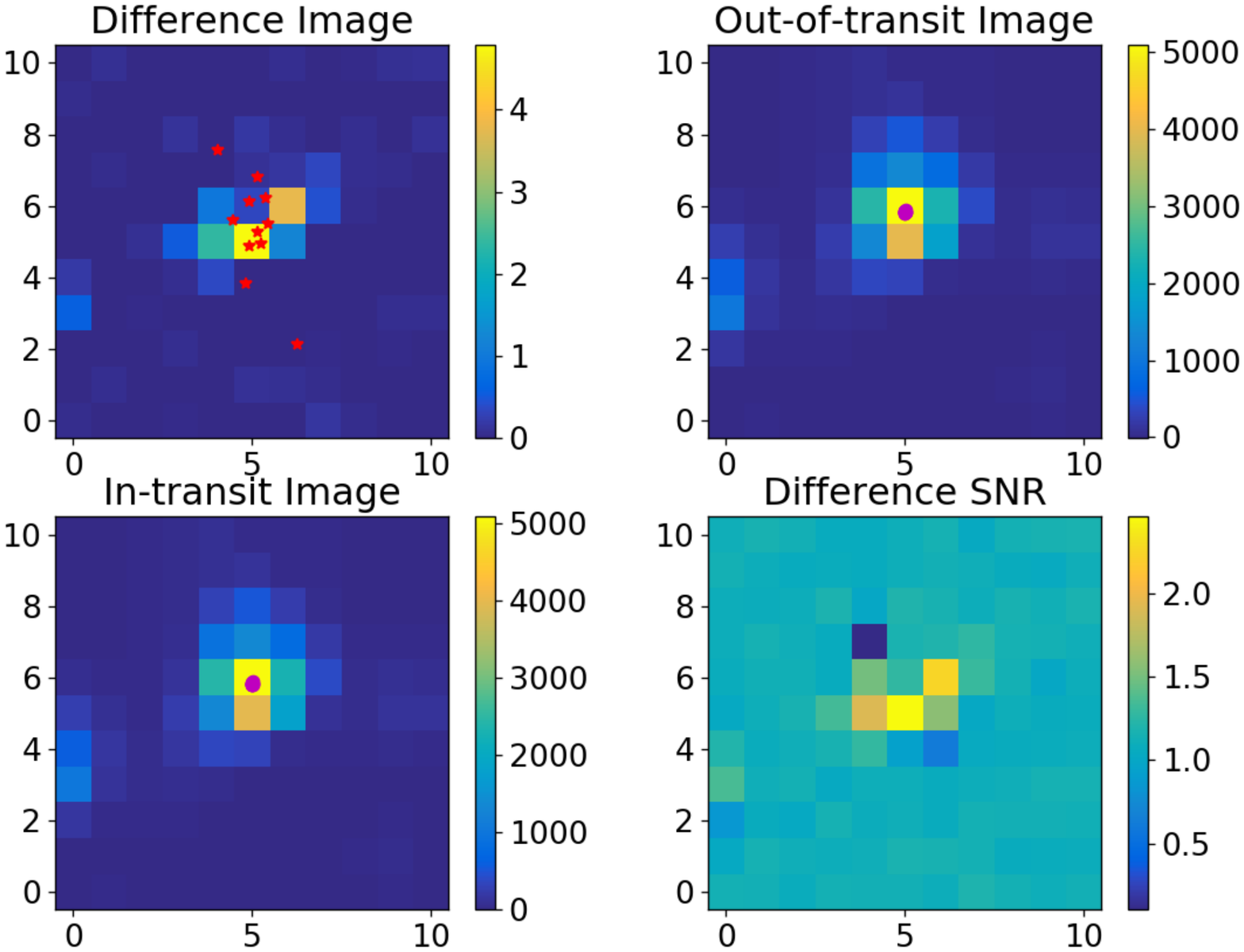}
\includegraphics[width=0.4\textwidth]{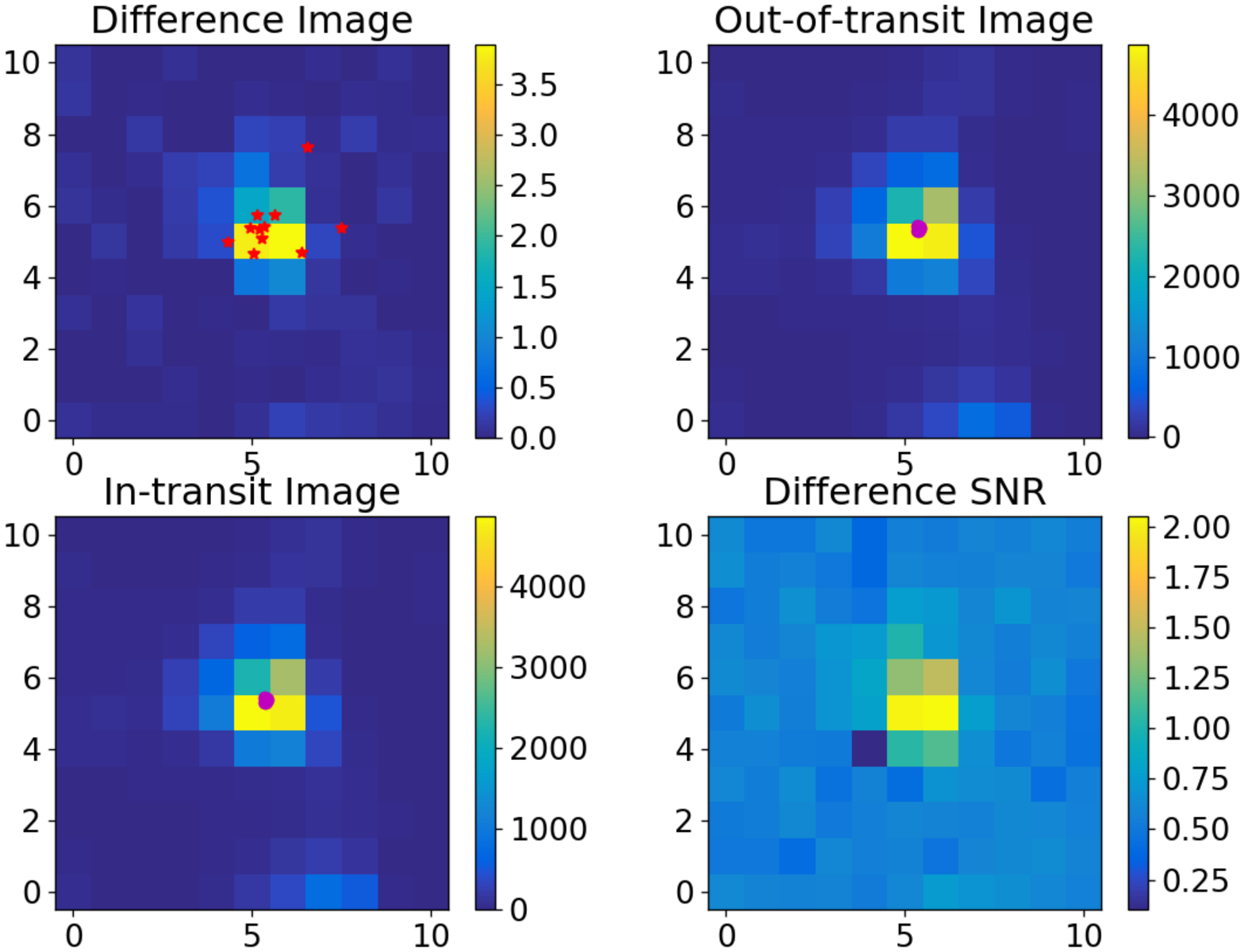}
\includegraphics[width=0.4\textwidth]{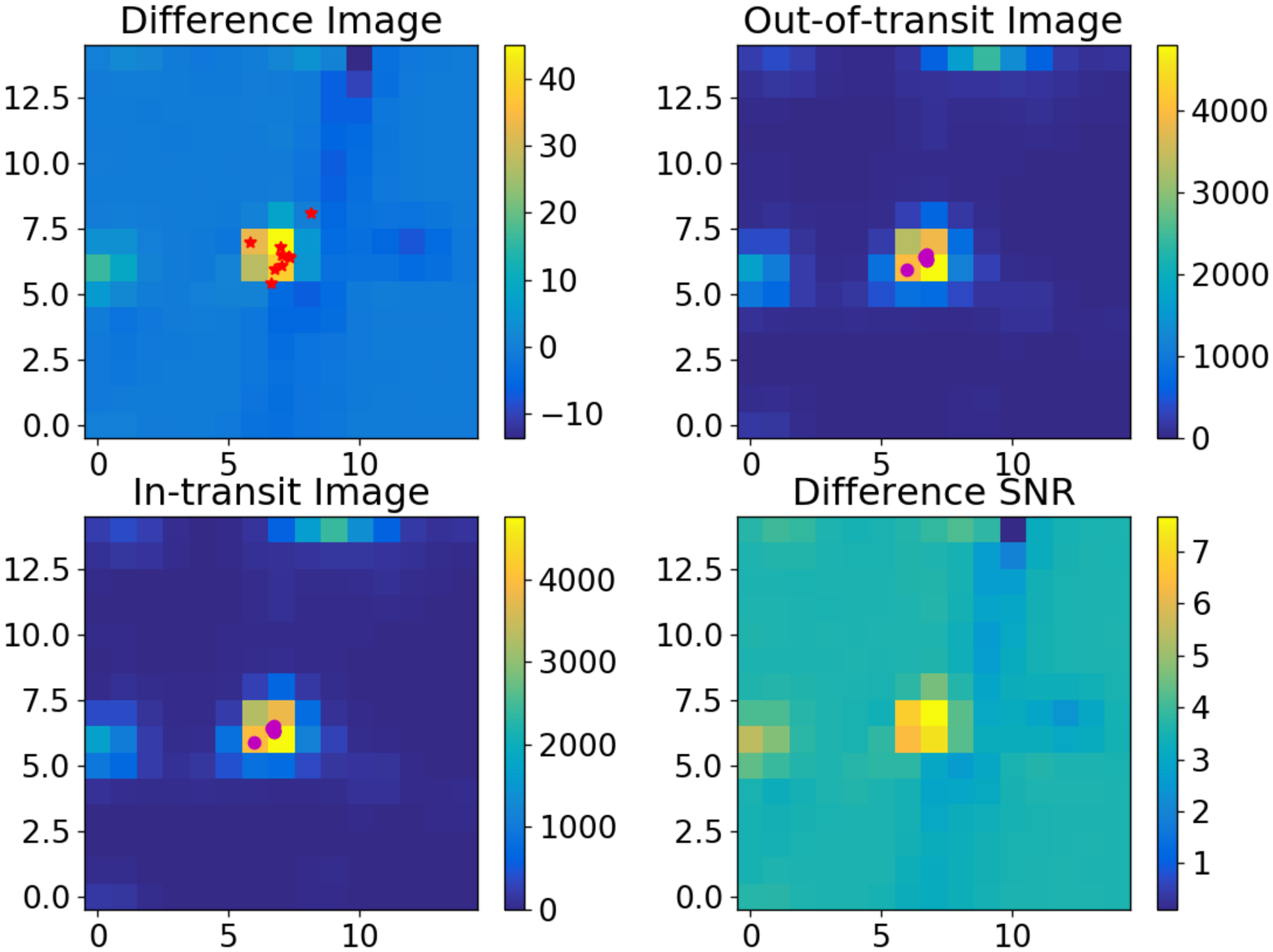}
    \caption{Same as Figure \ref{fig:DAVE_centroid_01} but for L 98-59 b.} 
    \label{fig:DAVE_centroid_03}
\end{figure}

To perform the (d) analysis for each candidate we used the photocenter module of \textsf{DAVE}, following the prescription of \cite{Bryson2013}. Specifically, for each candidate we: 1) create the mean in-transit and out-of-transit images for each transit (ignoring cadences with non-zero quality flags), where the latter are based on the same number of exposure cadences as the former, split evenly before and after the transit; 2) calculate the overall mean in-transit and out-of-transit images by averaging over all transits; 3) subtract the overall mean out-of-transit image from the overall in-transit image to produce the overall mean difference image; and 4) measure the center-of-light for each difference and out-of-transit image by calculating the corresponding $x$- and $y$-moments of the image. The measured photocenters for the three planet candidates are shown in Figures \ref{fig:DAVE_centroid_01}, \ref{fig:DAVE_centroid_02}, and \ref{fig:DAVE_centroid_03}, and listed in Table \ref{tab:centroids}. We detect no significant photocenter shifts between the respective difference images and out-of-transit images for any of the planet candidates (see Table \ref{tab:centroids}), which confirms that the target star is the source of the transits. We note that some of the individual difference images for L 98-59 b deviate from the expected Gaussian profile, and thus so does the mean difference image. 

\begin{deluxetable}{l c c c}[htbp]
\tablecaption{Photocenter Analysis of the three planet candidates \label{tab:centroids}}
\tablewidth{0pt}
\tablehead{
\colhead{Parameter} & \colhead{Row [pixels]} & \colhead{Column [pixels]}
}
\startdata
\multicolumn{3}{c}{} \\
{\bf L 98-59} b & & \\
Out of Transit Image Centroid & $338.83\pm0.02$ & $664.01\pm0.01$ \\
Difference Image Centroid &  $338.34\pm1.39$ & $664.04\pm0.55$ \\
Offset & $-0.49\pm1.39$ & $0.03\pm 0.55$ \\
Offset/$\sigma$ & 0.35 & 0.05 \\
\hline
{\bf L 98-59 c} & & \\
Out of Transit Image Centroid & $338.83\pm0.02$ & $664.01\pm0.01$ \\
Difference Image Centroid & $338.76\pm0.31$ & $664.11\pm0.24$ \\
Offset & $-0.07\pm0.31$ & $0.1\pm 0.24$ \\
Offset/$\sigma$ & 0.23 & 0.42 \\
\hline
{\bf L 98-59 d} & & \\
Out of Transit Image Centroid & $338.93\pm0.02$ & $664.02\pm0.02$ \\
Difference Image Centroid &  $339.27\pm0.21$ & $664.06\pm0.34$ \\
Offset & $0.34\pm0.21$ & $0.04\pm 0.34$ \\
Offset/$\sigma$ & 1.62 & 0.12 \\
\hline
\enddata
\end{deluxetable}

Overall, our \textsf{DAVE} results rule out false-positive features for all three planet candidates of L 98-59, are consistent with the analysis of the Data Validation Report, and indicate that the detected events are genuine transits associated with the star in question. We also note that while the automated vetting of Osborn et al. (2019) flagged L 98-59 with ``a high likelihood of being astrophysical false positives'' (their Table 3), their subsequent manual vetting lists the system as planet candidate.

\begin{figure}
    \includegraphics[width=0.5\textwidth]{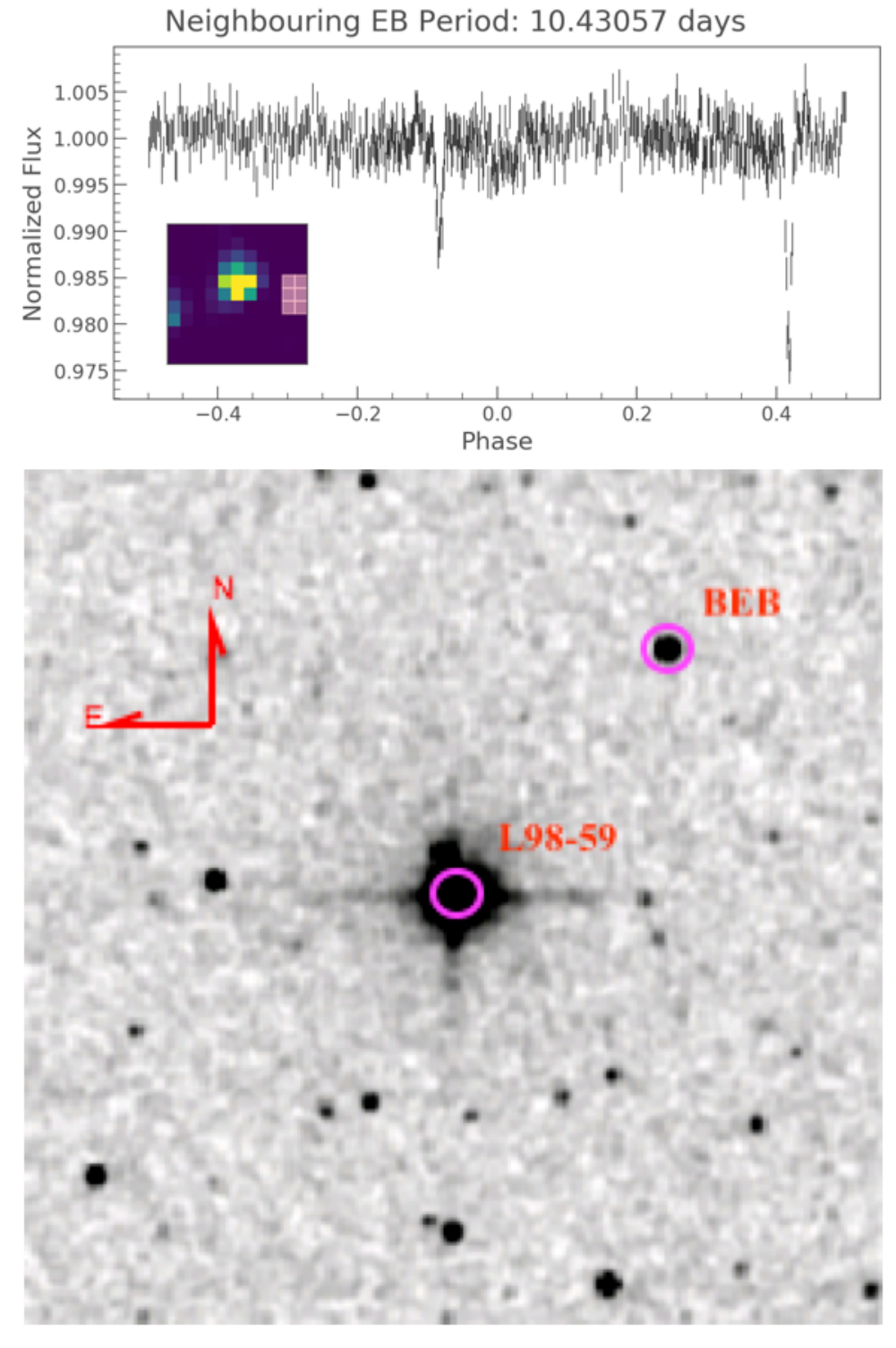}
    \caption{Upper panel: \textsf{lightkurve} analysis of the nearby field star 2MASS 08181825-6818430 showing that it is a background eclipsing binary (BEB) with a period of 10.43 days. The inset panels shows the pixel mask used to extract the lightcurve of the field star (near the right edge of the aperture) not physically associated with the target star. Lower panel: {\rm 4\arcmin x 4\arcmin}~(the size of the TESS aperture) 2MASS J-band image showing the position of the background eclipsing binary (BEB).}
    \label{fig:nearby_EB}
\end{figure}

Additionally, to investigate whether one or more of the transits associated with L 98-59 may result from nearby sources (e.g. a background eclipsing binary), we used \textsf{lightkurve} to extract lightcurves for nearby field stars. Our analysis revealed that a nearby field star ${\sim 80\arcsec}$ NW of L 98-59 (2MASS 08175808-6817459, TIC 307210817, Tmag = 13.45, i.e. $\approx4$ magnitudes fainter than L 98-59) is in fact an eclipsing binary (EB), manifesting both primary and secondary eclipses at a period of ${\approx10.43}$ days, with ${T_0 = 4.4309}$ (BJD-2,455,000) (see Figure \ref{fig:nearby_EB}). This field star could be associated with one of two sources in the Gaia catalog: Source 1 with RA = 124.49217149500, Dec = -68.29612321000, ID = 5271055685541797120, and parallax=0.1888 mas; and Source 2 with RA = 124.49126472300, Dec = -68.29602748080, ID = 5271055689840223744, and parallax = 0.9977 mas. Given the corresponding approximate distances of 1 and 5 kpc, neither of these targets can be physically associated with L 98-59 as they lie deep in the background. Regardless of which of these sources hosts the detected EB, the faintness of the host compared to L 98-59, the measured EB orbital parameters, and the dilution-corrected eclipse depths are inconsistent with the properties of the candidate planets and effectively rule it out as the potential source of any of these signals. 

\subsection{Follow-up Observations}

We pursued ground-based follow up of the three candidates to rule out potential sources of false positives and strengthen the evidence of their planetary nature. Our L 98-59 follow-up program was organized through the TESS Follow-up Observing Program (TFOP) Working Group (WG)\footnote{\url{https://tess.mit.edu/followup/}} which facilitates follow-up of TESS candidate systems. The primary goal of the TFOP WG is to provide follow-up observations that will advance the achievement of the TESS Level One Science Requirement to measure masses for 50 transiting planets smaller than 4 Earth radii. A secondary goal of the TFOP WG is to foster communication and coordination for any science coming out of TESS. Our L 98-59 follow-up was conducted by three TFOP Sub Groups (SGs): SG-1, seeing limited photometry; SG-2, reconaissance spectroscopy; and SG-3, high-resolution imaging.

\subsubsection{Seeing-Limited Photometry from the TFOP WG} 

Analysis of multi-planet systems from {\it Kepler} has shown that these have a higher probability of being real planets \cite[e.g.,][]{Lissauer2012}, lending credibility to the planetary nature of the transit events associated with L 98-59. However, the pixel scale of TESS is larger than Kepler's (${21\arcsec}$ for TESS vs. ${4\arcsec}$ for Kepler) and the point spread function of TESS could be as large as ${1\arcmin}$, both of which increase the probability of contamination by a nearby eclipsing binary (EB). For example, a deep eclipse in a nearby faint EB might mimic a shallow transit observed on the target star due to dilution. Thus it is critical to explore the potential contamination of relatively distant neighbors in order to confirm transit events detected on a TESS target.  

To identify potential false positives due to variable stars such as EBs up to ${2.5\arcmin}$ away from L 98-59, we made use of the TFOP SG1. Specifically, we observed the target with ground-based facilities at the predicted times of the planet transits to search for deep eclipses in nearby stars at higher spatial resolutions. We used the {\tt TESS Transit Finder}, which is a customized version of the {\tt Tapir} software package \citep{Jensen:2013}, to schedule photometric time-series follow-up observations. The facilities we used to collect TFOP SG1 data are: Las Cumbres Observatory (LCO) telescope network \citep{brown2013}; SPECULOOS South Observatory (SSO) \citep{laetitia2018spie,burdanov2018}; MEarth-South telescope array (MEarth) \citep{irwing2015}; and Siding Spring Observatory T17 (SSO T17). Detailed observation logs are provided in Table \ref{TFOP}.

\begin{table*}
\footnotesize
\caption{Observation Log}  
\label{TFOP}      
\centering 
\begin{tabular}{c l l c c c c c c c}     
\hline\hline       
\noalign{\smallskip}                     
\multirow{2}{*}{Planet} & Date & \multirow{2}{*}{Telescope}\tablenotemark{$\dag$} & \multirow{2}{*}{Filter} & ExpT & Exp & Dur. & Transit & Aperture  & FWHM\\
& (UTC) &  &  & (sec) & (N) & (min) & coverage & (arcsec) & (arcsec)   \\
\noalign{\smallskip} 

\hline 
\noalign{\smallskip} 
\multirow{8}{*}{b}
& 2018-10-19 & LCO-SSO${^\ddag}$ & r$^\prime$ & 120 & 41 & 147 & Ingr.+71$\%$ &  3.89 & 2.21  \\
& 2018-11-11 & SSO-Europa & r$^\prime$  & 15 & 764 & 315 & Full & 5.25 & 4.53 \\
& 2018-11-11 & LCO-CTIO-1  & i$^\prime$ & 25 & 120 & 154 & Full & 5.83 & 3.49  \\
& 2018-11-18 & LCO-SAAO-0.4${^\ddag}$ & i$^\prime$ & 70 & 125 & 175 & Full & 9.14 & 7.03  \\ 
& 2018-11-20 & LCO-CTIO-1  & r$^\prime$ & 30 & 146 & 178 & Full & 3.89 & 2.20  \\
& 2018-11-29 & LCO-CTIO-1  & r$^\prime$ & 30 & 186 & 223 & Full & 5.83 & 3.56  \\
& 2018-12-07 & SSO-Io & r$^\prime$  & 15 & 998 & 415 & Full & 5.60 & 4.06 \\
& 2019-01-26 & LCO-SAAO-1${^\ddag}$ & r$^\prime$ & 12 & 88 & 234 & Full & 3.89 & 3.08  \\ 
\hline  
\noalign{\smallskip}                  
\multirow{7}{*}{c}
& 2018-10-16 & LCO-CTIO-1 & i$^\prime$ & 20 & 63 & 70 & Ingr.+30$\%$ & 4.27 & 2.01 \\
& 2018-10-22${^\ddag}$ & SSO-T17 & clear & 30 & 122 & 86 & Ingr.+77$\%$ & 7.10 & 2.40 \\
& 2018-11-11 & SSO-Europa & r$^\prime$  & 15 & 764 & 315 & Full & 5.25 & 4.53 \\
& 2018-11-22 & MEarth  & RG715 & 45 & 1682 & 380 & Full &  20.16 & 8.00  \\ 
& 2018-12-25 & LCO-SSO-1  & i$^\prime$ & 22 & 108 & 113 & Full &  5.05 & 1.89  \\
& 2019-01-20 & LCO-CTIO-1  & g$^\prime$ & 100 & 85 & 197 & Full & 6.22 & 2.75  \\ 
& 2019-01-20 & LCO-CTIO-1  & zs & 30 & 170 & 197 & Full & 9.36 & 4.43  \\ 
\hline
\noalign{\smallskip} 
\multirow{5}{*}{d}
& 2018-11-07 & LCO-CTIO-0.4  & i$^\prime$ & 70 & 119 & 170 & Full & 6.85 & 4.48  \\
& 2018-11-22 & LCO-CTIO-1${^\ddag}$  & i$^\prime$ & 25 & 108 & 113 & Full & 6.22 & 5.05  \\ 
& 2018-11-22 & MEarth  & RG715 & 45 & 1682 & 380 & Full &  20.16 & 8.00  \\ 
& 2019-01-13 & LCO-CTIO-0.4  & i$^\prime$ & 14 & 143 & 73 & Full & 5.14 & 2.32  \\
& 2019-01-20 & LCO-CTIO-1  & r$^\prime$ & 30 & 132 & 151 & Full & 6.22 & 2.89  \\ 
& 2019-01-28 & LCO-CTIO-1  & g$^\prime$ & 50 & 153 & 223 & Full & 7.78 & 2.35  \\

\noalign{\smallskip}
\hline
\noalign{\smallskip}

\end{tabular}
\tablenotetext{$\dag$}{Telescopes: \\
             LCO-SSO-1: Las Cumbres Observatory - Siding Spring (1.0 m) \\
             LCO-CTIO-1: Las Cumbres Observatory - Cerro Tololo Interamerican Observatory (1.0 m) \\
             LCO-CTIO-0.4: Las Cumbres Observatory - Cerro Tololo Interamerican Observatory (0.4 m) \\
             LCO-SAAO-1: Las Cumbres Observatory - South African Astronomical Observatory (1.0 m) \\
             LCO-SAAO-0.4: Las Cumbres Observatory - South African Astronomical Observatory (0.4 m) \\
             SSO-Europa: SPECULOOS South Observatory - Europa (1.0 m) \\
             SSO-Io: SPECULOOS South Observatory - Io (1.0 m) \\
             SSO-T17: Siding Spring Observatory - T17 (0.4 m)\\
             MEarth: MEarth-South telescope array (0.4 m $\times$ 5 telescopes)
             }
             
\tablenotetext{$\ddag$}{Observations not shown in Figures \ref{fig:ground_175_1}, \ref{fig:ground_175_2}, and \ref{fig:ground_175_3} due to intrinsically high scatter in the light curve and/or because they were a deep exposure search for eclipsing binaries in nearby stars.}
            
\end{table*}

\begin{figure}
    \centering
    \plotone{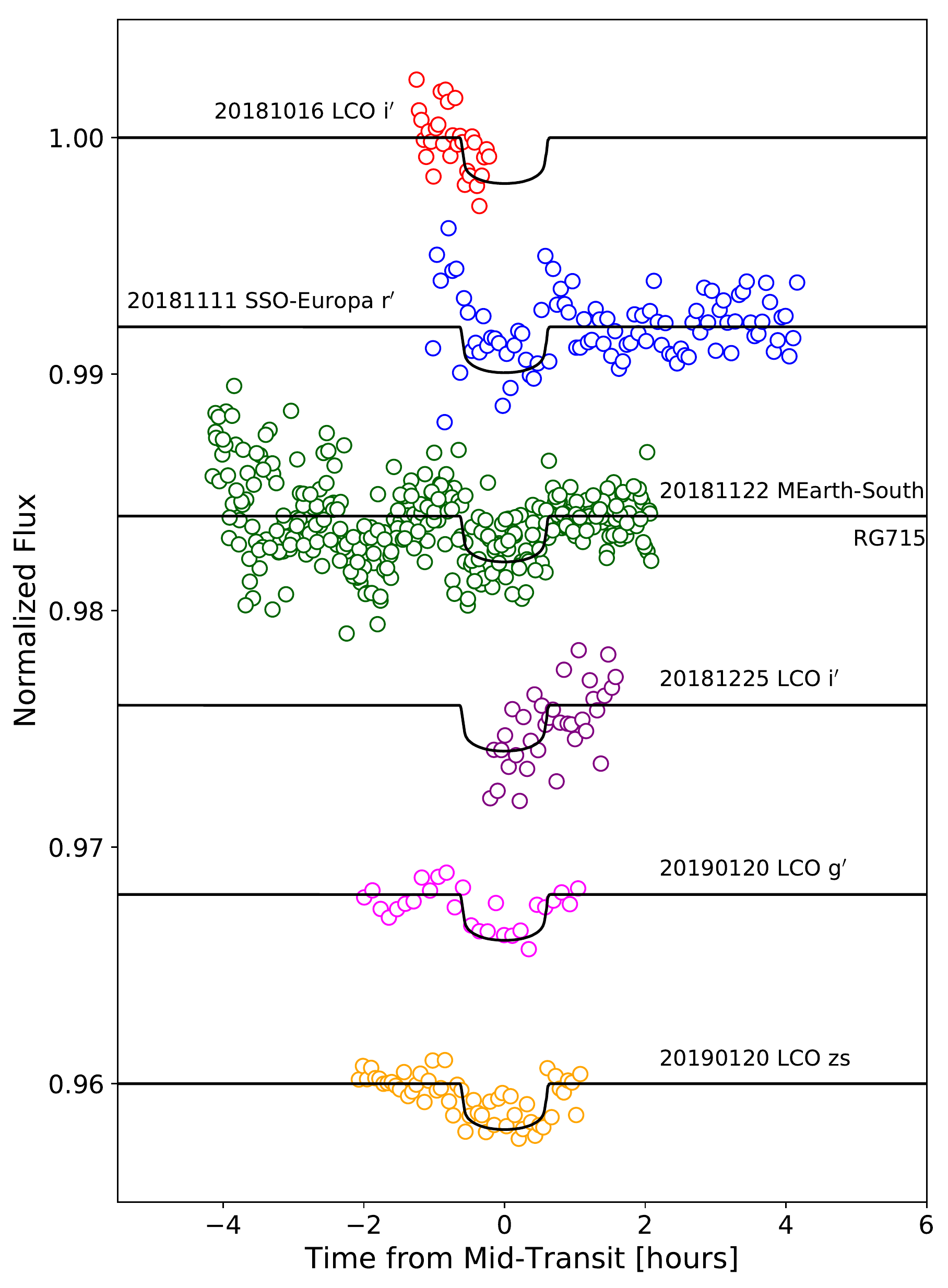}
    \caption{Ground-based follow-up observations of L 98-59 c. The date, facility, and filter used for each observation is marked, and each data set is offset for clarity. The black line represents the transit model based on the TESS data.}
    \label{fig:ground_175_1}
\end{figure}

\begin{figure}
    \centering
    \plotone{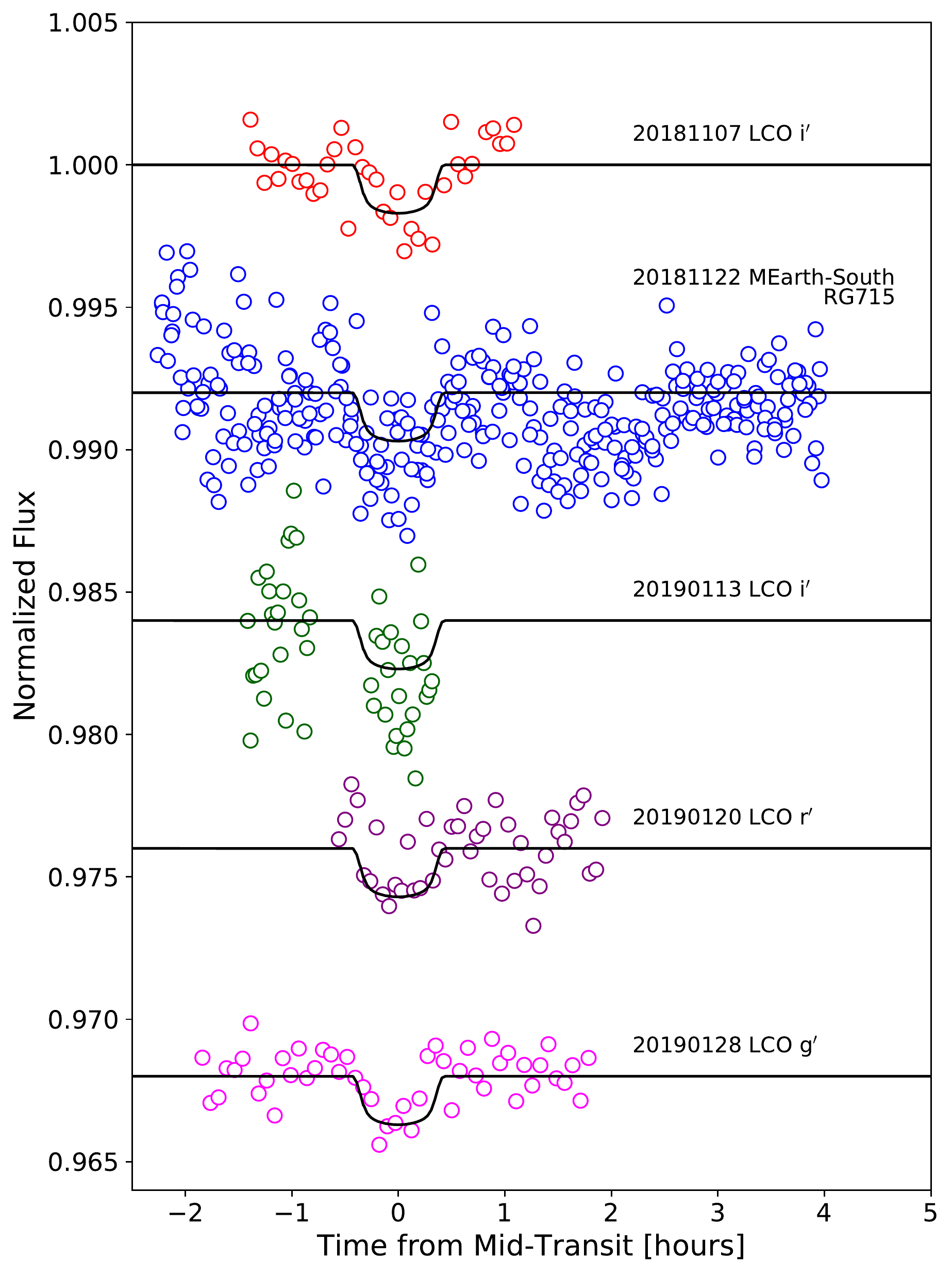}
    \caption{Same as Figure \ref{fig:ground_175_1} but for L 98-59 d.}
    \label{fig:ground_175_2}
\end{figure}

\begin{figure}
    \centering
    \plotone{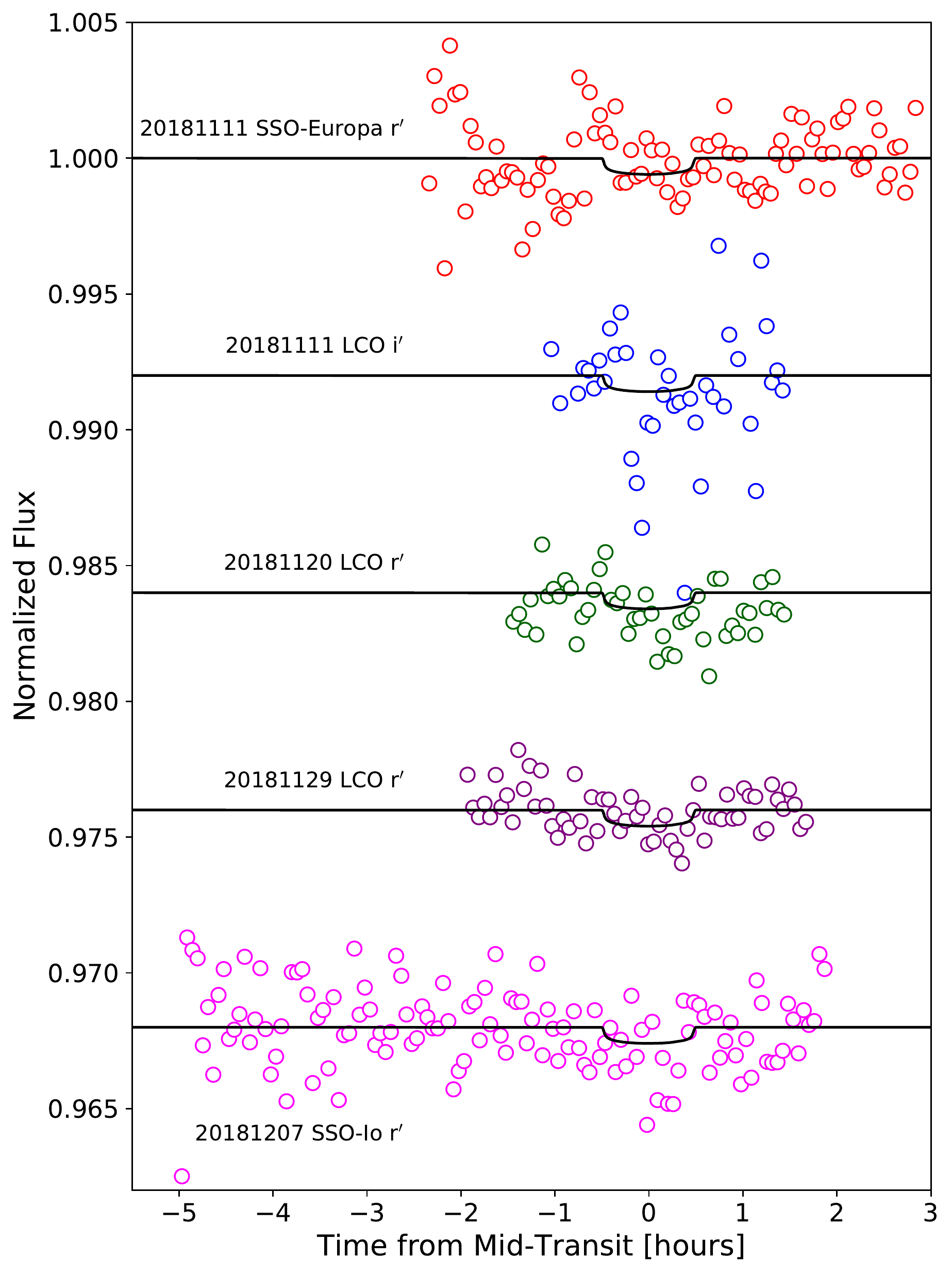}
    \caption{Same as Figure \ref{fig:ground_175_1} but for L 98-59 b.}
    \label{fig:ground_175_3}
\end{figure}

We used the AstroImageJ software package \citep{karen2017} for the data reduction and the aperture photometry in most of these follow-up photometric observations. For the observations carried out at SSO, the standard calibration of the images and the extraction of the stellar fluxes were performed using the IRAF/DAOPHOT aperture photometry software as described in \cite{gillon2013}. The results are shown in Figures \ref{fig:ground_175_1}, \ref{fig:ground_175_2} and \ref{fig:ground_175_3}.

For all three planet candidates, we confirmed that the target star is the source of the transits, and ruled out nearby EBs which could mimic the transits. In addition, observations of L 98-59 c and d in different filters showed no chromatic dependence, which strengthens the hypothesis that the candidates are real planets. Our follow-up did detect a nearby EB at a separation of 54\arcsec\ (TIC 307210845, Tmag = 16.042, i.e. $\sim7$ magnitudes fainter than TIC 307210830, producing no detectable eclipses in the lightcurve of the latter), and we used deep exposures to confirm that it is not the origin of the L 98-59 b transits, providing high level of confidence about the planetary nature of this candidate. 

The measured transit depths revealed by follow-up transit photometry are consistent with the transit depths measured from TESS. The differences in the follow-up and TESS transit depth measurements (in terms of $Rp/R*$) are listed in Table \ref{table:depths} as a function of wavelength, where we have included only the transits with scatter low enough to reasonably detect the events.

\begin{table*}[htb]
\centering
\caption{Follow-up ${R_p/R_*}$ minus TESS ${R_p/R_*}$ as a function of wavelength \label{table:depths}}
\begin{tabular}{l c c c c}
TOI & Date Obs & Filter (Observatory)  & ${R_p/R_*}$ minus TESS ${R_p/R_*}$\\
\hline
L 98-59 b & UT 2018-11-20 & Sloan r' (LCO) & $-0.0008^{+0.0054}_{-0.0085}$ \\
L 98-59 b & UT 2018-11-29 & Sloan r' (LCO) & $0.0018^{+0.0039}_{-0.0045}$ \\
\hline
L 98-59 c & UT 2019-01-20 & Sloan g' (LCO) & $-0.0033\pm0.003$\\
L 98-59 c & UT 2018-11-22 & RG715 (MEarth) & $-0.0007\pm0.002$ \\
L 98-59 c & UT 2019-01-20 & Sloan z' (LCO) & $-0.0056\pm0.0028$\\
\hline
L 98-59 d & UT 2019-01-28 & Sloan g' (LCO) & $0.0072^{+0.0078}_{-0.0064}$ \\
L 98-59 d & UT 2019-01-20 & Sloan r' (LCO) & $-0.003^{+0.01}_{-0.012}$ \\
L 98-59 d & UT 2018-11-07 & Sloan i' (LCO) & $0.0072^{+0.0088}_{-0.0072}$ \\
L 98-59 d & UT 2018-11-22 & RG715 (MEarth) & $0.0104^{+0.0044}_{-0.0043}$ \\
\hline
\end{tabular}
\end{table*}

\subsubsection{Reconnaissance Spectroscopy}

To investigate the magnetic activity and rotation of L 98-59 and rule out spectroscopic binary companions, we obtained two epochs of optical spectra of L~98-59 on UT 2018 February 12\footnote{The first epoch of spectroscopic data were obtained as part of the M dwarf spectroscopic program described in \citet{Winters2019}, before $TESS$ began observations.} and on UT 2018 November 20 using the slicer mode with the CTIO HIgh ResolutiON (CHIRON) spectrograph \citep{Tokovinin(2013)} (R $\simeq$ 80,000) on the Cerro Tololo Inter-American Observatory (CTIO)/ Small and Moderate Aperture Research Telescope System (SMARTS) 1.5m telescope. CHIRON has a spectral range of 410-870 nm. We obtained one 7.5-minute exposure for the first epoch and 3 $\times$ 2.5-minute exposures for the second observation, yielding a signal-to-noise ratio of roughly 13 in order 44 for both epochs. As described in \citet{Winters(2018)}, we used an observed template of Barnard's Star to derive a radial velocity of -5.8$\pm$0.1 km s$^{-1}$ using the TiO molecular bands at 7065--7165 \AA \footnote{We note that the total uncertainty on the systemic velocity should include the 0.5 km s$^{-1}$ uncertainty on the Barnard's Star template velocity.}. Our analyses of the spectra reveal no evidence of double lines. We see negligible rotational broadening (\vsini\ = $0.0 \pm 1.9$ km/s), and do not see H$\alpha$ in emission, providing evidence that the star is inactive and not host to unresolved, close-in, stellar companions. We are also able to rule out the presence of a brown dwarf companion to the host star. The radial velocity difference between the two observations, separated in time by roughly nine months, is 53$\pm$52 m s$^{-1}$. For comparison, a 13 Jupiter-mass companion in a circular, nine-month period would induce a velocity semi-amplitude of 863 m s$^{-1}$ on this star. This semi-amplitude is eight times larger than our velocity uncertainty and would have been readily detectable. Thus, it is highly unlikely that there is a low-mass stellar or brown dwarf companion around L~98-59 at periods shorter than nine months.

We note, as well, that the trigonometric distance of 10.623$\pm$0.003 pc from the {\it Gaia} second data release is in agreement with the photometric distance estimate of 12.6$\pm$1.9 pc reported in \citet{Winters2019}. Undetected equal-luminosity companions would contribute light to the system, making the system overluminous and resulting in an underestimated photometric distance estimate. Because the two distances are in agreement, this lends further support to the host star being a single star. 

\begin{figure}
    \includegraphics[width=0.5\textwidth]{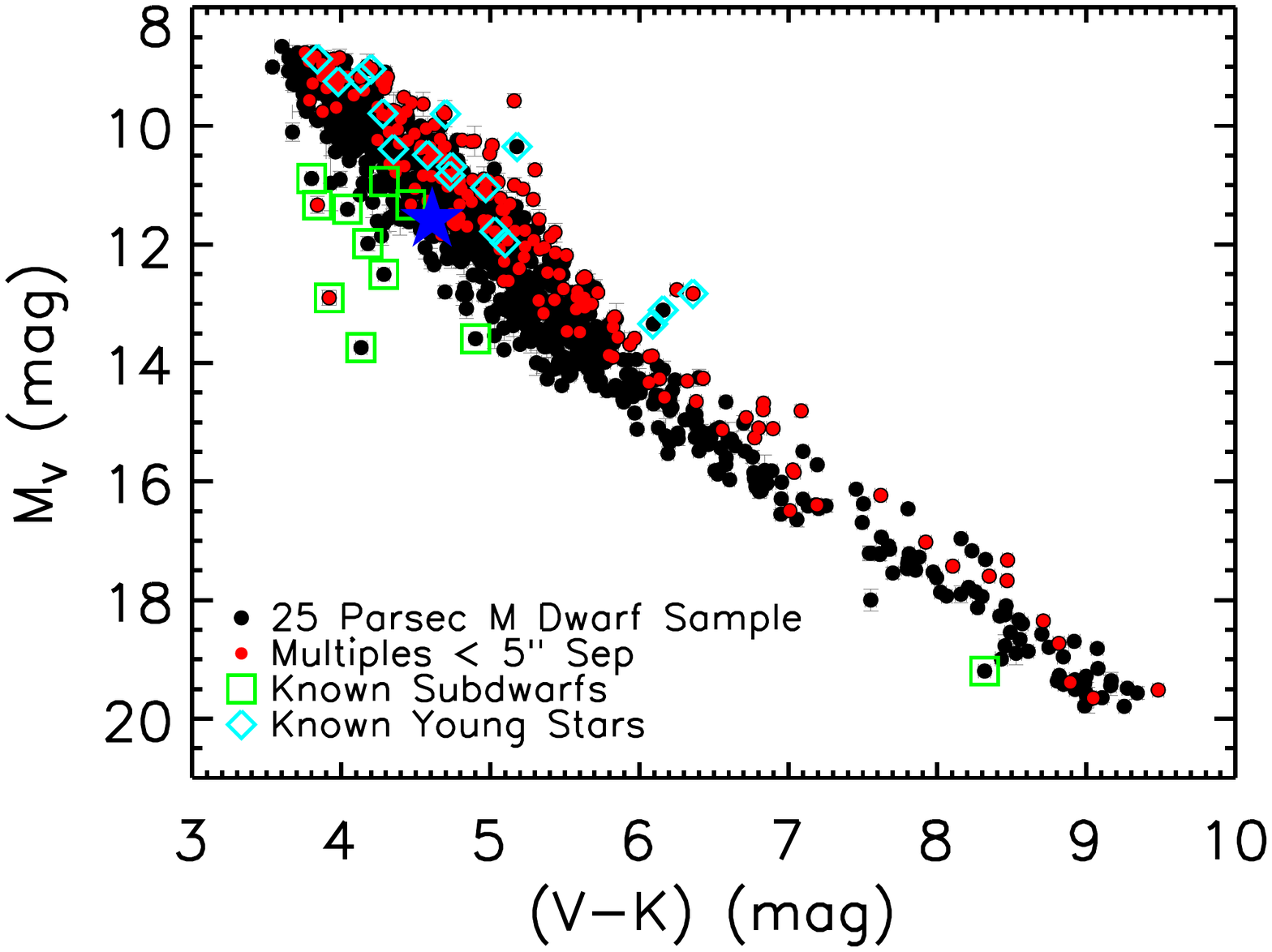}
    \caption{Observational Hertzsprung-Russel Diagram. The sample of 1120 M dwarf primaries within 25 pc from \citet{Winters2019} are plotted as black points. L~98-59 is noted as a blue star. For comparison, known close multiples with separations less than 5\arcsec~having blended photometry (red points), known cool subdwarfs (open green squares), and known young objects (open cyan diamonds) are noted. Error bars are shown in gray, and are smaller than the points, in most cases. }
    \label{fig:hrd}
\end{figure}

We also placed L~98-59 on an observational Hertzsprung-Russel color-magnitude Diagram (blue star; see Figure \ref{fig:hrd}). Because it is not elevated above the main sequence or among the blended photometry binary sequence (red points), even more strength is given to the argument of this star being single.

In addition, we obtained a near-IR spectrum of L~98-59 on 2018 December 22 with the Folded-port InfraRed Echellete (FIRE) spectrograph \citep{simcoe2008} on the 6.5 Baade Magellan telescope at Las Campanas observatory. FIRE covers the 0.8-2.5 micron band with a spectral resolution of $R = 6000$. The target was observed under favorable conditions, with an average seeing of $\sim 0\farcs6$. L~98-59 was observed twice in the ABBA nod patterns at 40s integration time for each frame using the $0\farcs6$ slit. Reductions and telluric corrections, using the nearby A0V standard HIP 41451, were completed with the FIREhose IDL package. We derived stellar parameters following the empirical methods derived by \cite{Newton2015}. For L~98-59, we infer: $T_{eff} = 3620 \pm 74$K, $R_{star} = 0.37 \pm 0.027 R_{\odot}$, and $L = 0.021 \pm 0.004 L_{\odot}$., consistent with the SED analysis.

\subsubsection{High Resolution Imaging}



\begin{figure*}
    \centering
    \plotone{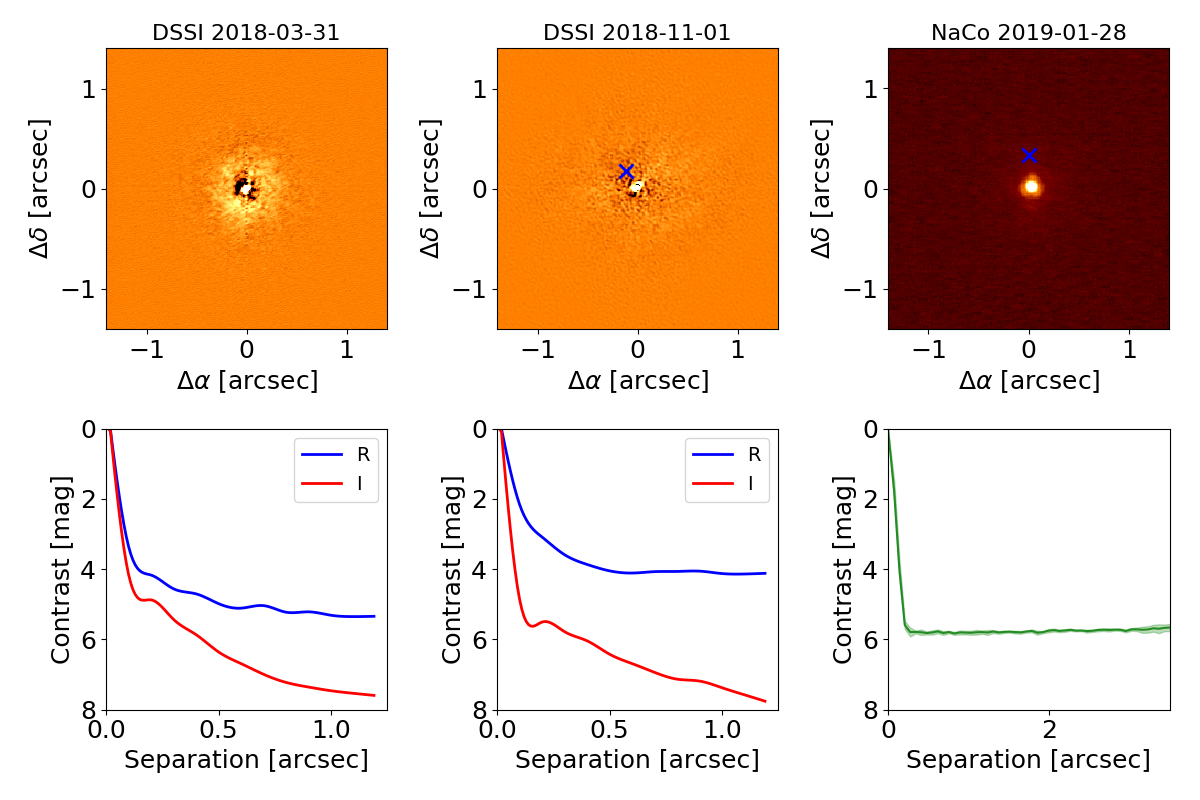}
    \caption{\textit{Top:} High resolution images of the target at each epoch. In the center and right panel, the blue cross indicates the position of the target on UT 2018 March 31, when the first Gemini/DSSI dataset was collected, and we show only the I band images here. No companions are detected here or anywhere in the field of view in any of the images. Only the central portion of the NaCo image is shown. \textit{Bottom:} 5-sigma sensitivity to companions as a function of separation from the host star, for each data epoch. Red and blue lines indicate the Gemini/DSSI R and I band observations respectively, which have central wavelengths 692nm and 880nm.}
    \label{fig:hri}
\end{figure*}

Photometric contamination from nearby sources can result in various false positive scenarios (e.g. background eclipsing binaries), and can bias the measured planetary radius from photometric analysis \citep[see e.g.][]{Ciardi2015,Furlan2017,Ziegler2018}. In this work we use several high resolution images to tightly constrain the possible background sources and companion stars present near L 98-59. Previous speckle observations of the target were collected with Gemini/DSSI on 2018 March 31 as part of the M dwarf speckle program described in \citet{Winters2019}. Once the candidate planets in this system had been identified by TESS, we collected additional speckle images with Gemini/DSSI \citep{Horch(2011b)} on 2018 November 01, and AO images with VLT/NaCo on 2019 January 28. Both epochs of Gemini/DSSI data are collected simultaneously through R and I band filters (692nm and 880nm respectively), while the VLT/NaCo data are collected in the Br$\gamma$ filter. At 0.5'' from the host, these data rule out companions 5.0, 6.6 and 5.8 magnitudes fainter than the host star in the R, I and Br$\gamma$ bands respectively. The 5-$\sigma$ contrast curves for each of these observations are presented in Figure \ref{fig:hri}.

Due to the high proper motion of the target (354~mas/yr), the target undergoes significant motion even over the relatively modest time baseline of these observations. The on-sky position of the target is displaced by 338~mas between the observations on 2018 March 31 and 2019 January 28. This motion is significantly more than the PSF width in the high resolution images, and we are therefore able to rule out the presence of stationary background objects within $\sim$6 magnitudes of the host at any separation: any background objects obscured by the target in the first observation would be clearly visible in the final observation and vice versa. The motion of the target is demonstrated in Figure \ref{fig:hri}. These data also allow tight constraints to be placed on the presence of co-moving companions beyond $\sim$150~mas (=1.6~AU at the distance of this target). An object with $\Delta$Br$\gamma\sim$6~mag and at the distance of L 98-59 would have a mass of $\sim75$M$_{\rm J}$ at an age of 10 Gyr \citep{Baraffe2003}\footnote{Given the difficulty in estimating M-dwarf ages, we use an age of 10 Gyr so as to calculate a conservative mass limit.}, and we can therefore rule out any stellar companions to this host with a projected separation greater than 1.6 AU, while stellar companions closer than this could be easily detected in radial velocity data with a sufficient baseline.




Overall, the follow-up efforts demonstrate that there are tentative transit detections of the various candidates from the ground but it is difficult to confirm them since they are all shallow events. More importantly, none of these detections  were identified as eclipsing binary scenarios (either on the host star or on nearby stars), supporting the planetary nature of the three transit candidates.


\section{Discussion}
\label{sec:sys_param}

Our lightcurve and false positive analyses, follow-up observations, and the multiplicity of the system provide strong evidence that the detected transit signals are planetary in nature. We consider the planets L 98-59b, c, and d to be a validated system of terrestrial planets orbiting a very nearby, bright M dwarf. Here we explore additional properties of the system to place constraints on the planet masses, orbital dynamics and evolution, and discuss their potential for future characterization. 

\subsection{Planet Mass Constraints}
\label{sec:masses}
In the absence of radial velocity measurements, we placed constraints on the masses of the planets using the \textsf{forecaster} package for probabilistic mass forecasting \citep{forecaster}. From the mean and standard deviation of each planet’s radius, we generated a grid of 5000 masses within the entire mass range of the conditioned model, which spans dwarf planets to high-mass Jovians, and sampled 50000 times from a truncated normal distribution. For each sampled radius, \textsf{forecaster} computes a vector of probabilities given each element in the mass grid and a randomly chosen set of hyper-parameters from the hyper-posteriors of the model (which include transition points and intrinsic dispersion in the mass-radius relation). From this vector, the package returns the median mass and $\pm1\sigma$ values. From the calculated radius values of 0.8 [0.05] ${R_\oplus}$ (L 98-59 b), 1.35 [0.07] ${R_\oplus}$ (L 98-59 c), and 1.57 [0.14] ${R_\oplus}$ (L 98-59 d), we determined mass values of 0.5 [+0.3, -0.2] ${M_\oplus}$,  2.4 [+1.8, -0.8] ${M_\oplus}$, and 3.4 [+2.7, -1.4] ${M_\oplus}$, respectively. The large errors on these values suggest that better constrained radii from continued follow-up observations, combined with precise radial velocity measurements, are necessary to constrain the true masses. We note that given the brightness of the host star, the L 98-59 planets should be great targets for mass measurements to establish the M-R relation for M-dwarf planets. Using the \textsf{forecaster} masses, the expected radial velocity semi-amplitude, K, for the three planets are 0.54, 2.22, and 2.48 m/s for L 98-58 b,  L 98-58 c, and  L 98-58 d respectively---the outer two comparable to the amplitude of the measured radial velocity signal produced by e.g. GJ 581 b \citep{Mayor2009} and Proxima Centauri b \citep{Anglada2016}.

\subsection{Dynamical stability and transit-timing variations}

\subsubsection{Long-Term Stability}
To examine the long-term dynamical stability and orbital evolution of the L 98-59 planets, we integrated the system using Rebound \citep{Rein2015} for 1 million orbits of the outer planet (${\rm P = 7.45}$ days). We used two sets of initial conditions -- planets on circular orbits, and on eccentric orbits with $e = 0.1$ -- and start all integrations with randomly-selected initial arguments of periastron. Given the large uncertainties on the \textsf{forecaster} masses, we also tested the dynamical stability for two sets of planetary masses: best-fit masses (i.e. 0.5 ${M_\oplus}$, 2.4 ${M_\oplus}$ and 3.4 ${M_\oplus}$ for L 98-58 b, L 98-58 c and L 98-58 d respectively), and best-fit+${1\sigma}$ masses (i.e. 0.58 ${M_\oplus}$, 4.2 ${M_\oplus}$ and 6.1 ${M_\oplus}$ for L 98-58 b, L 98-58 c and L 98-58 d respectively). Overall, we performed 1000 numerical simulations for each set of planetary masses, using the IAS15 non-symplectic integrator \citep{Rein2015} with a timestep of 0.01 the orbit of the inner planet (i.e. about 30 min).

\begin{figure*}
    \centering
    \plotone{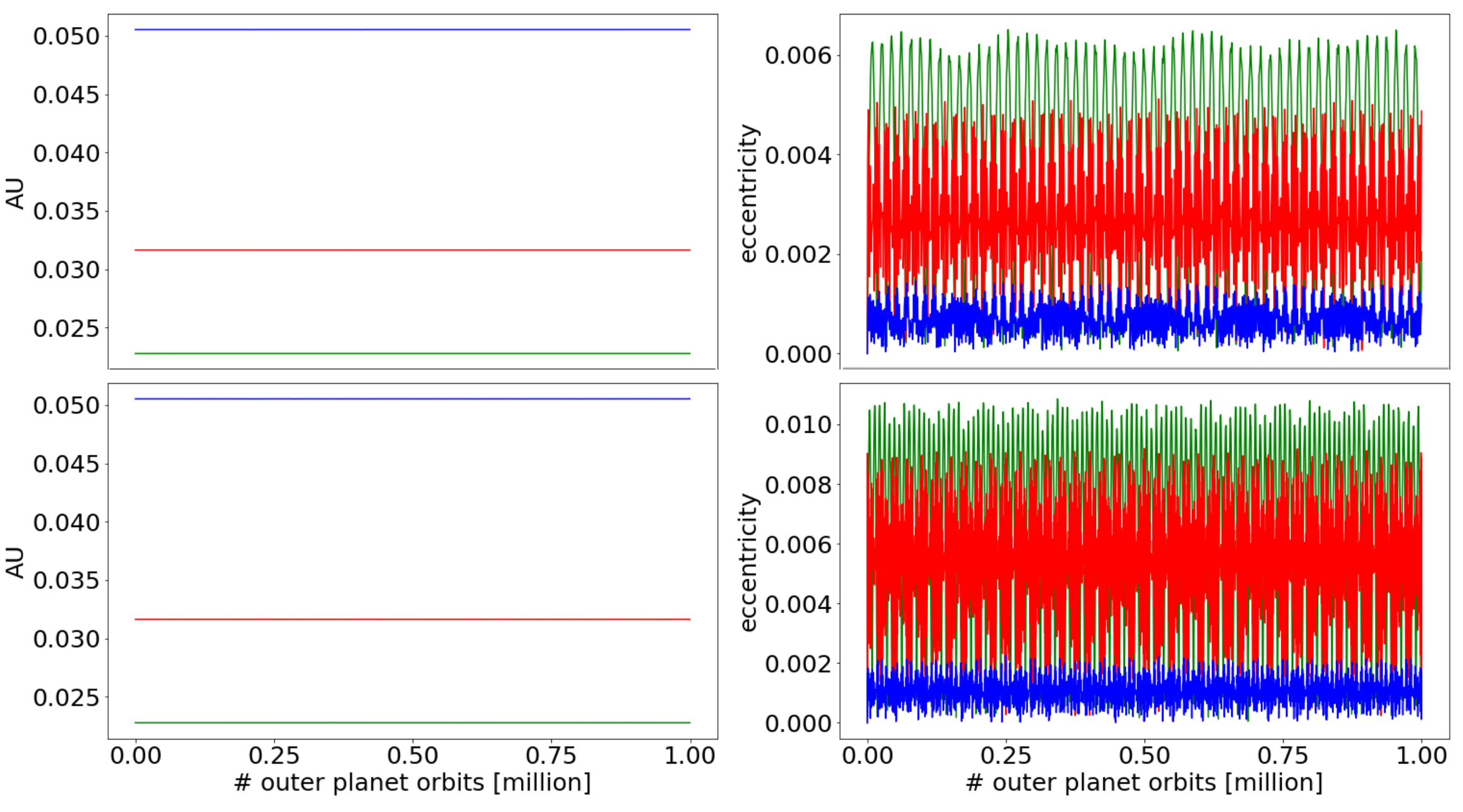}
    \caption{The evolution of the planets' semi-major axes (left panels) and eccentricities (right panels) for the corresponding best-fit (upper panels) and best-fit + ${1\sigma}$ (lower panels) masses for 1 million orbits of the outer planet (L 98-59 d), and assuming initially circular orbits. The orbital elements do not experience drastic variations and the system is dynamically stable for the duration of the integrations.}
    \label{fig:long_term_circ}
\end{figure*}

\begin{figure*}
    \centering
    \plotone{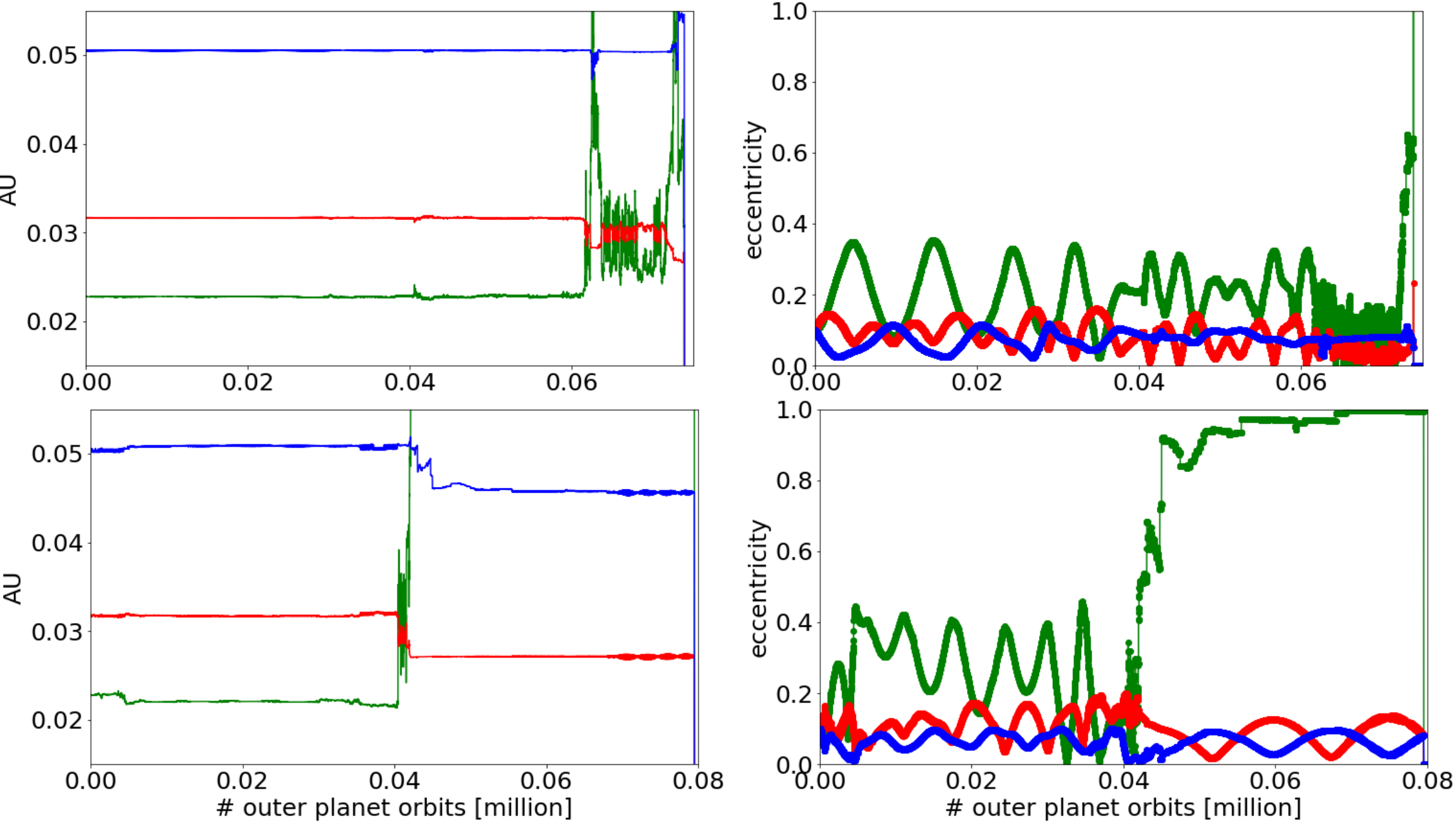}
    \caption{Same as Figure \ref{fig:long_term_circ} but for planets on initially eccentric orbits with $e = 0.1$. The system becomes dynamically unstable within a few thousand orbits of the outer planet in half of our simulations, both for the best-fit and for the best-fit + ${1\sigma}$ masses.}
    \label{fig:long_term_eccen}
\end{figure*}

Our simulations show that for initially circular orbits, the semi-major axes and eccentricities do not exhibit extreme variations, the system does not exhibit chaotic behavior for the duration of the numerical integrations for either set of planet masses, and the orbits remain practically circular (Figure \ref{fig:long_term_circ}). In contrast, for initially eccentric orbits with $e = 0.1$, the system becomes unstable in half of our simulations (with randomly-selected initial arguments of periastron), both for the best-fit and the best-fit+${1\sigma}$ planet masses (Figure \ref{fig:long_term_eccen}). Thus we consider orbits with non-negligible eccentricity as unlikely. This is consistent with other compact multiplanet systems where the orbital eccentricities are typically on the order of a few percent \citep[e.g.][]{Hadden2014}, and is in line with the L~98-59 planets being close to but not in resonance \citep[where the orbits may potentially be eccentric, e.g.][]{Charalambous2018}.

\begin{figure*}
    \centering
    \plotone{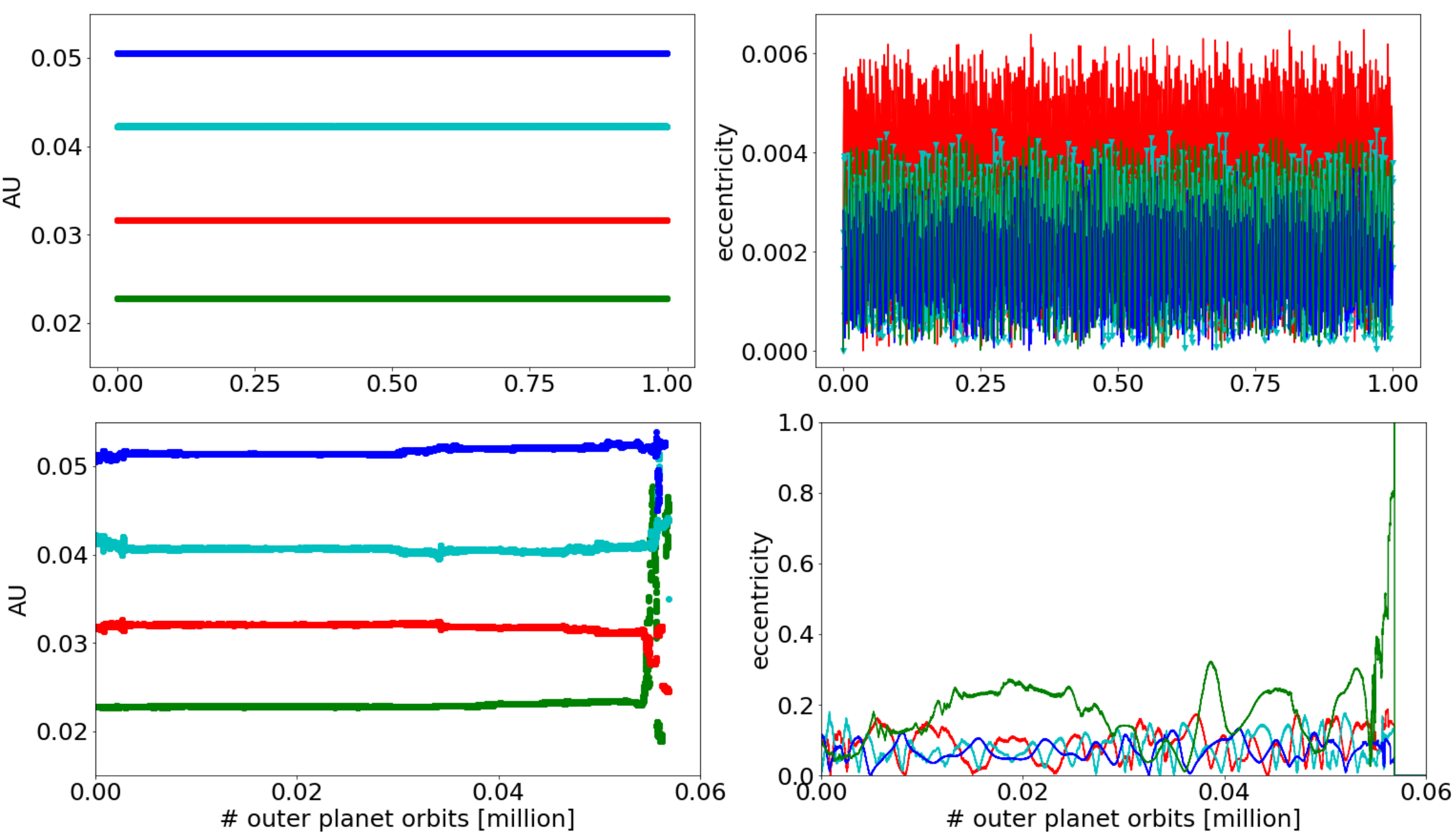}
    \caption{Same as Figure \ref{fig:long_term_circ} but for the three known planets (with best-fit masses) and a hypothetical fourth planet in-between L 98-59 c and L 98-59 d, with a mass of 2.5~${M_\oplus}$ and a period of 5.7 days. All four planets are on initially circular orbits (upper panel) or on initially eccentric orbits with $e = 0.1$ (lower panels). While the system is dynamically stable for the duration of the integrations in the former case, it quickly becomes unstable in the latter.}
    \label{fig:long_term_4planets}
\end{figure*}

Inspired by the closely-spaced multiplanet systems discovered by {\it Kepler} \citep{Muirhead2015} and other surveys (e.g. Gillon et al. 2017), we also explored the possibility of a fourth, non-transiting planet having a dynamically-stable orbit in-between L 98-59 c and L 98-59 d such that the four planets would form a near-resonant chain of 5:8:12:16 period commensurability similar to e.g. TRAPPIST-1 (Gillon et al. 2017). As an example, we tested a planet with a mass of ${\rm 2.5~M_\oplus}$ and a 5.7-day orbital period (${\rm \approx 1.55}$ and ${\rm \approx 0.77}$ times the period of L 98-59 c and L 98-59 d respectively), again for two cases of (a) initially circular orbits and (b) initially eccentric orbits with $e = 0.1$. For simplicity, we only used the best-fit masses for the three planet candidates. The system is dynamically stable in case (a) for the duration of the integrations (Figure \ref{fig:long_term_4planets}, upper panel), and becomes unstable within a few thousand orbits of the outer planet for case (b). Thus such a hypothetical planet is potentially possible if on a circular orbit. Overall, while a comprehensive dynamical analysis for the presence of additional planets is beyond the scope of this work, we will continuously monitor the system as data from future TESS sectors become available. 

\subsubsection{Transit Timing Variations}

If detected, deviations in the times of transits from a linear ephemeris can be a powerful method to constrain the masses and orbital eccentricities of planets in multiplanet systems (e.g.~Agol et al. 2005). We measured transit times for each individual transit using two different methods. First, we folded the transits on a linear ephemeris, fitting a transit model using the models of \citet{mandelagol02}. 
Next, we measured the time of each individual transit by, for each transit, sliding this model across a grid of potential transit midpoints with a time resolution of one second, and measuring the likelihood of each transit fit at each grid point. We then found the maximum likelihood transit time and a 68\% confidence interval on the same. 
The ability to measure TTV signals depends sensitively on our ability to measure precise transit times. For L 98-59 b, the scatter in measured transit times, suggestive of the ultimate transit timing precision we measure, is 5.1 minutes. For L 98-59 b, this is 2.1 minutes; and for L 98-59 d, 1.2 minutes. 
Our analysis showed that a linear ephemeris is sufficient to reproduce the transit times of the three planet candidates detected in Sector 2. We found no evidence for transit timing variations (TTVs), and no further constraints can be placed on the parameters of the system beyond those already provided by dynamical stability considerations. Given that L~98-59 will be observed in 7 of the 13 sectors that comprise the first year of the \textit{TESS} mission \citep{Mukai2017} (Sectors 2, 5, 8, 9, 10, 11, and 12), here we examine how continued TESS observations would affect the transit timing analysis of the system.

Specifically, we simulated continued observations following the nominal TESS schedule, assuming a linear ephemeris for future transits and that every planned sector will be observed as scheduled. We then used the {\textsf {TTVFast}} package \citep{Deck2014} to calculate predicted transit times for the three planets in various orbital configurations consistent with the current data and examine what can be ruled out by the data by the end of the mission.



\begin{figure}
    \includegraphics[width=0.5\textwidth]{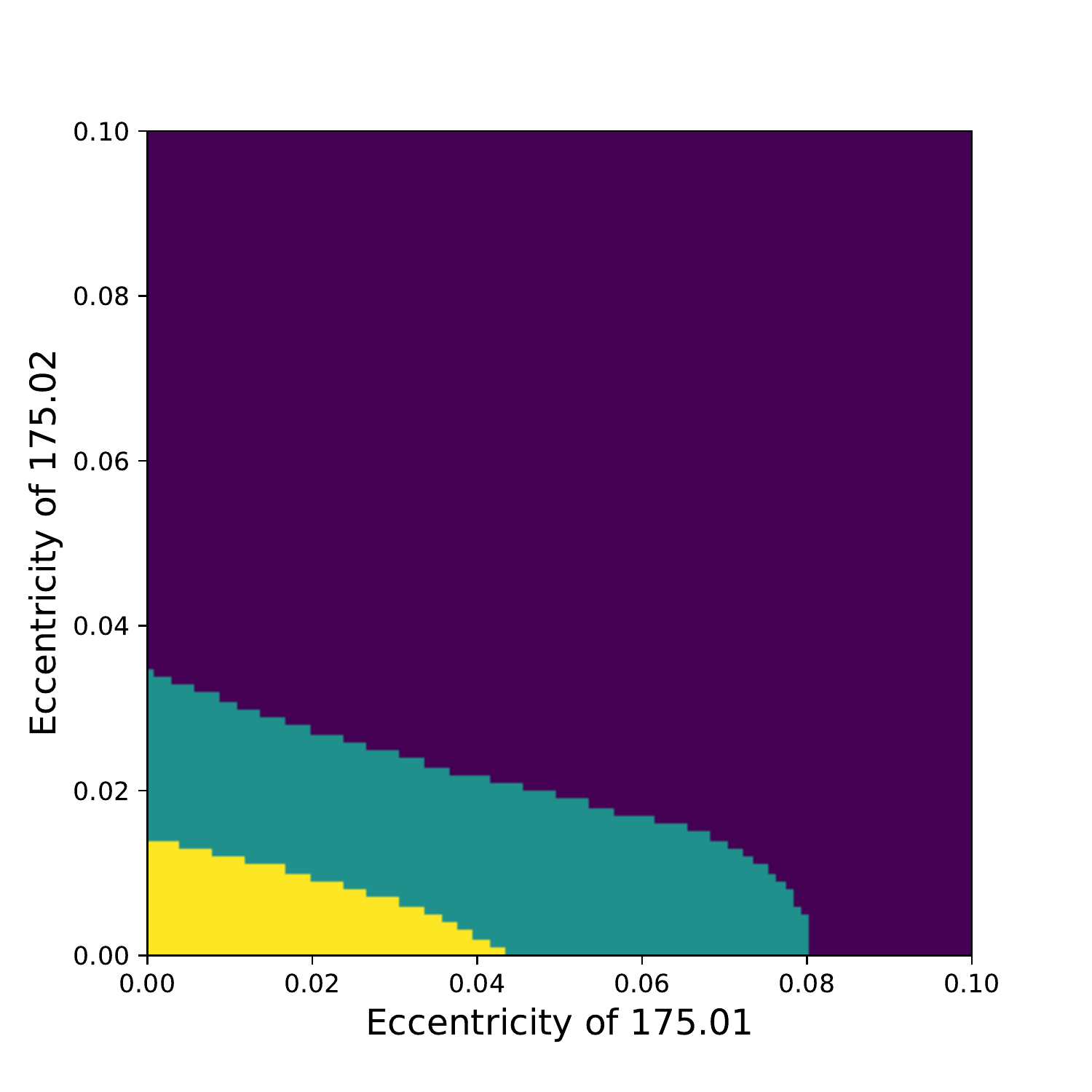}
    \caption{Parameter space in eccentricity (purple) where TTVs are likely to be detected for the L 98-59 system, assuming transit times can be measured to a precision of 2 minutes and using the {\textsf {forecaster}} planet masses. If transit times can instead be measured to a 1-minute precision, TTVs are likely to be detected if the planet eccentricities are in the green region.}
    \label{fig:detect_ttvs}
\end{figure}

The two outer planets (L 98-59 c and d) are close to first-order period commensurability (period ratio of 2.02) whereas the inner planet is not near a first-order resonance with either of the other planets (1.64 period ratio between L 98-59 b and L 98-59 c, and 3.31 period ratio between L 98-59 b and L 98-59 d). Thus the expected TTV signal for the former planet pair is stronger compared to that for the latter planet pair. Indeed, even for eccentricities of $\sim 0.1$, the expected TTV amplitude for the innermost planet is $\sim 90$ seconds, notably smaller than the observed precision on the measured times of transit. 
To evaluate the potential for measuring TTVs for the outer two planets, we performed numerical simulations of the system for the first year of the TESS mission (using \textsf{TTVFast}), thus covering all sectors it will be observed in. We allowed the planet eccentricities to vary, and assumed the maximum likelihood masses listed in Section~\ref{sec:masses}. Adopting a transit-timing precision of 1-2 minutes, we found that significant TTVs could be detected if the eccentricities of the outer two planets were larger than $\sim 0.03$ (as shown in Figure \ref{fig:detect_ttvs}).

Overall, as multiplanet systems typically have orbital eccentricities of a few percent \citep{Hadden2016}, it is unlikely that \textit{TESS} will reveal timing variations for this system during its primary mission; doing so would suggest either anomalously large eccentricities (which are unlikely based on dynamical stability consideration) or significantly larger planet masses/densities in this system relative to planets with similar radii in other systems.

\begin{figure}
    \plotone{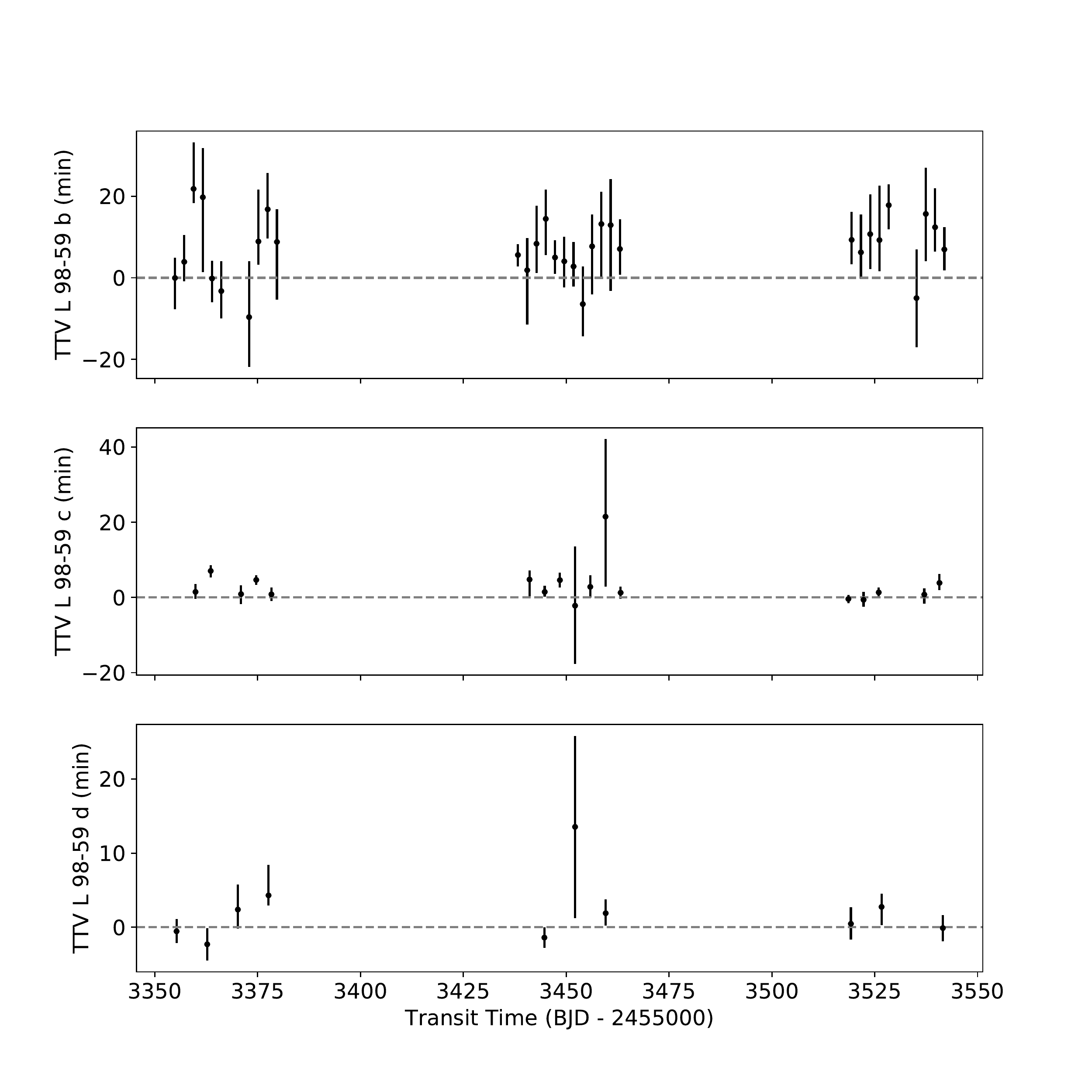}
    \caption{Observed TTVs of L 98-59. The vertical bars represent $\pm 1 \sigma$ uncertainties for the observed transits, and a horizontal line shows the zero point. The observed transit times do not meaningfully deviate from a circular orbit model.}
    \label{fig:toi_175_ttvs}
\end{figure}

To explore TTVs using an alternative software framework, we also used the \textsf{TTV2Fast2Furious} package \citep{Hadden2018} to project the expected TTV signals of the planets through Sector 12, again adopting the \textsf{forecaster} masses and, for simplicity, assuming circular orbits. Similar to the \textsf{TTVFast} analysis described above, our results show that it is unlikely TTVs are  measurable for this system during the TESS prime mission --- the maximum TTV amplitudes are $0.09$ minutes for L 98-59 b, $0.17$ minutes for L 98-59 c, and $0.56$ minutes for L 98-59 d (see Figure \ref{fig:toi_175_ttvs}).

Following the approach described in \citet{Hadden2018}, we also used \textsc{TTV2Fast2Furious} to project the precision of mass constraints derived from future TESS transit timing measurements. Planet mass constraints derived from TTVs depend on the precision of transit time measurements and we adopt the measured scatter in the transit time measurements taken through Sector 2, $\sigma_{t_{L 98-59 c}}$ = 2.1 minutes, and $\sigma_{t_{L 98-59 d}}$ = 1.2 minutes. The planets' masses are expected to be constrained with precisions $\sigma_{m,L 98-59 c}=13.1\,{\rm M}_{\oplus}$ and $\sigma_{m,L 98-59 d}=5\,{\rm M}_{\oplus}$ with transit timing data through Sector 12. \footnote{If the planets are restricted to circular orbits in the TTV model, e.g., under the assumption that eccentricities are damped away by tidal dissipation, the mass-eccentricity degeneracy \citep{Lithwick2012} is removed and the measured TTV signals therefore place tighter constraints on the planet masses. In particular, if the TTV model is restricted to circular orbits, the mass measurement precisions of $\sigma_{m,L 98-59 c}=3.4 M_{\oplus}$ and $\sigma_{m,L 98-59 d}=2.1 M_{\oplus}$ are projected.}

With three sectors of data available for this system at the time of writing (Sectors 2, 5, and 8), there is at present no evidence for TTVs. In line with the predictions of the previous paragraph, there is no significant mass constraint beyond what is given from plausible compositions of terrestrial planets. We use \textsc{TTVFast} to compare maximum-likelihood dynamical models of the orbits of the planets, assuming the masses listed in Section~\ref{sec:masses}. We also repeat this procedure holding the masses fixed at four times this nominal value, which would imply a density of approximately 20 g/cc. In both scenarios, there is a dynamical model which fits the observed transits.

\subsection{Potential for Atmospheric Characterization}
\label{atmoschar}

Owing to the small, bright host star the three planets of the L 98-59 system are promising targets for follow-up atmosphere characterization. The planets' small radii suggest that it is unlikely that they retain hydrogen-rich atmospheres \citep{Rogers2015, Fulton2017}, but secondary atmospheres could form from volcanic outgassing and/or delivery of volatiles from comets.

To investigate the feasibility of atmosphere studies for the L 98-59 planets, we compared the expected signal-to-noise of atmospheric features to that of GJ 1132b, another small planet around a nearby M-dwarf \citep{Berta2015}. \cite{Morley2017} found that a CO$_2$-dominated atmosphere could be detected for GJ 1132 with a modest number of JWST transits or eclipses (11 transits with NIRSpec/G235M or 2 eclipses with MIRI/LRS). To scale these estimates for L 98-59, we used the transmission and emission spectroscopy metrics from \cite{Kempton2018}, which calculate expected signal-to-noise for atmospheric features based on planet and star properties. We scaled the signal relative to each planet's transit/eclipse duration, estimated the brightness of the star based on its K-band magnitude, and assumed zero noise floor. We found that L 98-59 b, L 98-59 c and L 98-59 d have Transmission Spectroscopy Metric (TSM) values of 0.8, 1.4, and 1.0 that of GJ 1132b and Emission Spectroscopy Metric (ESM) values of 0.3, 0.4, and 0.7 respectively \citep{Kempton2018}. This implies that features in the transmission spectrum could be detected with 16, 6, or 11 transits, or 24, 13, or 4 eclipses for L 98-59 b, L 98-59 c and L 98-59 d. Provided that JWST observations reach the photon limit for stars as bright as L 98-59, this system is an exciting opportunity for studying comparative planetology of terrestrial exoplanet atmospheres.

\subsection{Planets in the Venus Zone}

It is worth noting the possibility that the planets in the system are analogs to Venus in terms of their atmospheric evolution. Venus shares several characteristics with Earth including its relative composition, size, and mass. Although Venus may have previously had temperate surface conditions \citep{Way2016}, Venus eventually diverged significantly from the habitable pathway of Earth and transitioned into a runaway greenhouse state. The planet now has a high pressure, high temperature, and a carbon dioxide dominated atmosphere. In our study of exoplanets and the search for life, it is vitally important that we understand why Earth is habitable and Venus is not \citep{Kane2014}. There is a need to discover planets that may have evolved into a post-runaway greenhouse state so that we can target their atmospheres for characterization with future facilities, such as JWST \citep{Ehrenreich2012}. However, most of the potential Venus analog candidates hitherto discovered orbit relatively faint stars \citep{Barclay2013,Kane2013,Angelo2017,Kane2018}.

The L 98-59 planets receive significantly more energy than the Earth receives from the Sun (a factor of between 4--22 more than Earth's insolation) and fall into the region that \citet{Kane2014} dubbed the Venus Zone. This is a region where the atmosphere of a planet like Earth would likely have been forced into a runaway greenhouse, producing conditions similar to those found on Venus. The range of incident fluxes within the Venus Zone corresponds to insolations of between 1--25 times that received by the Earth. Planets in the Venus Zone that can be spectroscopically characterized will become increasingly important in the realm of comparative planetology that aims to characterize the conditions for planetary habitability. In that respect, and considering the potential for atmospheric characterization discussed in Section~\ref{atmoschar}, L~98-59 could become a benchmark system.

\section{Conclusions}
\label{sec:end}

We presented the discovery of a system of three transiting, terrestrial-size planets orbiting L 98-59 (TESS Object of Interest TOI-175). The host star is a bright M3 dwarf (K = 7.1) at a distance of 10.6 pc, with ${\rm M_{*} = 0.313\pm0.014M_\odot}$, ${\rm R_* = 0.312\pm0.014R_\odot}$, and ${\rm T_{eff} = 3367\pm150 K}$. TFOP-led follow-up observations found no evidence of binarity or significant stellar activity. To thoroughly vet the transit signals detected in the TESS data, we used the software package DAVE. Our analysis ruled out significant secondary eclipses, odd-even differences or photocenter shifts during transits, verifying their planetary nature. Using \texttt{lightkurve}, we also discovered that the nearby field star 2MASS 08181825-6818430, inside the TESS aperture of L 98-59, is an eclipsing binary system with an orbital period of ${\sim 10.43}$ days, manifesting both primary and secondary eclipses. Utilizing {\it Gaia} data we confirmed that the eclipsing binary is a background object (likely a red giant) not associated with L 98-59. This battery of tests highlights the importance of comprehensive analysis of all sources inside the TESS aperture.  

The planets range in size from slightly smaller to slightly bigger than Earth, with radii of ${\rm 0.8\pm0.05 R_\oplus}$, ${\rm 1.35\pm0.07 R_\oplus}$, and ${\rm 1.59\pm0.23 R_\oplus}$ from inner to outer respectively. The planetary system is quite compact, with orbital periods of ${\rm 2.25~days}$, ${\rm 3.69~ days}$, and ${\rm 7.45~days}$ respectively. We estimated their masses using the forecast package for probabilistic mass forecasting, confirmed the dynamical stability of the system for circular orbits, and showed that there are no significant transit-timing variations. 

TESS will continue observing the system in upcoming sectors (9, 10, 11, 12), and it is also likely that the system will be observed during a TESS Extended Mission. These observations will allow for refinement of the known planet parameters, searches for additional planets, further investigations of the dynamics of the system, as well as long-term monitoring of the host star activity.

\acknowledgments{We thank the referee for the insightful comments that helped us improve this manuscript. This manuscript includes data collected by the TESS mission, which are publicly available from the Mikulski Archive for Space Telescopes (MAST). Funding for the TESS mission is provided by NASA's Science Mission directorate. We acknowledge the use of TESS Alerts data, as provided by the TESS Science Office. We acknowledge the use of public TESS Alert data from pipelines at the TESS Science Office and at the TESS Science Processing Operations Center. This research has made use of the Exoplanet Follow-up Observation Program website, which is operated by the California Institute of Technology, under contract with the National Aeronautics and Space Administration under the Exoplanet Exploration Program. JGW is supported by a grant from the John Templeton Foundation. Acquisition of the CHIRON data and the first epoch of DSSI data was made possible by a grant from the John Templeton Foundation. The opinions expressed in this publication are those of the authors and do not necessarily reflect the views of the John Templeton Foundation. We thank Leonardo Paredes, Hodari James, Rodrigo Hinojosa, and Todd Henry for their work in gathering and processing the CHIRON data, as well as for the management of CTIO / SMARTS 1.5m telescope. We are also grateful to the observer support staff at CTIO, at ESO/VLT (for program number 0102.C-0503(A)), and Gemini (for program number GS-2018B-LP-101).
This work has made use of data from the European Space Agency (ESA) mission
{\it Gaia} (\url{https://www.cosmos.esa.int/gaia}), processed by the {\it Gaia}
Data Processing and Analysis Consortium (DPAC,
\url{https://www.cosmos.esa.int/web/gaia/dpac/consortium}). Funding for the DPAC
has been provided by national institutions, in particular the institutions
participating in the {\it Gaia} Multilateral Agreement. This research made use of observations from the LCOGT network, and the AAVSO Photometric All-Sky Survey (APASS), funded by the Robert Martin Ayers Sciences Fund and NSF AST-1412587. The research leading to these results has received funding from the European Research Council under the European Union's Seventh Framework Programme (FP/2007-2013) ERC Grant Agreement n$\degr$ 336480, and from the ARC grant for Concerted Research Actions, financed by the Wallonia-Brussels Federation. M.G. and E.J. are FNRS Senior Research Associates.
Work by B.T.M. was performed in part under contract with the Jet Propulsion Laboratory (JPL) funded by NASA through the Sagan Fellowship Program executed by the NASA Exoplanet Science Institute. This work is partly supported by JSPS KAKENHI Grant Numbers JP18H01265 and 18H05439, and JST PRESTO Grant Number JPMJPR1775. KH acknowledges support from STFC grant ST/R000824/1.
}

\facilities{AAVSO, LCO, TESS}


\software{
AstroImageJ \citep{karen2017}, 
astropy \citep{exoplanet:astropy13,exoplanet:astropy18},
celerite \citep{exoplanet:foremanmackey17, exoplanet:foremanmackey18},
emcee \citep{foremanmackey12},
exoplanet \citep{exoplanet:exoplanet},
DAVE \citep{Kostov2019},
forecaster \citep{forecaster}, 
IPython \citep{ipython},
Jupyter \citep{jupyer},
Lightkurve \citep{lightkurve},
Matplotlib \citep{matplotlib}, 
NumPy \citep{numpy},
Pandas \citep{pandas},
PyMC3 \citep{exoplanet:pymc3},
SciPy \citep{scipy},
STARRY \citep{exoplanet:luger18},
Tapir  \citep{Jensen:2013},
Theano \citep{exoplanet:theano},
TTVFast \citep{Deck2014},
TTV2Fast2Furious \citep{Hadden2018},
}

\bibliography{bibliography}{}

\begin{thebibliography}{}
\expandafter\ifx\csname natexlab\endcsname\relax\def\natexlab#1{#1}\fi

\bibitem[{{Angelo} {et~al.}(2017){Angelo}, {Rowe}, {Howell}, {Quintana},
  {Still}, {Mann}, {Burningham}, {Barclay}, {Ciardi}, {Huber}, \&
  {Kane}}]{Angelo2017}
{Angelo}, I., {Rowe}, J.~F., {Howell}, S.~B., {et~al.} 2017, \aj, 153, 162

\bibitem[{{Anglada-Escud{\'e}} {et~al.}(2016){Anglada-Escud{\'e}}, {Amado},
  {Barnes}, {Berdi{\~n}as}, {Butler}, {Coleman}, {de La Cueva}, {Dreizler},
  {Endl}, {Giesers}, {Jeffers}, {Jenkins}, {Jones}, {Kiraga}, {K{\"u}rster},
  {L{\'o}pez-Gonz{\'a}lez}, {Marvin}, {Morales}, {Morin}, {Nelson}, {Ortiz},
  {Ofir}, {Paardekooper}, {Reiners}, {Rodr{\'\i}guez},
  {Rodr{\'\i}guez-L{\'o}pez}, {Sarmiento}, {Strachan}, {Tsapras}, {Tuomi}, \&
  {Zechmeister}}]{Anglada2016}
{Anglada-Escud{\'e}}, G., {Amado}, P.~J., {Barnes}, J., {et~al.} 2016, \nat,
  536, 437

\bibitem[{{Astropy Collaboration} {et~al.}(2013){Astropy Collaboration},
  {Robitaille}, {Tollerud}, {Greenfield}, {Droettboom}, {Bray}, {Aldcroft},
  {Davis}, {Ginsburg}, {Price-Whelan}, {Kerzendorf}, {Conley}, {Crighton},
  {Barbary}, {Muna}, {Ferguson}, {Grollier}, {Parikh}, {Nair}, {Unther},
  {Deil}, {Woillez}, {Conseil}, {Kramer}, {Turner}, {Singer}, {Fox}, {Weaver},
  {Zabalza}, {Edwards}, {Azalee Bostroem}, {Burke}, {Casey}, {Crawford},
  {Dencheva}, {Ely}, {Jenness}, {Labrie}, {Lim}, {Pierfederici}, {Pontzen},
  {Ptak}, {Refsdal}, {Servillat}, \& {Streicher}}]{exoplanet:astropy13}
{Astropy Collaboration}, {Robitaille}, T.~P., {Tollerud}, E.~J., {et~al.} 2013,
  \aap, 558, A33

\bibitem[{{Astropy Collaboration} {et~al.}(2018){Astropy Collaboration},
  {Price-Whelan}, {Sip{\H o}cz}, {G{\"u}nther}, {Lim}, {Crawford}, {Conseil},
  {Shupe}, {Craig}, {Dencheva}, {Ginsburg}, {VanderPlas}, {Bradley},
  {P{\'e}rez-Su{\'a}rez}, {de Val-Borro}, {Aldcroft}, {Cruz}, {Robitaille},
  {Tollerud}, {Ardelean}, {Babej}, {Bach}, {Bachetti}, {Bakanov}, {Bamford},
  {Barentsen}, {Barmby}, {Baumbach}, {Berry}, {Biscani}, {Boquien}, {Bostroem},
  {Bouma}, {Brammer}, {Bray}, {Breytenbach}, {Buddelmeijer}, {Burke},
  {Calderone}, {Cano Rodr{\'{\i}}guez}, {Cara}, {Cardoso}, {Cheedella},
  {Copin}, {Corrales}, {Crichton}, {D'Avella}, {Deil}, {Depagne}, {Dietrich},
  {Donath}, {Droettboom}, {Earl}, {Erben}, {Fabbro}, {Ferreira}, {Finethy},
  {Fox}, {Garrison}, {Gibbons}, {Goldstein}, {Gommers}, {Greco}, {Greenfield},
  {Groener}, {Grollier}, {Hagen}, {Hirst}, {Homeier}, {Horton}, {Hosseinzadeh},
  {Hu}, {Hunkeler}, {Ivezi{\'c}}, {Jain}, {Jenness}, {Kanarek}, {Kendrew},
  {Kern}, {Kerzendorf}, {Khvalko}, {King}, {Kirkby}, {Kulkarni}, {Kumar},
  {Lee}, {Lenz}, {Littlefair}, {Ma}, {Macleod}, {Mastropietro}, {McCully},
  {Montagnac}, {Morris}, {Mueller}, {Mumford}, {Muna}, {Murphy}, {Nelson},
  {Nguyen}, {Ninan}, {N{\"o}the}, {Ogaz}, {Oh}, {Parejko}, {Parley}, {Pascual},
  {Patil}, {Patil}, {Plunkett}, {Prochaska}, {Rastogi}, {Reddy Janga},
  {Sabater}, {Sakurikar}, {Seifert}, {Sherbert}, {Sherwood-Taylor}, {Shih},
  {Sick}, {Silbiger}, {Singanamalla}, {Singer}, {Sladen}, {Sooley},
  {Sornarajah}, {Streicher}, {Teuben}, {Thomas}, {Tremblay}, {Turner},
  {Terr{\'o}n}, {van Kerkwijk}, {de la Vega}, {Watkins}, {Weaver}, {Whitmore},
  {Woillez}, {Zabalza}, \& {Astropy Contributors}}]{exoplanet:astropy18}
{Astropy Collaboration}, {Price-Whelan}, A.~M., {Sip{\H o}cz}, B.~M., {et~al.}
  2018, \aj, 156, 123

\bibitem[{{Baraffe} {et~al.}(2003){Baraffe}, {Chabrier}, {Barman}, {Allard}, \&
  {Hauschildt}}]{Baraffe2003}
{Baraffe}, I., {Chabrier}, G., {Barman}, T.~S., {Allard}, F., \& {Hauschildt},
  P.~H. 2003, \aap, 402, 701

\bibitem[{{Barclay} {et~al.}(2018){Barclay}, {Pepper}, \&
  {Quintana}}]{Barclay2018}
{Barclay}, T., {Pepper}, J., \& {Quintana}, E.~V. 2018, \apjs, 239, 2

\bibitem[{{Barclay} {et~al.}(2015){Barclay}, {Quintana}, {Adams}, {Ciardi},
  {Huber}, {Foreman-Mackey}, {Montet}, \& {Caldwell}}]{Barclay2015}
{Barclay}, T., {Quintana}, E.~V., {Adams}, F.~C., {et~al.} 2015, \apj, 809, 7

\bibitem[{{Barclay} {et~al.}(2013){Barclay}, {Burke}, {Howell}, {Rowe},
  {Huber}, {Isaacson}, {Jenkins}, {Kolbl}, {Marcy}, {Quintana}, {Still},
  {Twicken}, {Bryson}, {Borucki}, {Caldwell}, {Ciardi}, {Clarke},
  {Christiansen}, {Coughlin}, {Fischer}, {Li}, {Haas}, {Hunter}, {Lissauer},
  {Mullally}, {Sabale}, {Seader}, {Smith}, {Tenenbaum}, {Kamal Uddin}, \&
  {Thompson}}]{Barclay2013}
{Barclay}, T., {Burke}, C.~J., {Howell}, S.~B., {et~al.} 2013, \apj, 768, 101

\bibitem[{{Benedict} {et~al.}(2016){Benedict}, {Henry}, {Franz}, {McArthur},
  {Wasserman}, {Jao}, {Cargile}, {Dieterich}, {Bradley}, {Nelan}, \&
  {Whipple}}]{Benedict(2016)}
{Benedict}, G.~F., {Henry}, T.~J., {Franz}, O.~G., {et~al.} 2016, \aj, 152, 141

\bibitem[{{Berta-Thompson} {et~al.}(2015){Berta-Thompson}, {Irwin},
  {Charbonneau}, {Newton}, {Dittmann}, {Astudillo-Defru}, {Bonfils}, {Gillon},
  {Jehin}, {Stark}, {Stalder}, {Bouchy}, {Delfosse}, {Forveille}, {Lovis},
  {Mayor}, {Neves}, {Pepe}, {Santos}, {Udry}, \& {W{\"u}nsche}}]{Berta2015}
{Berta-Thompson}, Z.~K., {Irwin}, J., {Charbonneau}, D., {et~al.} 2015, \nat,
  527, 204

\bibitem[{{Boyajian} {et~al.}(2012){Boyajian}, {von Braun}, {van Belle},
  {McAlister}, {ten Brummelaar}, {Kane}, {Muirhead}, {Jones}, {White},
  {Schaefer}, {Ciardi}, {Henry}, {L{\'o}pez-Morales}, {Ridgway}, {Gies}, {Jao},
  {Rojas-Ayala}, {Parks}, {Sturmann}, {Sturmann}, {Turner}, {Farrington},
  {Goldfinger}, \& {Berger}}]{Boyajian(2012)}
{Boyajian}, T.~S., {von Braun}, K., {van Belle}, G., {et~al.} 2012, \apj, 757,
  112

\bibitem[{{Brown} {et~al.}(2013){Brown}, {Baliber}, {Bianco}, {Bowman},
  {Burleson}, {Conway}, {Crellin}, {Depagne}, {De Vera}, {Dilday}, {Dragomir},
  {Dubberley}, {Eastman}, {Elphick}, {Falarski}, {Foale}, {Ford}, {Fulton},
  {Garza}, {Gomez}, {Graham}, {Greene}, {Haldeman}, {Hawkins}, {Haworth},
  {Haynes}, {Hidas}, {Hjelstrom}, {Howell}, {Hygelund}, {Lister}, {Lobdill},
  {Martinez}, {Mullins}, {Norbury}, {Parrent}, {Paulson}, {Petry}, {Pickles},
  {Posner}, {Rosing}, {Ross}, {Sand}, {Saunders}, {Shobbrook}, {Shporer},
  {Street}, {Thomas}, {Tsapras}, {Tufts}, {Valenti}, {Vander Horst}, {Walker},
  {White}, \& {Willis}}]{brown2013}
{Brown}, T.~M., {Baliber}, N., {Bianco}, F.~B., {et~al.} 2013, Publications of
  the Astronomical Society of the Pacific, 125, 1031

\bibitem[{{Bryson} {et~al.}(2013){Bryson}, {Jenkins}, {Gilliland}, {Twicken},
  {Clarke}, {Rowe}, {Caldwell}, {Batalha}, {Mullally}, {Haas}, \&
  {Tenenbaum}}]{Bryson2013}
{Bryson}, S.~T., {Jenkins}, J.~M., {Gilliland}, R.~L., {et~al.} 2013,
  Publications of the Astronomical Society of the Pacific, 125, 889

\bibitem[{{Burdanov} {et~al.}(2018){Burdanov}, {Delrez}, {Gillon}, \&
  {Jehin}}]{burdanov2018}
{Burdanov}, A., {Delrez}, L., {Gillon}, M., \& {Jehin}, E. 2018, {SPECULOOS
  Exoplanet Search and Its Prototype on TRAPPIST}, 130

\bibitem[{{Charalambous} {et~al.}(2018){Charalambous}, {Mart{\'\i}},
  {Beaug{\'e}}, \& {Ramos}}]{Charalambous2018}
{Charalambous}, C., {Mart{\'\i}}, J.~G., {Beaug{\'e}}, C., \& {Ramos}, X.~S.
  2018, \mnras, 477, 1414

\bibitem[{{Chen} \& {Kipping}(2017)}]{forecaster}
{Chen}, J., \& {Kipping}, D. 2017, \apj, 834, 17

\bibitem[{{Ciardi} {et~al.}(2015){Ciardi}, {Beichman}, {Horch}, \&
  {Howell}}]{Ciardi2015}
{Ciardi}, D.~R., {Beichman}, C.~A., {Horch}, E.~P., \& {Howell}, S.~B. 2015,
  \apj, 805, 16

\bibitem[{{Collins} {et~al.}(2017){Collins}, {Kielkopf}, {Stassun}, \&
  {Hessman}}]{karen2017}
{Collins}, K.~A., {Kielkopf}, J.~F., {Stassun}, K.~G., \& {Hessman}, F.~V.
  2017, \aj, 153, 77

\bibitem[{{Cutri} {et~al.}(2013){Cutri}, {Wright}, {Conrow}, {Fowler},
  {Eisenhardt}, {Grillmair}, {Kirkpatrick}, {Masci}, {McCallon}, {Wheelock},
  {Fajardo-Acosta}, {Yan}, {Benford}, {Harbut}, {Jarrett}, {Lake}, {Leisawitz},
  {Ressler}, {Stanford}, {Tsai}, {Liu}, {Helou}, {Mainzer}, {Gettings},
  {Gonzalez}, {Hoffman}, {Marsh}, {Padgett}, {Skrutskie}, {Beck}, {Papin}, \&
  {Wittman}}]{Cutri2013}
{Cutri}, R.~M., {Wright}, E.~L., {Conrow}, T., {et~al.} 2013, {Explanatory
  Supplement to the AllWISE Data Release Products}, Tech. rep.

\bibitem[{{Deck} {et~al.}(2014){Deck}, {Agol}, {Holman}, \&
  {Nesvorn{\'y}}}]{Deck2014}
{Deck}, K.~M., {Agol}, E., {Holman}, M.~J., \& {Nesvorn{\'y}}, D. 2014, \apj,
  787, 132

\bibitem[{{Delrez} {et~al.}(2018){Delrez}, {Gillon}, {Queloz}, {Demory},
  {Almleaky}, {de Wit}, {Jehin}, {Triaud}, {Barkaoui}, {Burdanov}, {Burgasser},
  {Ducrot}, {McCormac}, {Murray}, {Silva Fernandes}, {Sohy}, {Thompson}, {Van
  Grootel}, {Alonso}, {Benkhaldoun}, \& {Rebolo}}]{laetitia2018spie}
{Delrez}, L., {Gillon}, M., {Queloz}, D., {et~al.} 2018, in {Society of
  Photo-Optical Instrumentation Engineers (SPIE) Conference Series}, Vol.
  10700, {Ground-based and Airborne Telescopes VII}, 107001I

\bibitem[{{Deming} {et~al.}(2015){Deming}, {Knutson}, {Kammer}, {Fulton},
  {Ingalls}, {Carey}, {Burrows}, {Fortney}, {Todorov}, {Agol}, {Cowan},
  {Desert}, {Fraine}, {Langton}, {Morley}, \& {Showman}}]{Deming2015}
{Deming}, D., {Knutson}, H., {Kammer}, J., {et~al.} 2015, \apj, 805, 132

\bibitem[{{Dittmann} {et~al.}(2017){Dittmann}, {Irwin}, {Charbonneau},
  {Bonfils}, {Astudillo-Defru}, {Haywood}, {Berta-Thompson}, {Newton},
  {Rodriguez}, {Winters}, {Tan}, {Almenara}, {Bouchy}, {Delfosse}, {Forveille},
  {Lovis}, {Murgas}, {Pepe}, {Santos}, {Udry}, {W{\"u}nsche}, {Esquerdo},
  {Latham}, \& {Dressing}}]{Dittmann(2017a)}
{Dittmann}, J.~A., {Irwin}, J.~M., {Charbonneau}, D., {et~al.} 2017, \nat, 544,
  333

\bibitem[{{Dragomir} {et~al.}(2018){Dragomir}, {Teske}, {Gunther},
  {S{\'e}gransan}, {Burt}, {Huang}, {Vanderburg}, {Matthews}, {Dumusque},
  {Stassun}, {Pepper}, {Ricker}, {Vanderspek}, {Latham}, {Seager}, {Winn},
  {Jenkins}, {Beatty}, {Bouchy}, {Butler}, {Crane}, {Eastman}, {Francis},
  {Gaudi}, {Goeke}, {James}, {Klaus}, {Kuhn}, {Lovis}, {Lund}, {McDermott},
  {Paegert}, {Pepe}, {Rodriguez}, {Sha}, {Shectman}, {Siverd}, {Garcia Soto},
  {Stevens}, {Thompson}, {Twicken}, {Udry}, {Villanueva}, {Wang}, {Wohler},
  {Yao}, {Zhan}, \& {the TESS Team}}]{Dragomir2019}
{Dragomir}, D., {Teske}, J., {Gunther}, M.~N., {et~al.} 2018, arXiv e-prints,
  arXiv:1901.00051

\bibitem[{{Ehrenreich} {et~al.}(2012){Ehrenreich}, {Vidal-Madjar}, {Widemann},
  {Gronoff}, {Tanga}, {Barth{\'e}lemy}, {Lilensten}, {Lecavelier Des Etangs},
  \& {Arnold}}]{Ehrenreich2012}
{Ehrenreich}, D., {Vidal-Madjar}, A., {Widemann}, T., {et~al.} 2012, \aap, 537,
  L2

\bibitem[{{Fabrycky} {et~al.}(2014){Fabrycky}, {Lissauer}, {Ragozzine}, {Rowe},
  {Steffen}, {Agol}, {Barclay}, {Batalha}, {Borucki}, {Ciardi}, {Ford},
  {Gautier}, {Geary}, {Holman}, {Jenkins}, {Li}, {Morehead}, {Morris},
  {Shporer}, {Smith}, {Still}, \& {Van Cleve}}]{Fabrycky2014}
{Fabrycky}, D.~C., {Lissauer}, J.~J., {Ragozzine}, D., {et~al.} 2014, \apj,
  790, 146

\bibitem[{{Feinstein} {et~al.}(2019){Feinstein}, {Schlieder}, {Livingston},
  {Ciardi}, {Howard}, {Arnold}, {Barentsen}, {Bristow}, {Christiansen},
  {Crossfield}, {Dressing}, {Gonzales}, {Kosiarek}, {Lintott}, {Miller},
  {Morales}, {Petigura}, {Thackeray}, {Ault}, {Baeten}, {Jonkeren}, {Langley},
  {Moshinaly}, {Pearson}, {Tanner}, \& {Treasure}}]{feinstein2019}
{Feinstein}, A.~D., {Schlieder}, J.~E., {Livingston}, J.~H., {et~al.} 2019,
  \aj, 157, 40

\bibitem[{Foreman-Mackey(2018)}]{exoplanet:exoplanet}
Foreman-Mackey, D. 2018, exoplanet v0.1.3, , , doi:10.5281/zenodo.2536576

\bibitem[{{Foreman-Mackey}(2018)}]{exoplanet:foremanmackey18}
{Foreman-Mackey}, D. 2018, Research Notes of the American Astronomical Society,
  2, 31

\bibitem[{{Foreman-Mackey} {et~al.}(2017){Foreman-Mackey}, {Agol},
  {Ambikasaran}, \& {Angus}}]{exoplanet:foremanmackey17}
{Foreman-Mackey}, D., {Agol}, E., {Ambikasaran}, S., \& {Angus}, R. 2017, \aj,
  154, 220

\bibitem[{{Foreman-Mackey} {et~al.}(2013){Foreman-Mackey}, {Hogg}, {Lang}, \&
  {Goodman}}]{foremanmackey12}
{Foreman-Mackey}, D., {Hogg}, D.~W., {Lang}, D., \& {Goodman}, J. 2013, \pasp,
  125, 306

\bibitem[{{Fulton} {et~al.}(2017){Fulton}, {Petigura}, {Howard}, {Isaacson},
  {Marcy}, {Cargile}, {Hebb}, {Weiss}, {Johnson}, {Morton}, {Sinukoff},
  {Crossfield}, \& {Hirsch}}]{Fulton2017}
{Fulton}, B.~J., {Petigura}, E.~A., {Howard}, A.~W., {et~al.} 2017, \aj, 154,
  109

\bibitem[{{Furlan} \& {Howell}(2017)}]{Furlan2017}
{Furlan}, E., \& {Howell}, S.~B. 2017, \aj, 154, 66

\bibitem[{{Gaia Collaboration} {et~al.}(2018){Gaia Collaboration}, {Brown},
  {Vallenari}, {Prusti}, {de Bruijne}, {Babusiaux}, {Bailer-Jones}, {Biermann},
  {Evans}, {Eyer}, {Jansen}, {Jordi}, {Klioner}, {Lammers}, {Lindegren},
  {Luri}, {Mignard}, {Panem}, {Pourbaix}, {Randich}, {Sartoretti}, {Siddiqui},
  {Soubiran}, {van Leeuwen}, {Walton}, {Arenou}, {Bastian}, {Cropper},
  {Drimmel}, {Katz}, {Lattanzi}, {Bakker}, {Cacciari}, {Casta{\~n}eda},
  {Chaoul}, {Cheek}, {De Angeli}, {Fabricius}, {Guerra}, {Holl}, {Masana},
  {Messineo}, {Mowlavi}, {Nienartowicz}, {Panuzzo}, {Portell}, {Riello},
  {Seabroke}, {Tanga}, {Th{\'e}venin}, {Gracia-Abril}, {Comoretto},
  {Garcia-Reinaldos}, {Teyssier}, {Altmann}, {Andrae}, {Audard},
  {Bellas-Velidis}, {Benson}, {Berthier}, {Blomme}, {Burgess}, {Busso},
  {Carry}, {Cellino}, {Clementini}, {Clotet}, {Creevey}, {Davidson}, {De
  Ridder}, {Delchambre}, {Dell'Oro}, {Ducourant}, {Fern{\'a}ndez-
  Hern{\'a}ndez}, {Fouesneau}, {Fr{\'e}mat}, {Galluccio}, {Garc{\'\i}a-Torres},
  {Gonz{\'a}lez-N{\'u}{\~n}ez}, {Gonz{\'a}lez-Vidal}, {Gosset}, {Guy},
  {Halbwachs}, {Hambly}, {Harrison}, {Hern{\'a}ndez}, {Hestroffer}, {Hodgkin},
  {Hutton}, {Jasniewicz}, {Jean-Antoine-Piccolo}, {Jordan}, {Korn},
  {Krone-Martins}, {Lanzafame}, {Lebzelter}, {L{\"o}ffler}, {Manteiga},
  {Marrese}, {Mart{\'\i}n-Fleitas}, {Moitinho}, {Mora}, {Muinonen}, {Osinde},
  {Pancino}, {Pauwels}, {Petit}, {Recio-Blanco}, {Richards}, {Rimoldini},
  {Robin}, {Sarro}, {Siopis}, {Smith}, {Sozzetti}, {S{\"u}veges}, {Torra}, {van
  Reeven}, {Abbas}, {Abreu Aramburu}, {Accart}, {Aerts}, {Altavilla},
  {{\'A}lvarez}, {Alvarez}, {Alves}, {Anderson}, {Andrei}, {Anglada Varela},
  {Antiche}, {Antoja}, {Arcay}, {Astraatmadja}, {Bach}, {Baker},
  {Balaguer-N{\'u}{\~n}ez}, {Balm}, {Barache}, {Barata}, {Barbato}, {Barblan},
  {Barklem}, {Barrado}, {Barros}, {Barstow}, {Bartholom{\'e} Mu{\~n}oz},
  {Bassilana}, {Becciani}, {Bellazzini}, {Berihuete}, {Bertone}, {Bianchi},
  {Bienaym{\'e}}, {Blanco-Cuaresma}, {Boch}, {Boeche}, {Bombrun}, {Borrachero},
  {Bossini}, {Bouquillon}, {Bourda}, {Bragaglia}, {Bramante}, {Breddels},
  {Bressan}, {Brouillet}, {Br{\"u}semeister}, {Brugaletta}, {Bucciarelli},
  {Burlacu}, {Busonero}, {Butkevich}, {Buzzi}, {Caffau}, {Cancelliere},
  {Cannizzaro}, {Cantat-Gaudin}, {Carballo}, {Carlucci}, {Carrasco},
  {Casamiquela}, {Castellani}, {Castro-Ginard}, {Charlot}, {Chemin},
  {Chiavassa}, {Cocozza}, {Costigan}, {Cowell}, {Crifo}, {Crosta}, {Crowley},
  {Cuypers}, {Dafonte}, {Damerdji}, {Dapergolas}, {David}, {David}, {de
  Laverny}, {De Luise}, {De March}, {de Martino}, {de Souza}, {de Torres},
  {Debosscher}, {del Pozo}, {Delbo}, {Delgado}, {Delgado}, {Di Matteo},
  {Diakite}, {Diener}, {Distefano}, {Dolding}, {Drazinos}, {Dur{\'a}n},
  {Edvardsson}, {Enke}, {Eriksson}, {Esquej}, {Eynard Bontemps}, {Fabre},
  {Fabrizio}, {Faigler}, {Falc{\~a}o}, {Farr{\`a}s Casas}, {Federici},
  {Fedorets}, {Fernique}, {Figueras}, {Filippi}, {Findeisen}, {Fonti},
  {Fraile}, {Fraser}, {Fr{\'e}zouls}, {Gai}, {Galleti}, {Garabato},
  {Garc{\'\i}a-Sedano}, {Garofalo}, {Garralda}, {Gavel}, {Gavras}, {Gerssen},
  {Geyer}, {Giacobbe}, {Gilmore}, {Girona}, {Giuffrida}, {Glass}, {Gomes},
  {Granvik}, {Gueguen}, {Guerrier}, {Guiraud}, {Guti{\'e}rrez-S{\'a}nchez},
  {Haigron}, {Hatzidimitriou}, {Hauser}, {Haywood}, {Heiter}, {Helmi}, {Heu},
  {Hilger}, {Hobbs}, {Hofmann}, {Holland}, {Huckle}, {Hypki}, {Icardi},
  {Jan{\ss}en}, {Jevardat de Fombelle}, {Jonker}, {Juh{\'a}sz}, {Julbe},
  {Karampelas}, {Kewley}, {Klar}, {Kochoska}, {Kohley}, {Kolenberg},
  {Kontizas}, {Kontizas}, {Koposov}, {Kordopatis}, {Kostrzewa-Rutkowska},
  {Koubsky}, {Lambert}, {Lanza}, {Lasne}, {Lavigne}, {Le Fustec}, {Le
  Poncin-Lafitte}, {Lebreton}, {Leccia}, {Leclerc}, {Lecoeur-Taibi},
  {Lenhardt}, {Leroux}, {Liao}, {Licata}, {Lindstr{\o}m}, {Lister}, {Livanou},
  {Lobel}, {L{\'o}pez}, {Managau}, {Mann}, {Mantelet}, {Marchal}, {Marchant},
  {Marconi}, {Marinoni}, {Marschalk{\'o}}, {Marshall}, {Martino}, {Marton},
  {Mary}, {Massari}, {Matijevi{\v{c}}}, {Mazeh}, {McMillan}, {Messina},
  {Michalik}, {Millar}, {Molina}, {Molinaro}, {Moln{\'a}r}, {Montegriffo},
  {Mor}, {Morbidelli}, {Morel}, {Morris}, {Mulone}, {Muraveva}, {Musella},
  {Nelemans}, {Nicastro}, {Noval}, {O'Mullane}, {Ord{\'e}novic},
  {Ord{\'o}{\~n}ez-Blanco}, {Osborne}, {Pagani}, {Pagano}, {Pailler},
  {Palacin}, {Palaversa}, {Panahi}, {Pawlak}, {Piersimoni}, {Pineau}, {Plachy},
  {Plum}, {Poggio}, {Poujoulet}, {Pr{\v{s}}a}, {Pulone}, {Racero}, {Ragaini},
  {Rambaux}, {Ramos-Lerate}, {Regibo}, {Reyl{\'e}}, {Riclet}, {Ripepi}, {Riva},
  {Rivard}, {Rixon}, {Roegiers}, {Roelens}, {Romero-G{\'o}mez}, {Rowell},
  {Royer}, {Ruiz-Dern}, {Sadowski}, {Sagrist{\`a} Sell{\'e}s}, {Sahlmann},
  {Salgado}, {Salguero}, {Sanna}, {Santana- Ros}, {Sarasso}, {Savietto},
  {Schultheis}, {Sciacca}, {Segol}, {Segovia}, {S{\'e}gransan}, {Shih},
  {Siltala}, {Silva}, {Smart}, {Smith}, {Solano}, {Solitro}, {Sordo}, {Soria
  Nieto}, {Souchay}, {Spagna}, {Spoto}, {Stampa}, {Steele},
  {Steidelm{\"u}ller}, {Stephenson}, {Stoev}, {Suess}, {Surdej}, {Szabados},
  {Szegedi-Elek}, {Tapiador}, {Taris}, {Tauran}, {Taylor}, {Teixeira},
  {Terrett}, {Teyssandier}, {Thuillot}, {Titarenko}, {Torra Clotet}, {Turon},
  {Ulla}, {Utrilla}, {Uzzi}, {Vaillant}, {Valentini}, {Valette}, {van Elteren},
  {Van Hemelryck}, {van Leeuwen}, {Vaschetto}, {Vecchiato}, {Veljanoski},
  {Viala}, {Vicente}, {Vogt}, {von Essen}, {Voss}, {Votruba}, {Voutsinas},
  {Walmsley}, {Weiler}, {Wertz}, {Wevers}, {Wyrzykowski}, {Yoldas},
  {{\v{Z}}erjal}, {Ziaeepour}, {Zorec}, {Zschocke}, {Zucker}, {Zurbach}, \&
  {Zwitter}}]{gaia2018}
{Gaia Collaboration}, {Brown}, A.~G.~A., {Vallenari}, A., {et~al.} 2018, \aap,
  616, A1

\bibitem[{{Gaidos} {et~al.}(2014){Gaidos}, {Mann}, {L{\'e}pine}, {Buccino},
  {James}, {Ansdell}, {Petrucci}, {Mauas}, \& {Hilton}}]{gaidos2014}
{Gaidos}, E., {Mann}, A.~W., {L{\'e}pine}, S., {et~al.} 2014, \mnras, 443, 2561

\bibitem[{{Gandolfi} {et~al.}(2018){Gandolfi}, {Barrag{\'a}n}, {Livingston},
  {Fridlund}, {Justesen}, {Redfield}, {Fossati}, {Mathur}, {Grziwa}, {Cabrera},
  {Garc{\'\i}a}, {Persson}, {Van Eylen}, {Hatzes}, {Hidalgo}, {Albrecht},
  {Bugnet}, {Cochran}, {Csizmadia}, {Deeg}, {Eigm{\"u}ller}, {Endl}, {Erikson},
  {Esposito}, {Guenther}, {Korth}, {Luque}, {Monta{\~n}es Rodr{\'\i}guez},
  {Nespral}, {Nowak}, {P{\"a}tzold}, \& {Prieto- Arranz}}]{Gandolfi2018}
{Gandolfi}, D., {Barrag{\'a}n}, O., {Livingston}, J.~H., {et~al.} 2018, \aap,
  619, L10

\bibitem[{{Gillon} {et~al.}(2013){Gillon}, {Anderson}, {Collier-Cameron},
  {Doyle}, {Fumel}, {Hellier}, {Jehin}, {Lendl}, {Maxted}, {Montalb{\'a}n},
  {Pepe}, {Pollacco}, {Queloz}, {S{\'e}gransan}, {Smith}, {Smalley},
  {Southworth}, {Triaud}, {Udry}, \& {West}}]{gillon2013}
{Gillon}, M., {Anderson}, D.~R., {Collier-Cameron}, A., {et~al.} 2013, \aap,
  552, A82

\bibitem[{{Gillon} {et~al.}(2017){Gillon}, {Triaud}, {Demory}, {Jehin}, {Agol},
  {Deck}, {Lederer}, {de Wit}, {Burdanov}, {Ingalls}, {Bolmont}, {Leconte},
  {Raymond}, {Selsis}, {Turbet}, {Barkaoui}, {Burgasser}, {Burleigh}, {Carey},
  {Chaushev}, {Copperwheat}, {Delrez}, {Fernandes}, {Holdsworth}, {Kotze}, {Van
  Grootel}, {Almleaky}, {Benkhaldoun}, {Magain}, \& {Queloz}}]{Gillon2017}
{Gillon}, M., {Triaud}, A.~H.~M.~J., {Demory}, B.-O., {et~al.} 2017, \nat, 542,
  456

\bibitem[{{Hadden} {et~al.}(2018){Hadden}, {Barclay}, {Payne}, \&
  {Holman}}]{Hadden2018}
{Hadden}, S., {Barclay}, T., {Payne}, M.~J., \& {Holman}, M.~J. 2018, arXiv
  e-prints, arXiv:1811.01970

\bibitem[{{Hadden} \& {Lithwick}(2014)}]{Hadden2014}
{Hadden}, S., \& {Lithwick}, Y. 2014, \apj, 787, 80

\bibitem[{{Hadden} \& {Lithwick}(2016)}]{Hadden2016}
---. 2016, \apj, 828, 44

\bibitem[{{Hauschildt} {et~al.}(1999){Hauschildt}, {Allard}, \&
  {Baron}}]{Hauschildt1999}
{Hauschildt}, P.~H., {Allard}, F., \& {Baron}, E. 1999, \apj, 512, 377

\bibitem[{{Henden} {et~al.}(2016){Henden}, {Templeton}, {Terrell}, {Smith},
  {Levine}, \& {Welch}}]{Henden2016}
{Henden}, A.~A., {Templeton}, M., {Terrell}, D., {et~al.} 2016, VizieR Online
  Data Catalog, 2336

\bibitem[{Hoffman \& Gelman(2014)}]{NUTS}
Hoffman, M.~D., \& Gelman, A. 2014, Journal of Machine Learning Research, 15,
  1593

\bibitem[{{Horch} {et~al.}(2011){Horch}, {van Altena}, {Howell}, {Sherry}, \&
  {Ciardi}}]{Horch(2011b)}
{Horch}, E.~P., {van Altena}, W.~F., {Howell}, S.~B., {Sherry}, W.~H., \&
  {Ciardi}, D.~R. 2011, The Astronomical Journal, 141, 180

\bibitem[{{Huang} {et~al.}(2018{\natexlab{a}}){Huang}, {Shporer}, {Dragomir},
  {Fausnaugh}, {Levine}, {Morgan}, {Nguyen}, {Ricker}, {Wall}, {Woods}, \&
  {Vanderspek}}]{Huang2018a}
{Huang}, C.~X., {Shporer}, A., {Dragomir}, D., {et~al.} 2018{\natexlab{a}},
  arXiv e-prints, arXiv:1807.11129

\bibitem[{{Huang} {et~al.}(2018{\natexlab{b}}){Huang}, {Burt}, {Vanderburg},
  {G{\"u}nther}, {Shporer}, {Dittmann}, {Winn}, {Wittenmyer}, {Sha}, {Kane},
  {Ricker}, {Vanderspek}, {Latham}, {Seager}, {Jenkins}, {Caldwell}, {Collins},
  {Guerrero}, {Smith}, {Quinn}, {Udry}, {Pepe}, {Bouchy}, {S{\'e}gransan},
  {Lovis}, {Ehrenreich}, {Marmier}, {Mayor}, {Wohler}, {Haworth}, {Morgan},
  {Fausnaugh}, {Ciardi}, {Christiansen}, {Charbonneau}, {Dragomir}, {Deming},
  {Glidden}, {Levine}, {McCullough}, {Yu}, {Narita}, {Nguyen}, {Morton},
  {Pepper}, {P{\'a}l}, {Rodriguez}, {Stassun}, {Torres}, {Sozzetti}, {Doty},
  {Christensen-Dalsgaard}, {Laughlin}, {Clampin}, {Bean}, {Buchhave}, {Bakos},
  {Sato}, {Ida}, {Kaltenegger}, {Palle}, {Sasselov}, {Butler}, {Lissauer},
  {Ge}, \& {Rinehart}}]{Huang2018}
{Huang}, C.~X., {Burt}, J., {Vanderburg}, A., {et~al.} 2018{\natexlab{b}},
  \apjl, 868, L39

\bibitem[{Hunter(2007)}]{matplotlib}
Hunter, J.~D. 2007, Computing In Science \& Engineering, 9, 90

\bibitem[{{Irwin} {et~al.}(2015){Irwin}, {Berta-Thompson}, {Charbonneau},
  {Dittmann}, {Falco}, {Newton}, \& {Nutzman}}]{irwing2015}
{Irwin}, J.~M., {Berta-Thompson}, Z.~K., {Charbonneau}, D., {et~al.} 2015, in
  {Cambridge Workshop on Cool Stars, Stellar Systems, and the Sun}, Vol.~18,
  {18th Cambridge Workshop on Cool Stars, Stellar Systems, and the Sun},
  767--772

\bibitem[{{Jenkins} {et~al.}(2017){Jenkins}, {Seader}, \&
  {Burke}}]{Jenkins2017}
{Jenkins}, J.~M., {Seader}, S., \& {Burke}, C.~J. 2017, {Kepler Data Processing
  Handbook: A Statistical Bootstrap Test}, Tech. rep.

\bibitem[{{Jenkins} {et~al.}(2016){Jenkins}, {Twicken}, {McCauliff},
  {Campbell}, {Sanderfer}, {Lung}, {Mansouri-Samani}, {Girouard}, {Tenenbaum},
  {Klaus}, {Smith}, {Caldwell}, {Chacon}, {Henze}, {Heiges}, {Latham},
  {Morgan}, {Swade}, {Rinehart}, \& {Vanderspek}}]{Jenkins2016}
{Jenkins}, J.~M., {Twicken}, J.~D., {McCauliff}, S., {et~al.} 2016, in
  \procspie, Vol. 9913, Software and Cyberinfrastructure for Astronomy IV,
  99133E

\bibitem[{{Jensen}(2013)}]{Jensen:2013}
{Jensen}, E. 2013, {Tapir: A web interface for transit/eclipse observability},
  Astrophysics Source Code Library, , , ascl:1306.007

\bibitem[{{Kane} {et~al.}(2013){Kane}, {Barclay}, \& {Gelino}}]{Kane2013}
{Kane}, S.~R., {Barclay}, T., \& {Gelino}, D.~M. 2013, \apj, 770, L20

\bibitem[{{Kane} {et~al.}(2018){Kane}, {Ceja}, {Way}, \& {Quintana}}]{Kane2018}
{Kane}, S.~R., {Ceja}, A.~Y., {Way}, M.~J., \& {Quintana}, E.~V. 2018, \apj,
  869, 46

\bibitem[{{Kane} {et~al.}(2014){Kane}, {Kopparapu}, \&
  {Domagal-Goldman}}]{Kane2014}
{Kane}, S.~R., {Kopparapu}, R.~K., \& {Domagal-Goldman}, S.~D. 2014, \apj, 794,
  L5

\bibitem[{{Kempton} {et~al.}(2018){Kempton}, {Bean}, {Louie}, {Deming}, {Koll},
  {Mansfield}, {Christiansen}, {L{\'o}pez-Morales}, {Swain}, {Zellem},
  {Ballard}, {Barclay}, {Barstow}, {Batalha}, {Beatty}, {Berta-Thompson},
  {Birkby}, {Buchhave}, {Charbonneau}, {Cowan}, {Crossfield}, {de Val-Borro},
  {Doyon}, {Dragomir}, {Gaidos}, {Heng}, {Hu}, {Kane}, {Kreidberg}, {Mallonn},
  {Morley}, {Narita}, {Nascimbeni}, {Pall{\'e}}, {Quintana}, {Rauscher},
  {Seager}, {Shkolnik}, {Sing}, {Sozzetti}, {Stassun}, {Valenti}, \& {von
  Essen}}]{Kempton2018}
{Kempton}, E.~M.-R., {Bean}, J.~L., {Louie}, D.~R., {et~al.} 2018, \pasp, 130,
  114401

\bibitem[{{Kipping}(2013{\natexlab{a}})}]{exoplanet:kipping13}
{Kipping}, D.~M. 2013{\natexlab{a}}, \mnras, 435, 2152

\bibitem[{{Kipping}(2013{\natexlab{b}})}]{Kipping2013}
---. 2013{\natexlab{b}}, \mnras, 434, L51

\bibitem[{Kluyver {et~al.}(2016)Kluyver, Ragan-Kelley, P{\'e}rez, Granger,
  Bussonnier, Frederic, Kelley, Hamrick, Grout, Corlay, Ivanov, Avila, Abdalla,
  Willing, \& development~team [Unknown]}]{jupyer}
Kluyver, T., Ragan-Kelley, B., P{\'e}rez, F., {et~al.} 2016, in Positioning and
  Power in Academic Publishing: Players, Agents and Agendas, ed. F.~Loizides \&
  B.~Scmidt (IOS Press), 87--90

\bibitem[{{Kostov} {et~al.}(2019){Kostov}, {Mullally}, {Quintana}, {Coughlin},
  {Mullally}, {Barclay}, {Colon}, {Schlieder}, {Barentsen}, \&
  {Burke}}]{Kostov2019}
{Kostov}, V.~B., {Mullally}, S.~E., {Quintana}, E.~V., {et~al.} 2019, arXiv
  e-prints, arXiv:1901.07459

\bibitem[{{Li} {et~al.}(2019){Li}, {Tenenbaum}, {Twicken}, {Burke}, {Jenkins},
  {Quintana}, {Rowe}, \& {Seader}}]{Li2019}
{Li}, J., {Tenenbaum}, P., {Twicken}, J.~D., {et~al.} 2019, \pasp, 131, 024506

\bibitem[{{Lightkurve Collaboration} {et~al.}(2018){Lightkurve Collaboration},
  {Cardoso}, {Hedges}, {Gully-Santiago}, {Saunders}, {Cody}, {Barclay}, {Hall},
  {Sagear}, {Turtelboom}, {Zhang}, {Tzanidakis}, {Mighell}, {Coughlin}, {Bell},
  {Berta- Thompson}, {Williams}, {Dotson}, \& {Barentsen}}]{lightkurve}
{Lightkurve Collaboration}, {Cardoso}, J. V. d.~M., {Hedges}, C., {et~al.}
  2018, {Lightkurve: Kepler and TESS time series analysis in Python}, , ,
  ascl:1812.013

\bibitem[{{Lissauer} {et~al.}(2011){Lissauer}, {Ragozzine}, {Fabrycky},
  {Steffen}, {Ford}, {Jenkins}, {Shporer}, {Holman}, {Rowe}, {Quintana},
  {Batalha}, {Borucki}, {Bryson}, {Caldwell}, {Carter}, {Ciardi}, {Dunham},
  {Fortney}, {Gautier}, {Howell}, {Koch}, {Latham}, {Marcy}, {Morehead}, \&
  {Sasselov}}]{Lissauer2011}
{Lissauer}, J.~J., {Ragozzine}, D., {Fabrycky}, D.~C., {et~al.} 2011, \apjs,
  197, 8

\bibitem[{{Lissauer} {et~al.}(2012){Lissauer}, {Marcy}, {Rowe}, {Bryson},
  {Adams}, {Buchhave}, {Ciardi}, {Cochran}, {Fabrycky}, {Ford}, {Fressin},
  {Geary}, {Gilliland}, {Holman}, {Howell}, {Jenkins}, {Kinemuchi}, {Koch},
  {Morehead}, {Ragozzine}, {Seader}, {Tanenbaum}, {Torres}, \&
  {Twicken}}]{Lissauer2012}
{Lissauer}, J.~J., {Marcy}, G.~W., {Rowe}, J.~F., {et~al.} 2012, \apj, 750, 112

\bibitem[{{Lithwick} {et~al.}(2012){Lithwick}, {Xie}, \& {Wu}}]{Lithwick2012}
{Lithwick}, Y., {Xie}, J., \& {Wu}, Y. 2012, \apj, 761, 122

\bibitem[{{Livingston} {et~al.}(2018){Livingston}, {Crossfield}, {Petigura},
  {Gonzales}, {Ciardi}, {Beichman}, {Christiansen}, {Dressing}, {Henning},
  {Howard}, {Isaacson}, {Fulton}, {Kosiarek}, {Schlieder}, {Sinukoff}, \&
  {Tamura}}]{Livingston2018}
{Livingston}, J.~H., {Crossfield}, I. J.~M., {Petigura}, E.~A., {et~al.} 2018,
  \aj, 156, 277

\bibitem[{{Louie} {et~al.}(2018){Louie}, {Deming}, {Albert}, {Bouma}, {Bean},
  \& {Lopez-Morales}}]{Louie2018}
{Louie}, D.~R., {Deming}, D., {Albert}, L., {et~al.} 2018, Publications of the
  Astronomical Society of the Pacific, 130, 044401

\bibitem[{{Luger} {et~al.}(2018){Luger}, {Agol}, {Foreman-Mackey}, {Fleming},
  {Lustig-Yaeger}, \& {Deitrick}}]{exoplanet:luger18}
{Luger}, R., {Agol}, E., {Foreman-Mackey}, D., {et~al.} 2018, ArXiv e-prints,
  1810.06559

\bibitem[{{Luger} {et~al.}(2016){Luger}, {Agol}, {Kruse}, {Barnes}, {Becker},
  {Foreman-Mackey}, \& {Deming}}]{Luger2016}
{Luger}, R., {Agol}, E., {Kruse}, E., {et~al.} 2016, \aj, 152, 100

\bibitem[{{Mandel} \& {Agol}(2002)}]{mandelagol02}
{Mandel}, K., \& {Agol}, E. 2002, \apjl, 580, L171

\bibitem[{{Mann} {et~al.}(2015){Mann}, {Feiden}, {Gaidos}, {Boyajian}, \& {von
  Braun}}]{Mann2015}
{Mann}, A.~W., {Feiden}, G.~A., {Gaidos}, E., {Boyajian}, T., \& {von Braun},
  K. 2015, \apj, 804, 64

\bibitem[{{Mayor} {et~al.}(2009){Mayor}, {Bonfils}, {Forveille}, {Delfosse},
  {Udry}, {Bertaux}, {Beust}, {Bouchy}, {Lovis}, {Pepe}, {Perrier}, {Queloz},
  \& {Santos}}]{Mayor2009}
{Mayor}, M., {Bonfils}, X., {Forveille}, T., {et~al.} 2009, \aap, 507, 487

\bibitem[{McKinney(2010)}]{pandas}
McKinney, W. 2010, in Proceedings of the 9th Python in Science Conference, ed.
  S.~van~der Walt \& J.~Millman, 51 -- 56

\bibitem[{{Ment} {et~al.}(2018){Ment}, {Dittmann}, {Astudillo-Defru},
  {Charbonneau}, {Irwin}, {Bonfils}, {Murgas}, {Almenara}, {Forveille}, {Agol},
  {Ballard}, {Berta-Thompson}, {Bouchy}, {Cloutier}, {Delfosse}, {Doyon},
  {Dressing}, {Esquerdo}, {Haywood}, {Kipping}, {Latham}, {Lovis}, {Newton},
  {Pepe}, {Rodriguez}, {Santos}, {Tan}, {Udry}, {Winters}, \&
  {W{\"u}nsche}}]{Ment(2018a)}
{Ment}, K., {Dittmann}, J.~A., {Astudillo-Defru}, N., {et~al.} 2018, ArXiv
  e-prints, arXiv:1808.00485

\bibitem[{{Morley} {et~al.}(2017){Morley}, {Kreidberg}, {Rustamkulov},
  {Robinson}, \& {Fortney}}]{Morley2017}
{Morley}, C.~V., {Kreidberg}, L., {Rustamkulov}, Z., {Robinson}, T., \&
  {Fortney}, J.~J. 2017, \apj, 850, 121

\bibitem[{{Morton} {et~al.}(2016){Morton}, {Bryson}, {Coughlin}, {Rowe},
  {Ravichandran}, {Petigura}, {Haas}, \& {Batalha}}]{Morton2016}
{Morton}, T.~D., {Bryson}, S.~T., {Coughlin}, J.~L., {et~al.} 2016, \apj, 822,
  86

\bibitem[{{Muirhead} {et~al.}(2018){Muirhead}, {Dressing}, {Mann},
  {Rojas-Ayala}, {L{\'e}pine}, {Paegert}, {De Lee}, \&
  {Oelkers}}]{Muirhead2018}
{Muirhead}, P.~S., {Dressing}, C.~D., {Mann}, A.~W., {et~al.} 2018, \aj, 155,
  180

\bibitem[{{Muirhead} {et~al.}(2015){Muirhead}, {Mann}, {Vanderburg}, {Morton},
  {Kraus}, {Ireland}, {Swift}, {Feiden}, {Gaidos}, \& {Gazak}}]{Muirhead2015}
{Muirhead}, P.~S., {Mann}, A.~W., {Vanderburg}, A., {et~al.} 2015, \apj, 801,
  18

\bibitem[{{Mukai} \& {Barclay}(2017)}]{Mukai2017}
{Mukai}, K., \& {Barclay}, T. 2017, {tvguide: A tool for determining whether
  stars and galaxies are observable by TESS.}, v.1.0.1,  Zenodo,
  doi:10.5281/zenodo.823357

\bibitem[{{Newton} {et~al.}(2015){Newton}, {Charbonneau}, {Irwin}, \&
  {Mann}}]{Newton2015}
{Newton}, E.~R., {Charbonneau}, D., {Irwin}, J., \& {Mann}, A.~W. 2015, \apj,
  800, 85

\bibitem[{{Nielsen} {et~al.}(2018){Nielsen}, {Bouchy}, {Turner}, {Giles},
  {Suarez Mascareno}, {Lovis}, {Marmier}, {Pepe}, {Segransan}, {Udry}, {Otegi},
  {Ottoni}, {Stalport}, {Ricker}, {Vanderspek}, {Latham}, {Seager}, {Winn},
  {Jenkins}, {Wittenmyer}, {Kane}, {Cartwright}, {Collins}, {Francis},
  {Guerrero}, {Huang}, {Matthews}, {Pepper}, {Rose}, {Villasenor}, {Wohler},
  {Stassun}, {Crossfield}, {Howell}, {Ciardi}, {Gonzales}, {Matson},
  {Beighman}, \& {Schlieder}}]{Nielsen2018}
{Nielsen}, L.~D., {Bouchy}, F., {Turner}, O., {et~al.} 2018, arXiv e-prints,
  arXiv:1811.01882

\bibitem[{Oliphant(2007)}]{scipy}
Oliphant, T.~E. 2007, Computing in Science Engineering, 9, 10

\bibitem[{{Pecaut} \& {Mamajek}(2013)}]{Pecaut(2013)}
{Pecaut}, M.~J., \& {Mamajek}, E.~E. 2013, \apjs, 208, 9

\bibitem[{Perez \& Granger(2007)}]{ipython}
Perez, F., \& Granger, B.~E. 2007, Computing in Science Engineering, 9, 21

\bibitem[{{Quinn} {et~al.}(2019){Quinn}, {Becker}, {Rodriguez}, {Hadden},
  {Huang}, {Morton}, {Adams}, {Armstrong}, {Eastman}, {Horner}, {Kane},
  {Lissauer}, {Twicken}, {Vanderburg}, {Wittenmyer}, {Ricker}, {Vanderspek},
  {Latham}, {Seager}, {Winn}, {Jenkins}, {Agol}, {Barkaoui}, {Beichman},
  {Bouchy}, {Bouma}, {Burdanov}, {Campbell}, {Carlino}, {Cartwright},
  {Charbonneau}, {Christiansen}, {Ciardi}, {Collins}, {Collins}, {Conti},
  {Crossfield}, {Daylan}, {Dittmann}, {Doty}, {Dragomir}, {Ducrot}, {Gillon},
  {Glidden}, {Goeke}, {Gonzales}, {He{\l}miniak}, {Horch}, {Howell}, {Jehin},
  {Jensen}, {Kielkopf}, {Kristiansen}, {Law}, {Mann}, {Marmier}, {Matson},
  {Matthews}, {Mazeh}, {Mori}, {Murgas}, {Murray}, {Narita}, {Nielsen},
  {Ottoni}, {Palle}, {Paw{\l}aszek}, {Pepe}, {de Leon}, {Pozuelos}, {Relles},
  {Schlieder}, {Sebastian}, {S{\'e}gransan}, {Shporer}, {Stassun}, {Tamura},
  {Tokovinin}, {Udry}, {Waite}, \& {Ziegler}}]{Quinn2019}
{Quinn}, S.~N., {Becker}, J.~C., {Rodriguez}, J.~E., {et~al.} 2019, arXiv
  e-prints, arXiv:1901.09092

\bibitem[{{Quintana} {et~al.}(2014){Quintana}, {Barclay}, {Raymond}, {Rowe},
  {Bolmont}, {Caldwell}, {Howell}, {Kane}, {Huber}, {Crepp}, {Lissauer},
  {Ciardi}, {Coughlin}, {Everett}, {Henze}, {Horch}, {Isaacson}, {Ford},
  {Adams}, {Still}, {Hunter}, {Quarles}, \& {Selsis}}]{Quintana2014}
{Quintana}, E.~V., {Barclay}, T., {Raymond}, S.~N., {et~al.} 2014, Science,
  344, 277

\bibitem[{{Rein} \& {Spiegel}(2015)}]{Rein2015}
{Rein}, H., \& {Spiegel}, D.~S. 2015, \mnras, 446, 1424

\bibitem[{{Ricker} {et~al.}(2015){Ricker}, {Winn}, {Vanderspek}, {Latham},
  {Bakos}, {Bean}, {Berta-Thompson}, {Brown}, {Buchhave}, {Butler}, {Butler},
  {Chaplin}, {Charbonneau}, {Christensen-Dalsgaard}, {Clampin}, {Deming},
  {Doty}, {De Lee}, {Dressing}, {Dunham}, {Endl}, {Fressin}, {Ge}, {Henning},
  {Holman}, {Howard}, {Ida}, {Jenkins}, {Jernigan}, {Johnson}, {Kaltenegger},
  {Kawai}, {Kjeldsen}, {Laughlin}, {Levine}, {Lin}, {Lissauer}, {MacQueen},
  {Marcy}, {McCullough}, {Morton}, {Narita}, {Paegert}, {Palle}, {Pepe},
  {Pepper}, {Quirrenbach}, {Rinehart}, {Sasselov}, {Sato}, {Seager},
  {Sozzetti}, {Stassun}, {Sullivan}, {Szentgyorgyi}, {Torres}, {Udry}, \&
  {Villasenor}}]{Ricker2015}
{Ricker}, G.~R., {Winn}, J.~N., {Vanderspek}, R., {et~al.} 2015, Journal of
  Astronomical Telescopes, Instruments, and Systems, 1, 014003

\bibitem[{{Rodriguez} {et~al.}(2019){Rodriguez}, {Quinn}, {Huang},
  {Vanderburg}, {Penev}, {Brahm}, {Jord{\'a}n}, {Ikwut-Ukwa}, {Tsirulik},
  {Latham}, {Stassun}, {Shporer}, {Ziegler}, {Matthews}, {Eastman}, {Gaudi},
  {Collins}, {Guerrero}, {Relles}, {Barclay}, {Batalha}, {Berlind}, {Bieryla},
  {Bouma}, {Boyd}, {Burt}, {Calkins}, {Christiansen}, {Ciardi}, {Col{\'o}n},
  {Conti}, {Crossfield}, {Daylan}, {Dittmann}, {Dragomir}, {Dynes}, {Espinoza},
  {Esquerdo}, {Essack}, {Soto}, {Glidden}, {G{\"u}nther}, {Henning}, {Jenkins},
  {Kielkopf}, {Krishnamurthy}, {Law}, {Levine}, {Lewin}, {Mann}, {Morgan},
  {Morris}, {Oelkers}, {Paegert}, {Pepper}, {Quintana}, {Ricker}, {Rowden},
  {Seager}, {Sarkis}, {Schlieder}, {Sha}, {Tokovinin}, {Torres}, {Vanderspek},
  {Villanueva Jr.}, {Villase{\~n}or}, {Winn}, {Wohler}, {Wong}, {Yahalomi},
  {Yu}, {Zhan}, \& {Zhou}}]{Rodriguez2019}
{Rodriguez}, J.~E., {Quinn}, S.~N., {Huang}, C.~X., {et~al.} 2019, arXiv
  e-prints, arXiv:1901.09950

\bibitem[{{Rogers}(2015)}]{Rogers2015}
{Rogers}, L.~A. 2015, \apj, 801, 41

\bibitem[{{Rowe} {et~al.}(2014){Rowe}, {Bryson}, {Marcy}, {Lissauer},
  {Jontof-Hutter}, {Mullally}, {Gilliland}, {Issacson}, {Ford}, {Howell},
  {Borucki}, {Haas}, {Huber}, {Steffen}, {Thompson}, {Quintana}, {Barclay},
  {Still}, {Fortney}, {Gautier}, {Hunter}, {Caldwell}, {Ciardi}, {Devore},
  {Cochran}, {Jenkins}, {Agol}, {Carter}, \& {Geary}}]{Rowe2014}
{Rowe}, J.~F., {Bryson}, S.~T., {Marcy}, G.~W., {et~al.} 2014, \apj, 784, 45

\bibitem[{Salvatier {et~al.}(2016)Salvatier, Wiecki, \&
  Fonnesbeck}]{exoplanet:pymc3}
Salvatier, J., Wiecki, T.~V., \& Fonnesbeck, C. 2016, PeerJ Computer Science,
  2, e55

\bibitem[{{Shporer} {et~al.}(2018){Shporer}, {Wong}, {Huang}, {Line},
  {Stassun}, {Fetherolf}, {Kane}, {Ricker}, {Latham}, {Seager}, {Winn},
  {Jenkins}, {Glidden}, {Berta-Thompson}, {Ting}, {Li}, \&
  {Haworth}}]{Shporer2018}
{Shporer}, A., {Wong}, I., {Huang}, C.~X., {et~al.} 2018, arXiv e-prints,
  arXiv:1811.06020

\bibitem[{{Simcoe} {et~al.}(2008){Simcoe}, {Burgasser}, {Bernstein}, {Bigelow},
  {Fishner}, {Forrest}, {McMurtry}, {Pipher}, {Schechter}, \&
  {Smith}}]{simcoe2008}
{Simcoe}, R.~A., {Burgasser}, A.~J., {Bernstein}, R.~A., {et~al.} 2008, in
  Society of Photo-Optical Instrumentation Engineers (SPIE) Conference Series,
  Vol. 7014, Ground-based and Airborne Instrumentation for Astronomy II, 70140U

\bibitem[{{Skrutskie} {et~al.}(2006){Skrutskie}, {Cutri}, {Stiening},
  {Weinberg}, {Schneider}, {Carpenter}, {Beichman}, {Capps}, {Chester},
  {Elias}, {Huchra}, {Liebert}, {Lonsdale}, {Monet}, {Price}, {Seitzer},
  {Jarrett}, {Kirkpatrick}, {Gizis}, {Howard}, {Evans}, {Fowler}, {Fullmer},
  {Hurt}, {Light}, {Kopan}, {Marsh}, {McCallon}, {Tam}, {Van Dyk}, \&
  {Wheelock}}]{Skrutskie2006}
{Skrutskie}, M.~F., {Cutri}, R.~M., {Stiening}, R., {et~al.} 2006, \aj, 131,
  1163

\bibitem[{{Smith} {et~al.}(2012){Smith}, {Stumpe}, {Van Cleve}, {Jenkins},
  {Barclay}, {Fanelli}, {Girouard}, {Kolodziejczak}, {McCauliff}, {Morris}, \&
  {Twicken}}]{Smith2012}
{Smith}, J.~C., {Stumpe}, M.~C., {Van Cleve}, J.~E., {et~al.} 2012, \pasp, 124,
  1000

\bibitem[{{Stassun} {et~al.}(2017){Stassun}, {Collins}, \&
  {Gaudi}}]{Stassun:2017}
{Stassun}, K.~G., {Collins}, K.~A., \& {Gaudi}, B.~S. 2017, \aj, 153, 136

\bibitem[{{Stassun} \& {Torres}(2016)}]{StassunTorres:2016}
{Stassun}, K.~G., \& {Torres}, G. 2016, \aj, 152, 180

\bibitem[{{Stassun} \& {Torres}(2018)}]{StassunTorres:2018}
---. 2018, \apj, 862, 61

\bibitem[{{Stassun} {et~al.}(2018){Stassun}, {Oelkers}, {Pepper}, {Paegert},
  {De Lee}, {Torres}, {Latham}, {Charpinet}, {Dressing}, {Huber}, {Kane},
  {L{\'e}pine}, {Mann}, {Muirhead}, {Rojas-Ayala}, {Silvotti}, {Fleming},
  {Levine}, \& {Plavchan}}]{Stassun_CTL_2018}
{Stassun}, K.~G., {Oelkers}, R.~J., {Pepper}, J., {et~al.} 2018, \aj, 156, 102

\bibitem[{{Stumpe} {et~al.}(2014){Stumpe}, {Smith}, {Catanzarite}, {Van Cleve},
  {Jenkins}, {Twicken}, \& {Girouard}}]{Stumpe2014}
{Stumpe}, M.~C., {Smith}, J.~C., {Catanzarite}, J.~H., {et~al.} 2014, \pasp,
  126, 100

\bibitem[{{Theano Development Team}(2016)}]{exoplanet:theano}
{Theano Development Team}. 2016, arXiv e-prints, abs/1605.02688

\bibitem[{{Tokovinin} {et~al.}(2013){Tokovinin}, {Fischer}, {Bonati},
  {Giguere}, {Moore}, {Schwab}, {Spronck}, \& {Szymkowiak}}]{Tokovinin(2013)}
{Tokovinin}, A., {Fischer}, D.~A., {Bonati}, M., {et~al.} 2013, \pasp, 125,
  1336

\bibitem[{{Twicken} {et~al.}(2018){Twicken}, {Catanzarite}, {Clarke},
  {Girouard}, {Jenkins}, {Klaus}, {Li}, {McCauliff}, {Seader}, {Tenenbaum},
  {Wohler}, {Bryson}, {Burke}, {Caldwell}, {Haas}, {Henze}, \&
  {Sanderfer}}]{Twicken2018}
{Twicken}, J.~D., {Catanzarite}, J.~H., {Clarke}, B.~D., {et~al.} 2018, \pasp,
  130, 064502

\bibitem[{van~der Walt {et~al.}(2011)van~der Walt, Colbert, \&
  Varoquaux}]{numpy}
van~der Walt, S., Colbert, S.~C., \& Varoquaux, G. 2011, Computing in Science
  Engineering, 13, 22

\bibitem[{{Vanderspek} {et~al.}(2018){Vanderspek}, {Huang}, {Vanderburg},
  {Ricker}, {Latham}, {Seager}, {Winn}, {Jenkins}, {Burt}, {Dittmann},
  {Newton}, {Quinn}, {Shporer}, {Charbonneau}, {Irwin}, {Ment}, {Winters},
  {Collins}, {Evans}, {Gan}, {Hart}, {Jensen}, {Kielkopf}, {Mao}, {Waalkes},
  {Bouchy}, {Marmier}, {Nielsen}, {Ottoni}, {Pepe}, {S{\'e}gransan}, {Udry},
  {Henry}, {Paredes}, {James}, {Hinojosa}, {Silverstein}, {Palle},
  {Berta-Thompson}, {Davies}, {Fausnaugh}, {Glidden}, {Pepper}, {Morgan},
  {Rose}, {Twicken}, {Villase{\~n}or}, \& {the TESS team}}]{Vanderspek2018}
{Vanderspek}, R., {Huang}, C.~X., {Vanderburg}, A., {et~al.} 2018, arXiv
  e-prints, arXiv:1809.07242

\bibitem[{{Wang} {et~al.}(2018){Wang}, {Jones}, {Shporer}, {Fulton}, {Paredes},
  {Trifonov}, {Kossakowski}, {Eastman}, {Gunther}, {Huang}, {Millholland},
  {Seligman}, {Fischer}, {Brahm}, {Wang}, {Cruz}, {James}, {Addison}, {Henry},
  {Liang}, {Davis}, {Tronsgaard}, {Worku}, {Brewer}, {Kurster}, {Beichman},
  {Bieryla}, {Brown}, {Christiansen}, {Ciardi}, {Collins}, {Esquerdo},
  {Howard}, {Isaacson}, {Latham}, {Mazeh}, {Petigura}, {Quinn}, {Shahaf},
  {Siverd}, {Ricker}, {Vanderspek}, {Seager}, {Winn}, {Jenkins}, {Boyd},
  {Furesz}, {Henze}, {Levine}, {Morris}, {Paegert}, {Stassun}, {Ting}, {Vezie},
  \& {Laughlin}}]{Wang2018}
{Wang}, S., {Jones}, M., {Shporer}, A., {et~al.} 2018, arXiv e-prints,
  arXiv:1810.02341

\bibitem[{{Way} {et~al.}(2016){Way}, {Del Genio}, {Kiang}, {Sohl}, {Grinspoon},
  {Aleinov}, {Kelley}, \& {Clune}}]{Way2016}
{Way}, M.~J., {Del Genio}, A.~D., {Kiang}, N.~Y., {et~al.} 2016, \grl, 43, 8376

\bibitem[{{Welsh} {et~al.}(2011){Welsh}, {Orosz}, {Aerts}, {Brown},
  {Brugamyer}, {Cochran}, {Gilliland}, {Guzik}, {Kurtz}, {Latham}, {Marcy},
  {Quinn}, {Zima}, {Allen}, {Batalha}, {Bryson}, {Buchhave}, {Caldwell},
  {Gautier}, {Howell}, {Kinemuchi}, {Ibrahim}, {Isaacson}, {Jenkins}, {Prsa},
  {Still}, {Street}, {Wohler}, {Koch}, \& {Borucki}}]{Welsh2011}
{Welsh}, W.~F., {Orosz}, J.~A., {Aerts}, C., {et~al.} 2011, \apjs, 197, 4

\bibitem[{{Winters} {et~al.}(2015){Winters}, {Henry}, {Lurie}, {Hambly}, {Jao},
  {Bartlett}, {Boyd}, {Dieterich}, {Finch}, {Hosey}, {Ianna}, {Riedel},
  {Slatten}, \& {Subasavage}}]{Winters2015}
{Winters}, J.~G., {Henry}, T.~J., {Lurie}, J.~C., {et~al.} 2015, \aj, 149, 5

\bibitem[{{Winters} {et~al.}(2018){Winters}, {Irwin}, {Newton}, {Charbonneau},
  {Latham}, {Han}, {Muirhead}, {Berlind}, {Calkins}, \&
  {Esquerdo}}]{Winters(2018)}
{Winters}, J.~G., {Irwin}, J., {Newton}, E.~R., {et~al.} 2018, \aj, 155, 125

\bibitem[{{Winters} {et~al.}(2019){Winters}, {Henry}, {Jao}, {Subasavage},
  {Chatelain}, {Slatten}, {Riedel}, {Silverstein}, \& {Payne}}]{Winters2019}
{Winters}, J.~G., {Henry}, T.~J., {Jao}, W.-C., {et~al.} 2019, arXiv e-prints,
  arXiv:1901.06364

\bibitem[{{Zacharias} {et~al.}(2017){Zacharias}, {Finch}, \&
  {Frouard}}]{Zacharias2017}
{Zacharias}, N., {Finch}, C., \& {Frouard}, J. 2017, \aj, 153, 166

\bibitem[{{Ziegler} {et~al.}(2018){Ziegler}, {Law}, {Baranec}, {Riddle},
  {Duev}, {Howard}, {Jensen-Clem}, {Kulkarni}, {Morton}, \&
  {Salama}}]{Ziegler2018}
{Ziegler}, C., {Law}, N.~M., {Baranec}, C., {et~al.} 2018, \aj, 155, 161

\end{thebibliography}
\bibliographystyle{aasjournal}

\end{document}